\documentclass[fleqn,usenatbib]{mnras}

\usepackage{newtxtext,newtxmath}
\usepackage[T1]{fontenc}
\usepackage{ae,aecompl}
\usepackage{graphicx}
\usepackage{lscape}
\usepackage{longtable}

\defcitealias{gon2015}{Paper~I}
\defcitealias{dor2016a}{Paper~II}
\defcitealias{dor2016b}{Paper~III}

\title[RSG population in the Perseus arm]{The red supergiant population in the Perseus arm}

\author[Dorda et al.]{
R. Dorda,$^{1}$\thanks{E-mail: ricardo.dorda@ua.es}
I. Negueruela$^{1}$ and
C. Gonz\'alez-Fern\'andez$^{2}$
\\
$^{1}$Departamento de F\'{\i}sica, Ingenier\'{\i}a de Sistemas y Teor\'{\i}a de la Se\~nal, Universidad de Alicante, Carretera de San Vicente s/n,\\
San Vicente del Raspeig E03690, Alicante, Spain\\
$^{2}$Institute of Astronomy, University of Cambridge, Madingley Road, Cambridge CB3 0HA, United Kingdom\\
}

\date{Accepted XXX. Received YYY; in original form ZZZ}

\pubyear{2017}

\begin{document}
\label{firstpage}
\pagerange{\pageref{firstpage}--\pageref{lastpage}}
\maketitle

\begin{abstract}
We present a new catalogue of cool supergiants in a section of the Perseus arm, most of which had not been previously identified. To generate it, we have used a set of well-defined photometric criteria to select a large number of candidates (637) that were later observed at intermediate resolution in the the Infrared Calcium Triplet spectral range, using a long-slit spectrograph. To separate red supergiants from luminous red giants, we used a statistical method, developed in previous works and improved in the present paper. We present a method to assign probabilities of being a red supergiant to a given spectrum and use the properties of a population to generate clean samples, without contamination from lower-luminosity stars. We compare our identification with a classification done using classical criteria and discuss their respective efficiencies and contaminations as identification methods. We confirm that our method is as efficient at finding supergiants as the best classical methods, but with a far lower contamination by red giants than any other method. The result is a catalogue with 197 cool supergiants, 191 of which did not appear in previous lists of red supergiants. This is the largest coherent catalogue of cool supergiants in the Galaxy.
\end{abstract}


\section{Introduction}

The section of the Perseus arm visible from the Northern hemisphere is a Galactic region rich in young stars, with many OB associations and young open clusters \citep{hum1978}. Given its proximity to the Sun \citep[with typical distances ranging between 3~kpc at $l\sim100\degr$ to 2~kpc at $l\sim140\degr$;][]{choi2014}, it offers important advantages for the study of stellar populations over other Galactic regions. Located towards the outskirts of the Milky Way, it presents a moderately low reddening, which makes young blue stars easily accessible. In consequence, the high-mass population in Perseus has been widely studied for decades \citep[e.g.][]{hum1978}. Among the young stars in Perseus, there are also many red supergiant (RSG) stars \citep[$>70$;][]{hum1978,lev2005}. These stars possess moderately-high mass ($\sim10$ to $\sim40\:$M$_{\odot}$), high luminosity \citep[$\log(L/L_{\sun})\sim4.5$\,--\,5.8;][]{hum1979}, low temperature\footnote{The temperature scale of RSGs is still an open question. Over the last decade, different works (\citealt{lev2007}, \citealt{dav2013}, and Tabernero et al. submitted) have reported quite different temperature ranges. In all cases, though, the effective temperatures of these stars are well below 4\,500\:K.}, and late (K or M) spectral type (SpT). Although they have evolved off the main sequence, RSGs are still young stars \citep[with ages between $\sim8$ and $\sim25\:$Ma;][]{eks2013}. In consequence, they are associated to regions of recent stellar formation.

The correct characterisation of the RSG phase plays a major role in the understanding of the evolution and final fate of high-mass stars \citep[e.g.][]{eks2013}. Despite this pivotal position, there are still many critical questions about them that remain without definitive answers; among them, the definition of a temperature scale and its relation with luminosity, as discussed in \citet[from now on Paper~II]{dor2016a}. To bring some light to these questions, we started an ambitious observational programme on RSGs, aimed at characterising their properties by using statistically significant samples. In \citet[from now on, Paper~I]{gon2015} we presented the largest spectroscopic sample to date of cool supergiants\footnote{"Cool supergiants" is a denomination that includes all red and some yellow supergiants. In \citetalias{gon2015}, we showed that G-type SGs in the SMC (and presumably other low-metallicity environments) are part of the the same population as RSGs. This is not the case in the Milky Way, but a few luminous G-type supergiants are part of our calibration sample. Thus, we use the term CSG to make reference to the present sample. Despite this, the term RSG is used in many cases, in reference to the samples of K and M supergiants studied in previous works \citep[e.g.\ ][]{hum1978,lev2005}.} (CSGs) from the Magellanic Clouds (MCs). By combining this large sample with an important number of well-characterised Milky Way RSGs, in \citetalias{dor2016a} we could present firm statistical confirmation of a correlation between SpT and temperature, or the relation between SpT, luminosity, and mass loss. Taking advantage of this sample, in \citet[Paper~III]{dor2016b} we developed an automated method for the identification of CSGs using the atomic and molecular features in the spectral range around the infrared Calcium Triplet (CaT). Finally, Tabernero et al. (submitted) have calculated the effective temperatures for the sample in \citetalias{gon2015} and studied the temperature scales of the RSGs from the MCs.

The present work is the next step in our study of CSGs. After analysing the CSG population from the MCs, we extend our study to the Milky Way population of CSGs. As many of the properties of a given CSG population (e.g.\ its typical SpT and temperatures) depend on its metalliticy \citep[][]{eli1985}, we selected a specific region of the Galaxy where we can expect rather uniform (typically solar) metallicities: the section of the Perseus arm between $l=97\degr$ and~$150\degr$, with Galactocentric distances in the $\sim8$ to $10\:$kpc range. This region was chosen because of the many RSGs that were previously known and well characterised, but also because its CSGs have very low apparent magnitudes and can be observed efficiently with long-slit spectrographs. A systematic search for CSGs in an area that is considered well studied allows a good estimation of the incompleteness of previous samples. Moreover, as the extinction towards the Perseus arm is relatively low, its blue population is well known. In consequence, the relation between OB stars and CSGs can be studied. This analysis would be specially interesting, because many clusters and OB associations in the Perseus arm have total masses and ages coherent with the presence of CSGs.

In this paper, we apply the methods developed in \citetalias{dor2016b} to a sample of candidate RSGs from the Perseus arm, to test their reliability and obtain a statistically significant sample of CSGs in the area. In addition, we develop a method to compute the likelihood that a given star is indeed a supergiant and estimate the reliability of our identification. We also study some basic properties of the CSG population at solar metallicities, such as its SpT distribution and its relation with the luminosity class (LC). In a future work, we will carry out a deeper study of the astrophysical properties of the the CSG sample found here, analysing its spatial and kinematic distributions, as well as its connection to the known population of high-mass stars close to the main sequence.

\section{Observations and measurements}
\label{Per_arm}

\subsection{Target selection}
\label{targ_sel}
To identify RSG candidates in the Perseus arm, we performed a comprehensive photometric search in the Galactic Plane ($b=+6\degr$ to $-6\degr$, and $l=97\degr$ to~$150\degr$). We used as a guide the works of \cite{hum1970,hum1978}. The selection is the result of the following steps:

\begin{itemize}
\item From \cite{hum1978} we selected those regions with detected RSGs and distance moduli (DM) coherent with being part of the Perseus arm.
\item Using these DM, along with the measured $A_{\mathrm{V}}$, we selected from 2MASS those sources with $K$~band magnitudes bright enough to be a RSG, assuming a lower limit for their intrinsic brightness at $M_{K}=-5$. This may seem a very low limit, as for example in \citetalias{gon2015} there are no CSGs below $M_\mathrm{K}\sim-7$, but it allows for large errors in DM and/or extinction while keeping the CSG candidate sample as complete as possible. This step gets rid of most of the foreground and background undesired populations, as the expected density profile of the Galaxy along this line of sight allows us to adopt a low luminosity threshold without risking too much contamination (more distant RSGs will likely be also included, but they are expected to be rare in the outer reaches of the Galaxy and will be of interest for future studies). This leaves only nearby dwarfs and giants with types later than M3 as main interlopers.
\item The filtered sample was then cross-correlated with well known catalogues of optical photometry, such as USNO-B1 \citep{mon2003} and UCAC3 \citep{zac2010}, obtaining $I$~band magnitudes and proper motions. Candidates are required to have $(I-K_{\textrm{S}})_{0} > 2$ (roughly, the colour of a K0 star) and proper motions similar to those of the blue and red supergiants already known in the field. This step cleans the sample of most of the foreground stars, as they have higher motions.
\item The remaining catalogue was then submitted to SIMBAD and all the stars with confirmed SpTs were removed, although we kept $51$ previously-studied RSGs, for a number of reasons: check spectral variations, test the efficiency of our methods and provide a comparison sample. In fact, 43 of these objects with reliable SpT or marked as MK standards were used for the calibration sample used in \citetalias{dor2016b}. In consequence, we are not considering these 43 SGs as part of the test sample, but we include them to calculate the efficiency of the photometric selection in Sect.~\ref{phot_eff}.
\end{itemize}

\subsection{Observations}
\label{obs}

The targets were observed during two different campaigns. The first one was done in 2011, on the nights of October 16th, 17th, and 18th. The second campaign was carried out in 2012, from September 3rd to 7th. We used the Intermediate Dispersion Spectrograph (IDS), mounted on the 2.5~m Isaac Newton Telescope (INT) in La Palma (Spain). We used the \textit{Red}+2~CCD with its 4096-pixel axis along the wavelength direction. The grating  employed was R1200R, which covers an unvignetted spectral range $572\:$\AA{} wide, centred on $8500\:$\AA{} (i.e.\ the spectral region around the infrared Calcium Triplet, CaT). This configuration, together with a slit width of $1^{\prime\prime}$, provides a resolving power of $R\sim10\,500$ in the spectral region observed. This $R$ is very similar to the resolution of the data used in \citetalias{gon2015} ($R\sim11\,000$). The reduction was carried out in the standard manner, using the {\sc IRAF} facility\footnote{IRAF is distributed by the National Optical Astronomy Observatories, which are operated by the Association of Universities for Research in Astronomy, Inc., under cooperative agreement with the National Science Foundation}.

In total, we observed 637 unique targets, 102 in 2011 and 535 in 2012, without any overlap between epochs. As discussed above, 43 of them are CSGs with well determined SpTs (all but one observed in the 2012 run) that were included in the calibration sample of \citetalias{dor2016b} (see appendix~B in that work). These objects are not considered part of the Perseus sample studied here. This leaves 594 targets in our sample, which are detailed in Table~\ref{cat_perseo}.

\subsection{Manual classification and spectral measurements}
\label{clas_meas}

We performed a visual classification for all the stars in the sample, using the classical criteria for the CaT spectral region explained in \cite{neg2012}. All the carbon stars found (46) were marked and removed from later calculations. Thus, we do not use them in the present work, but they are included in our complete catalogue (see Table~\ref{cat_perseo}). Without the carbon stars, our sample has 548 targets. 

For the analysis of our sample, we used the principal component analysis (PCA) method described in \citetalias{dor2016b}. This method begins with the automated measurement of the main spectral features in the CaT spectral region. We measured all the features needed to calculate the principal components (PCs) of our stars (i.e.\ those marked as shortened input list in table C.1 from \citetalias{dor2016b}). The method to measure these features is the same as for the calibration sample in \citetalias{dor2016b}. Although the resolution of our sample is not exactly the same as in the calibration sample, it is close enough not to introduce any significant difference in the result, as explained in \citetalias{dor2016b}. 

Finally we combined linearly the PCA coefficients (tables~D.1 and D.2 in \citetalias{dor2016b}) with the spectral measurements of each star in our sample, obtaining their corresponding PCs. We also calculated their uncertainties, propagating the uncertainties of the EWs and PC coefficients through a lineal combination.

\section{Analysis}
\label{analysis}

\subsection{Estimating the probability of being a CSG}
\label{individual_prob}

In \citetalias{dor2016b} we revisited the main criteria classically used to identify RSGs, discussing the advantages and limitations of each one. We also proposed an original method, based on the PCs calculated through a large calibration sample and the use of Support Vector Machines (SVM). All the classical criteria, as well as the PCA method, use boundaries between the SGs and non-SGs as separators (our method uses many boundaries in a multidimensional space, but it is qualitatively the same in concept). Thus, they provide a binary classification for the targets (each of them is classified as either SG or non-SG), but without any direct estimation of the reliability of their classifications. 

In \citetalias{dor2016b} we also defined two useful concepts for our analysis: efficiency and contamination. Efficiency is the fraction of all SGs that is identified as such by a given criterion, while contamination is the fraction of the stars selected as SGs by a given criterion that are not really SGs. Efficiencies and contaminations obtained for the calibration sample are based on the statistics of the whole sample, and give a good idea of the reliability of each method when it is applied to a large number of candidates. However, it is not a good measurement of the reliability of the individual classification of each target: the result is the same for a star that lies close to the boundary as for one that is far away from it. In consequence, we wanted to measure the reliability of each individual identification. For this, we used a Montecarlo process that delivers the individual probability of each target being a SG ($P(\mathrm{SG})$). We detail the process and the results for the calibration sample in the following Section~\ref{cal_prob}. Later, after testing the method in the calibration sample, we calculate the probabilities for the test sample of this work in Section~\ref{prob_per}.

\subsubsection{Calculation}
\label{cal_prob}

For each one of the three classification methods described in the following paragraph, we obtained uncertainties through a  Montecarlo process using each target in the calibration sample from  \citetalias{dor2016b}. We took the variables needed for each method and their errors, and we drew a new value for the variable from a random normal distribution, with the original measurement as centre and the error as its standard deviation. For each target, we sampled 1\,000 draws, and so we obtained 1\,000 different sets of derived variables. To these we applied the corresponding classification methods, and checked how many times the target was classified as a SG or not in each draw. The $P(\mathrm{SG})_{\mathrm{method}}$ of a target is the fraction of realizations which resulted in a positive identification.

\begin{figure*}
   \centering
   \includegraphics[trim=1cm 0.5cm 2.4cm 1.2cm,clip,width=\columnwidth]{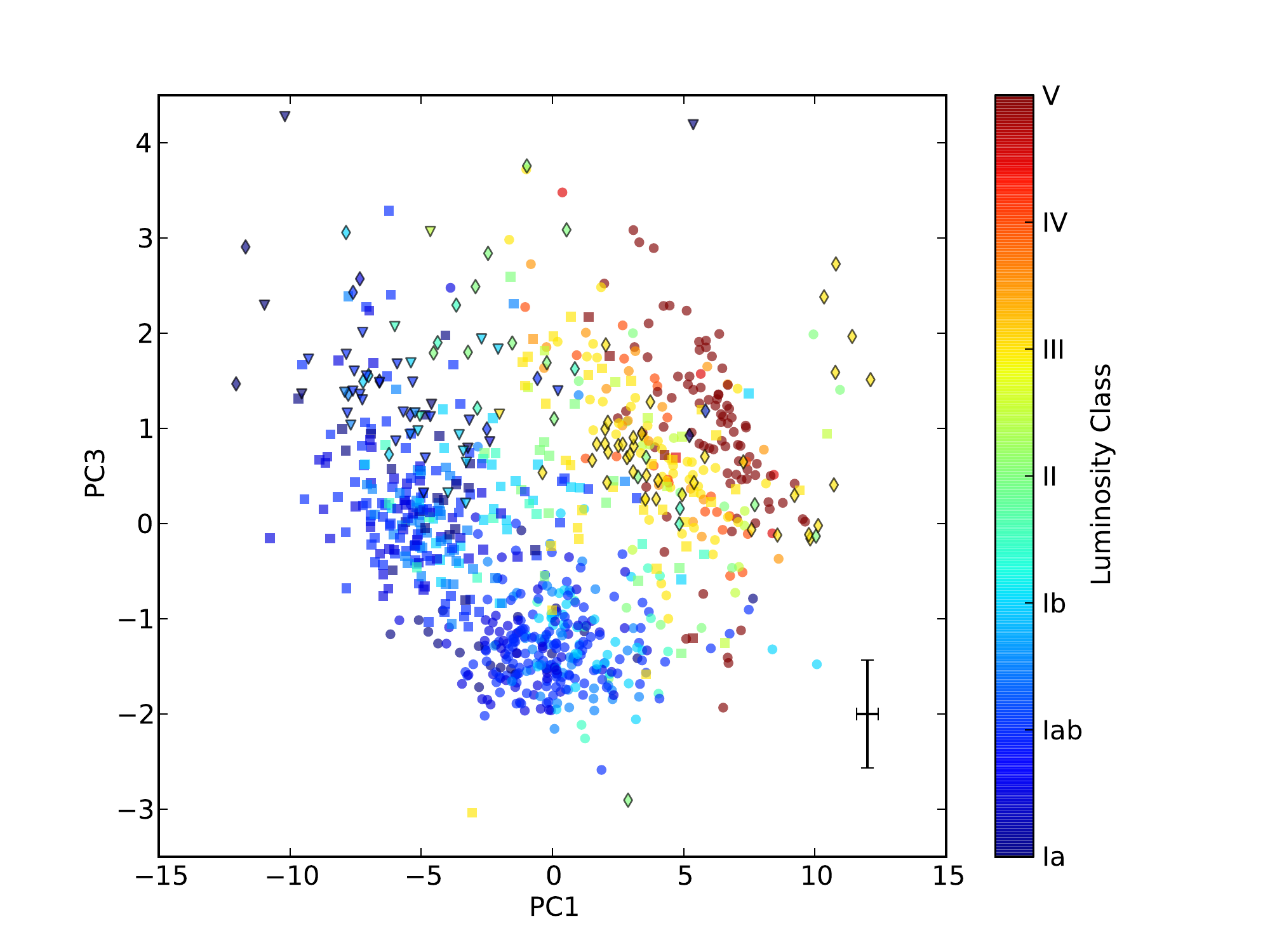}
   \includegraphics[trim=1cm 0.5cm 2.4cm 1.2cm,clip,width=\columnwidth]{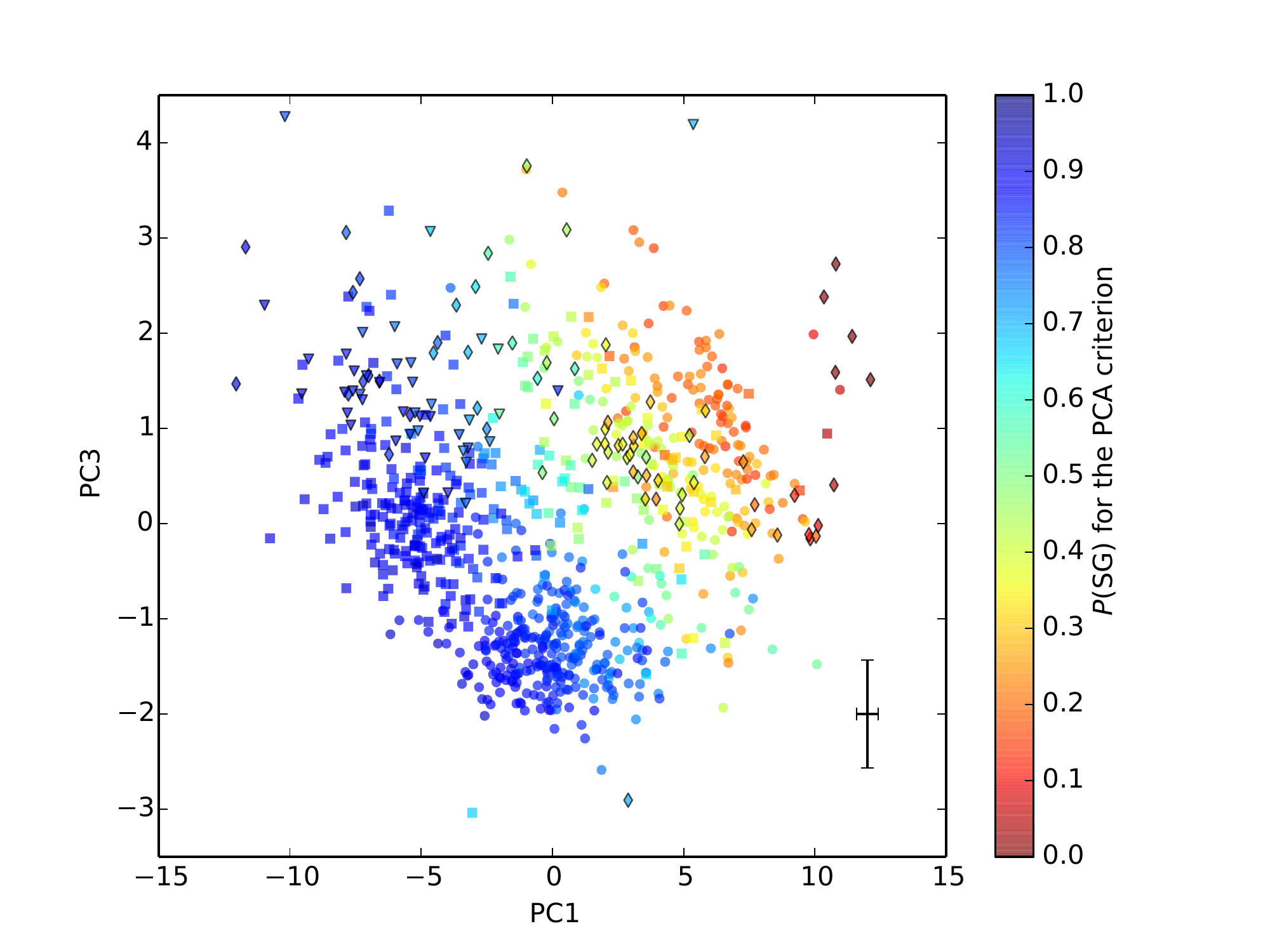}
   \caption{PC1 versus PC3 diagram for the calibration sample. The shapes indicate their origin: circles are from the SMC survey, squares are from the LMC survey, diamonds are Galactic standard stars, and inverted triangles are the stars from the Perseus arm survey used as part of the calibration sample (see Section~\ref{clas_meas}). The cross indicates the median uncertainties, which have been calculated by propagating the uncertainties through the lineal combination of the input data (EWs and bandheads) with the coefficients calculated in \citetalias{dor2016b}.
   {\bf Left (\ref{PC1_PC3_prob}a):} The colour indicates LC (identical to figure~7b in \citetalias{dor2016b}).
   {\bf Right (\ref{PC1_PC3_prob}b):} The colour indicates the probability of being a SG (see~\ref{cal_prob}).}
   \label{PC1_PC3_prob}
\end{figure*}

\begin{figure*}
   \centering
      \includegraphics[trim=1cm 0.5cm 2.5cm 1.2cm,clip,width=\columnwidth]{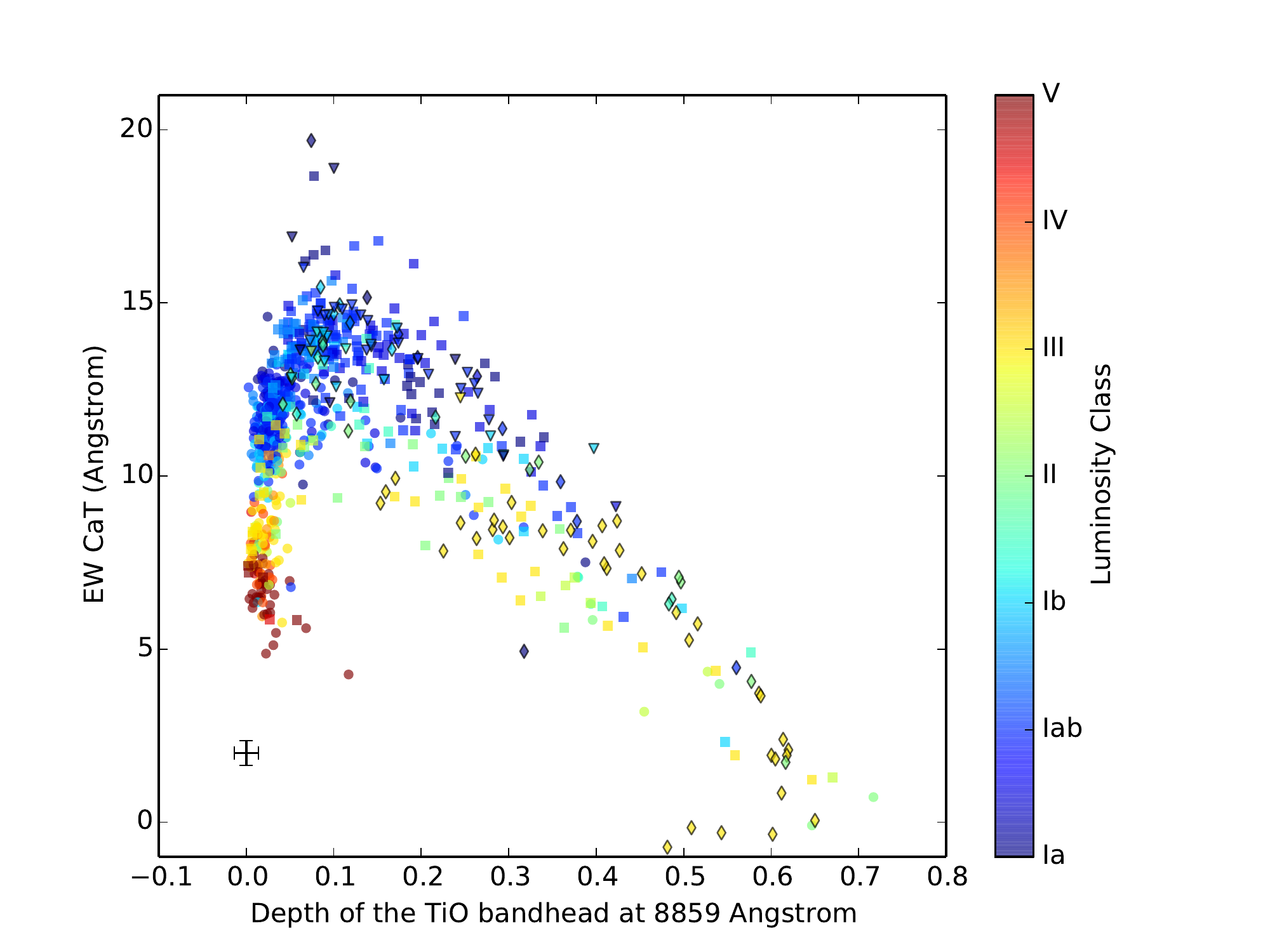}
   \includegraphics[trim=1cm 0.4cm 2.4cm 1.2cm,clip,width=\columnwidth]{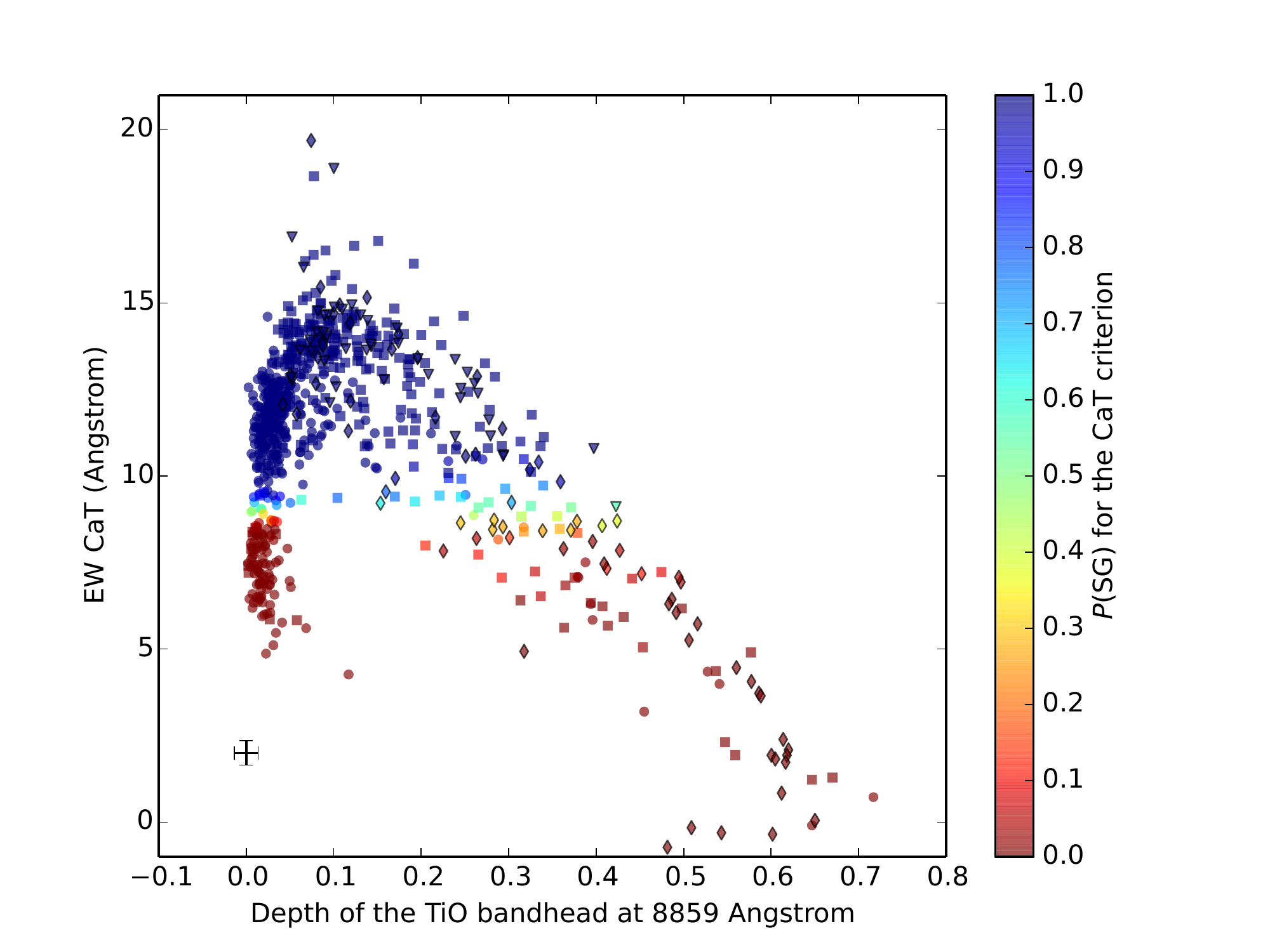}
   \caption{Depth of TiO bandhead at $8859$\:\AA{} versus total equivalent width of the Calcium Triplet ($8498$\:\AA{}, $8542$\:\AA{}, and $8662\:$\AA{}), for the calibration sample. The strength of the TiO~$8859\:$~\AA{} bandhead is simply an indicator of the spectral sequence for early to mid-M stars (see Section~4.3.4 in \citetalias{dor2016b}) and is included here simply to display the measurements in a 2D graphs, so that the CaT criterion is easily visualised. Symbol shapes are the same as in Fig.~\ref{PC1_PC3_prob}. The black cross indicates the median uncertainties. In these panels the probability of being a SG (see~\ref{individual_prob}) can be compared to the actual LC classification.
   {\bf Left (\ref{CaT_prob}a):} The colour indicates LC.
   {\bf Right (\ref{CaT_prob}b):} The colour indicates the probability of being a SG (see~\ref{cal_prob}). }
   \label{CaT_prob}
\end{figure*}

\begin{figure*}
   \centering
   \includegraphics[trim=0.80cm 0.4cm 2.4cm 1.2cm,clip,width=\columnwidth]{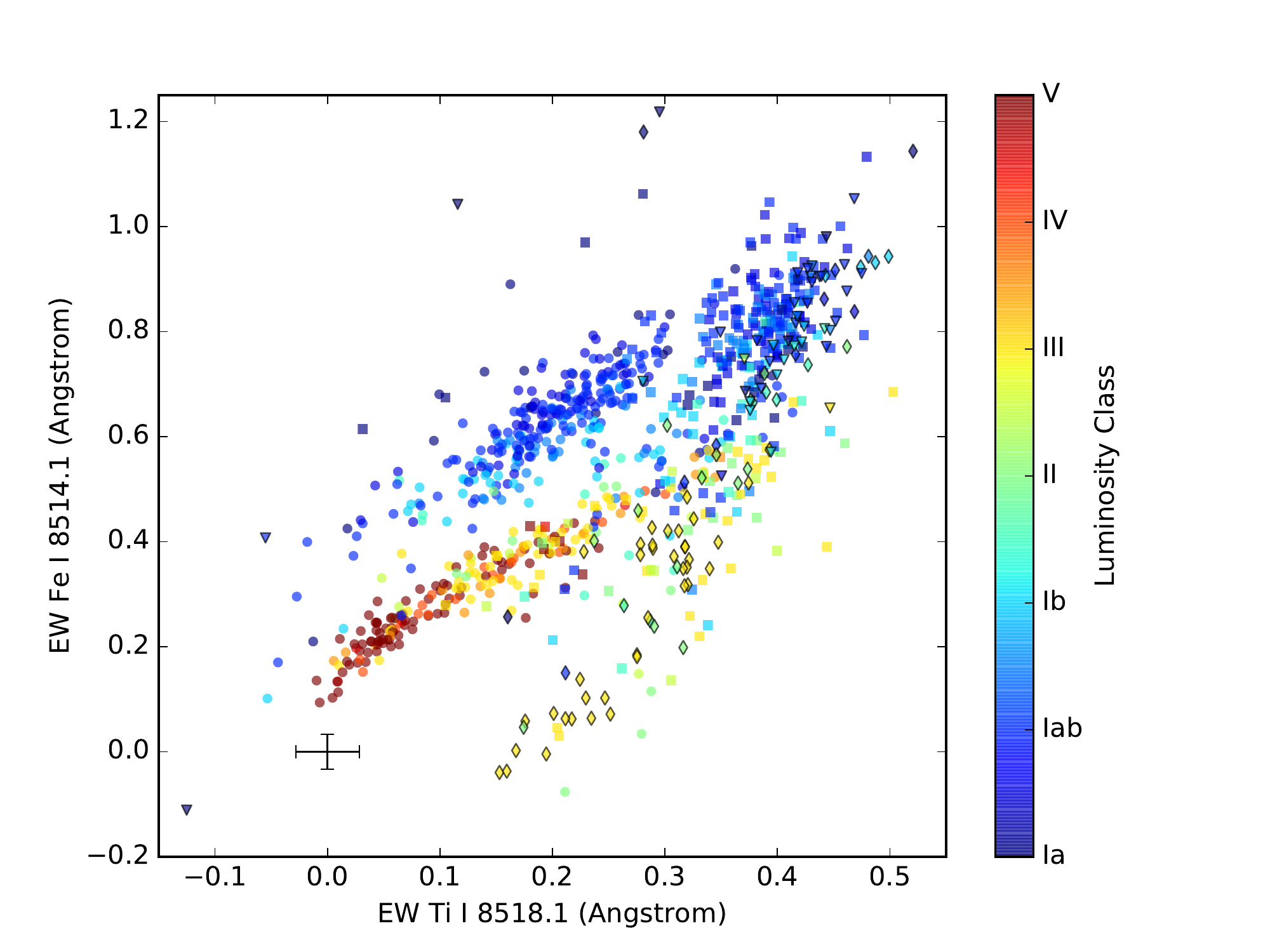}
   \includegraphics[trim=0.80cm 0.4cm 2.4cm 1.2cm,clip,width=\columnwidth]{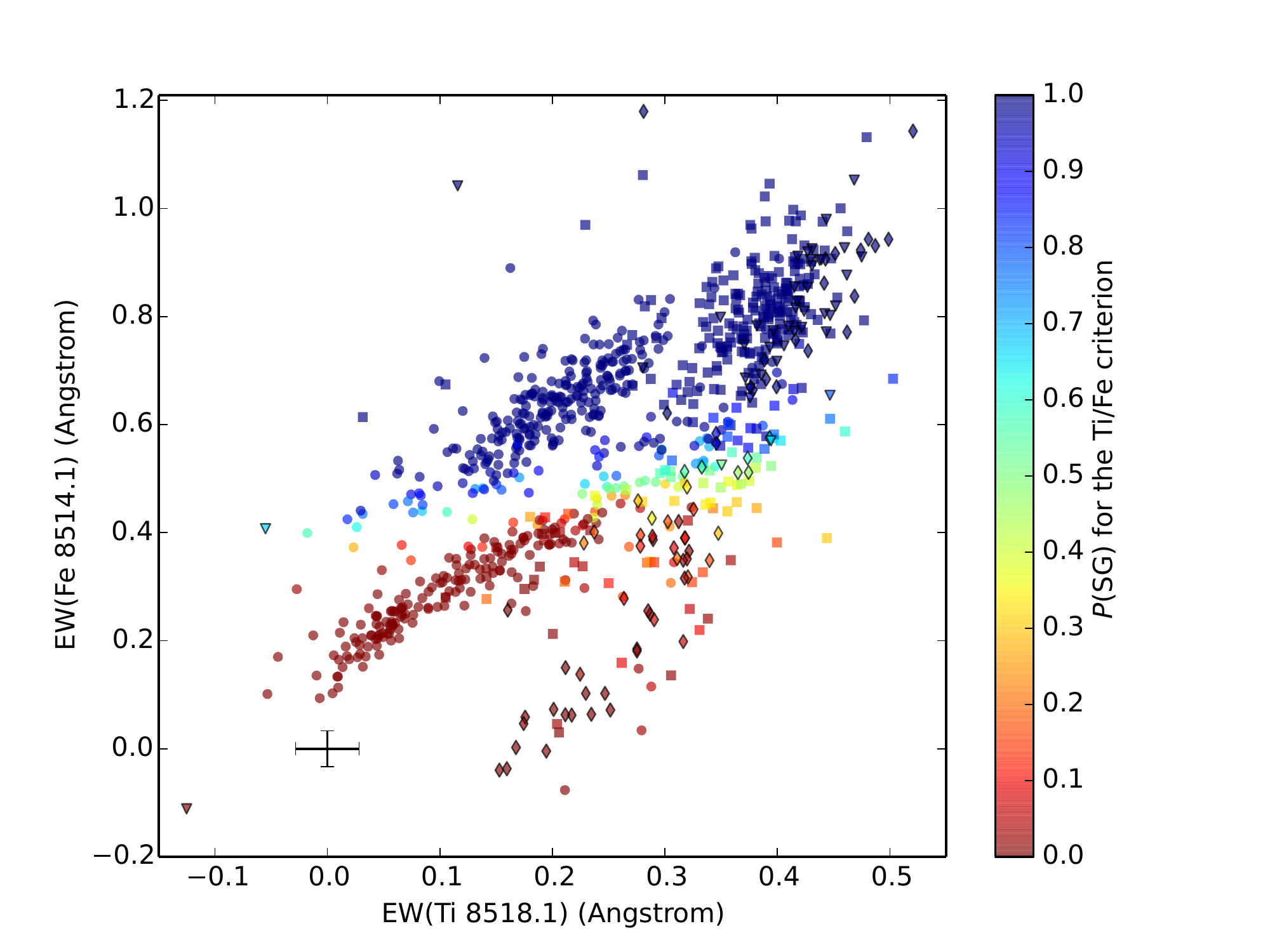}
   \caption{EWs of the lines Fe\,{\sc{i}}~$8514\:$\AA{} and Ti\,{\sc{i}}~$8518\:$\AA{} for the calibration sample. Symbol shapes are the same as in Fig.~\ref{PC1_PC3_prob}. The cross indicates the median uncertainties. In these panels the probability of being a SG (see~\ref{individual_prob}) can be compared with the actual LC classification.
    {\bf Left (\ref{CaT_prob}a):} The colour indicates LC (Equivalent to Fig.~12b from \citetalias{dor2016b}).
   {\bf Right (\ref{CaT_prob}b):} The colour indicates the probability of being a SG (see~\ref{cal_prob}). 
   }
   \label{Ti_Fe_prob}
\end{figure*}

For what we call the PCA method ($P(\mathrm{SG})_{\mathrm{PCA}}$), we used the first 15 PCs (which contain 98\% of the accumulated variance), and the SVM calculation defined in \citetalias{dor2016b} (using a putative boundary at M0; see \citetalias{dor2016b}), obtaining the $P(\mathrm{SG})_{\mathrm{PCA}}$ for each target. The results of this procedure are shown in Fig.~\ref{PC1_PC3_prob}. We also calculated the $P(\mathrm{SG})_{\mathrm{CaT}}$ for the criterium based on the strength of the CaT (a target is identified as a SG if the sum of the EWs of its three Ca lines is equal to or higher than $9\:$\AA{}), and $P(\mathrm{SG})_{\mathrm{Ti/Fe}}$ for the Ti/Fe method (which uses as boundary the line ($\mathrm{EW}(8514.1)=0.37\cdot~\mathrm{EW}(8518.1)+0.388$ in the Fe\,{\sc{i}}~$8514\:$\AA{} vs.\ Ti\,{\sc{i}}~$8518\:$\AA{} diagram). The results are shown in Figs.~\ref{CaT_prob} and~\ref{Ti_Fe_prob}. The other classical criteria considered in \citetalias{dor2016b}, based on the strength of the blend at $8468$\:\AA{} and the EWs of only the two strongest lines of the CaT, have been not used here because of their low efficiency or high contamination.

\subsubsection{Identification based on individual probabilities}
\label{ident_indiv_prob}

With the classical criteria studied, based on the CaT and on the Ti/Fe ratio, a large fraction of the SGs ($>0.85$ and $>0.70$) in the sample have $P(\mathrm{SG})=1$ and most non-SGs have $P(\mathrm{SG})=0$. Only those stars close to the boundary used by these methods present intermediate values of $P(\mathrm{SG})$. Since the boundaries between SGs and non-SGs in these diagrams are straight lines, a given star can be identified as a SG if it has $P(\mathrm{SG})\geq0.5$ -- this is equivalent to the simple assignment to one of the two categories. On the other hand, in the PCA method there are not many targets with their $P(\mathrm{SG})$ equal to $1$ or to $0$. This is because the PCA uses many boundaries in the multidimensional space of the PCs, not a single boundary in a two dimensional diagram, as is the case of the classic criteria. Thus it is more difficult to stay far away from every boundary and the probabilities tend to have intermediate values.

\begin{figure}
   \centering
   \includegraphics[trim=0.8cm 0.3cm 1.2cm 1.2cm,clip,width=\columnwidth]{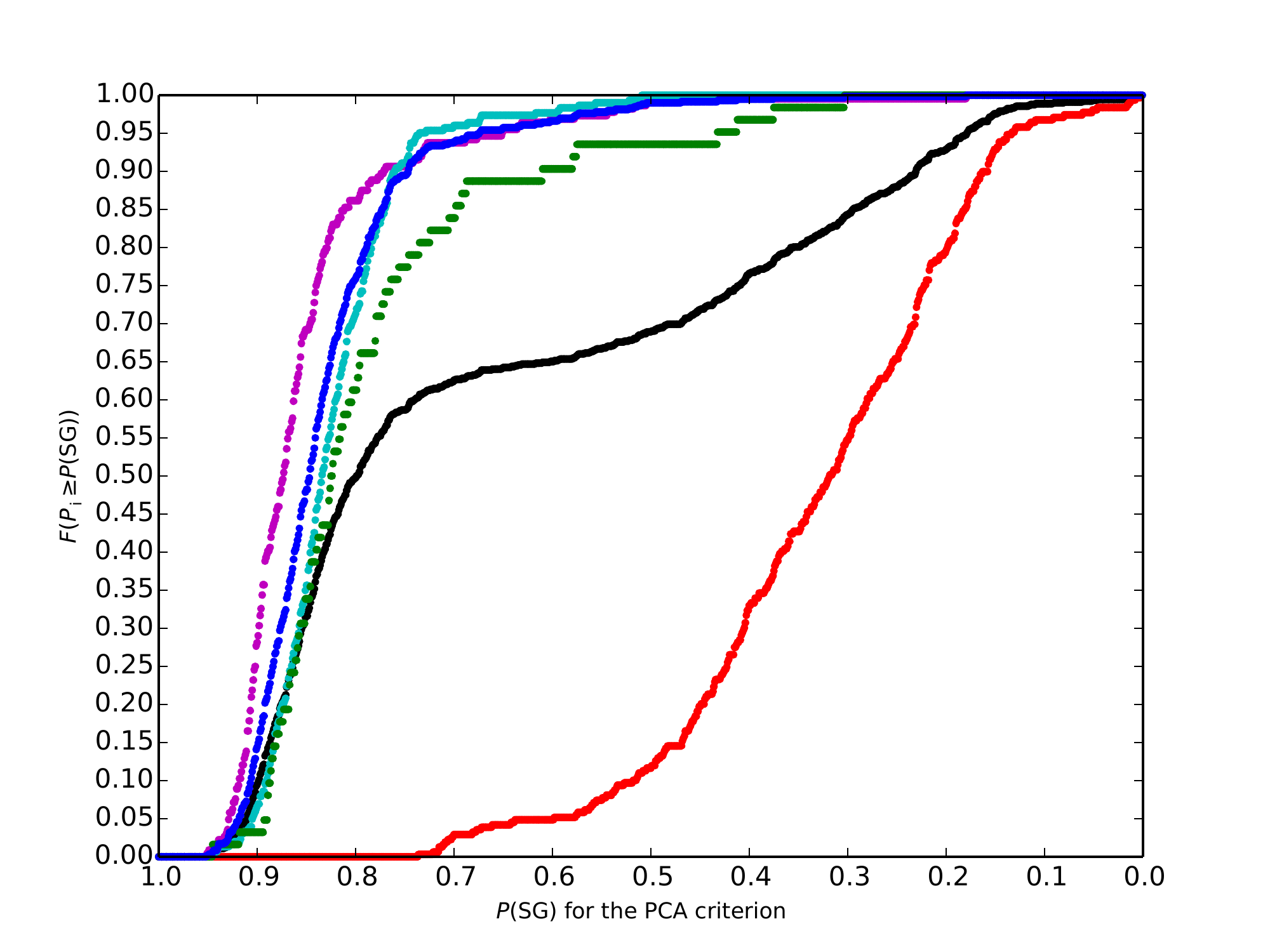}
   \caption{Fraction of the calibration sample that has a probability of being a SG (calculated through the PCA method) equal to or higher than the corresponding $x$-axis value. The colours indicate the subsample: black for whole sample, red for non-SGs, blue for all SGs, magenta for SMC SGs, cyan for LMC SGs, and green for MW SGs. Each fraction is calculated with respect to the size of its own subsample.}
   \label{all_pca}
\end{figure}

\begin{figure}
   \centering
   \includegraphics[trim=0.8cm 0.3cm 1.2cm 1.2cm,clip,width=\columnwidth]{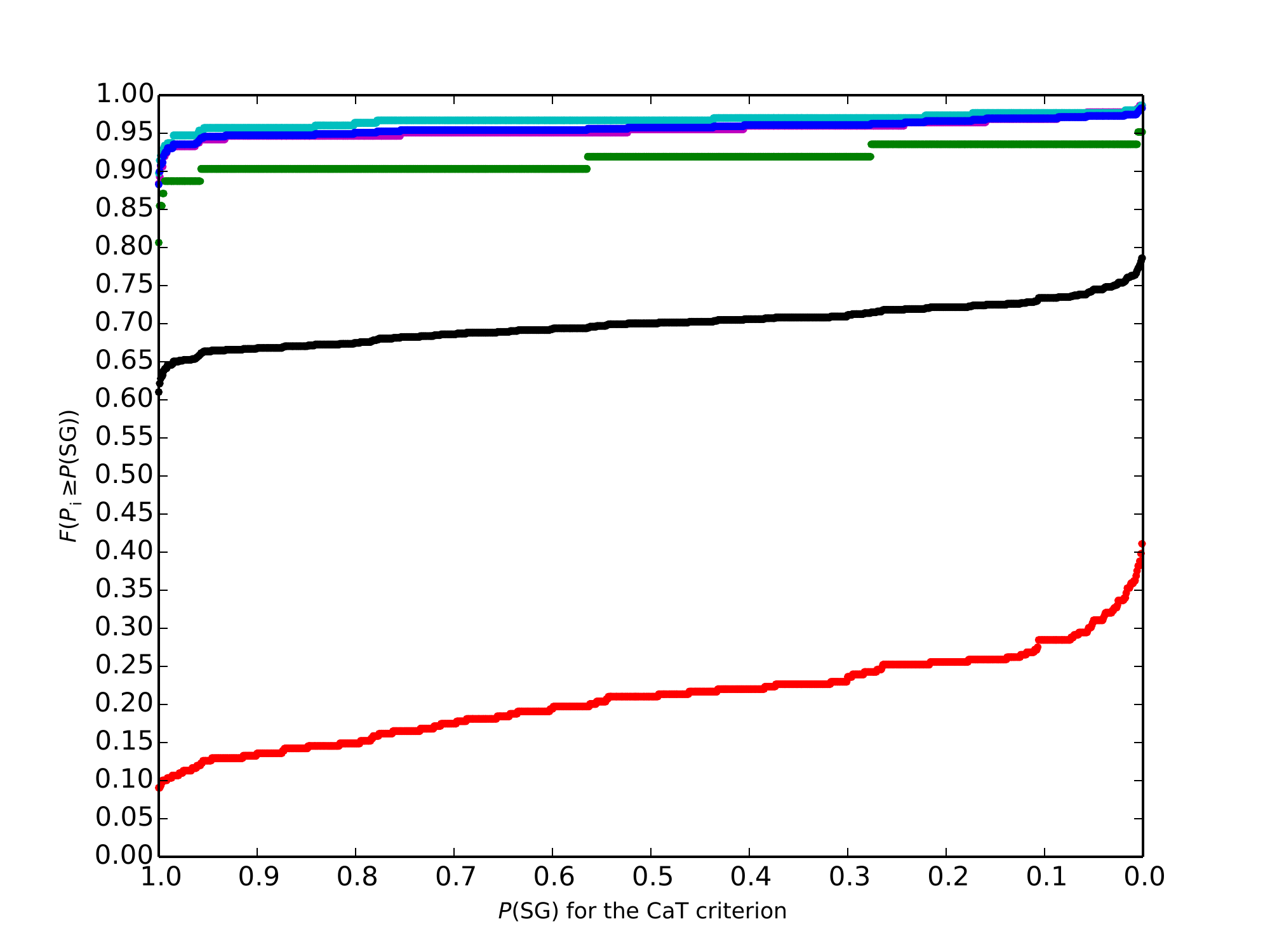}
   \caption{Fraction of the calibration sample that has a probability of being a SG (calculated through the CaT method) equal to or higher than the corresponding $x$-axis value. The colours indicate the subsample, as explained in Fig.~\ref{all_pca}. Each fraction is calculated with respect to the size of its own subsample.}
   \label{all_cat}
\end{figure}

\begin{figure}
   \centering
   \includegraphics[trim=0.8cm 0.3cm 1.2cm 1.2cm,clip,width=\columnwidth]{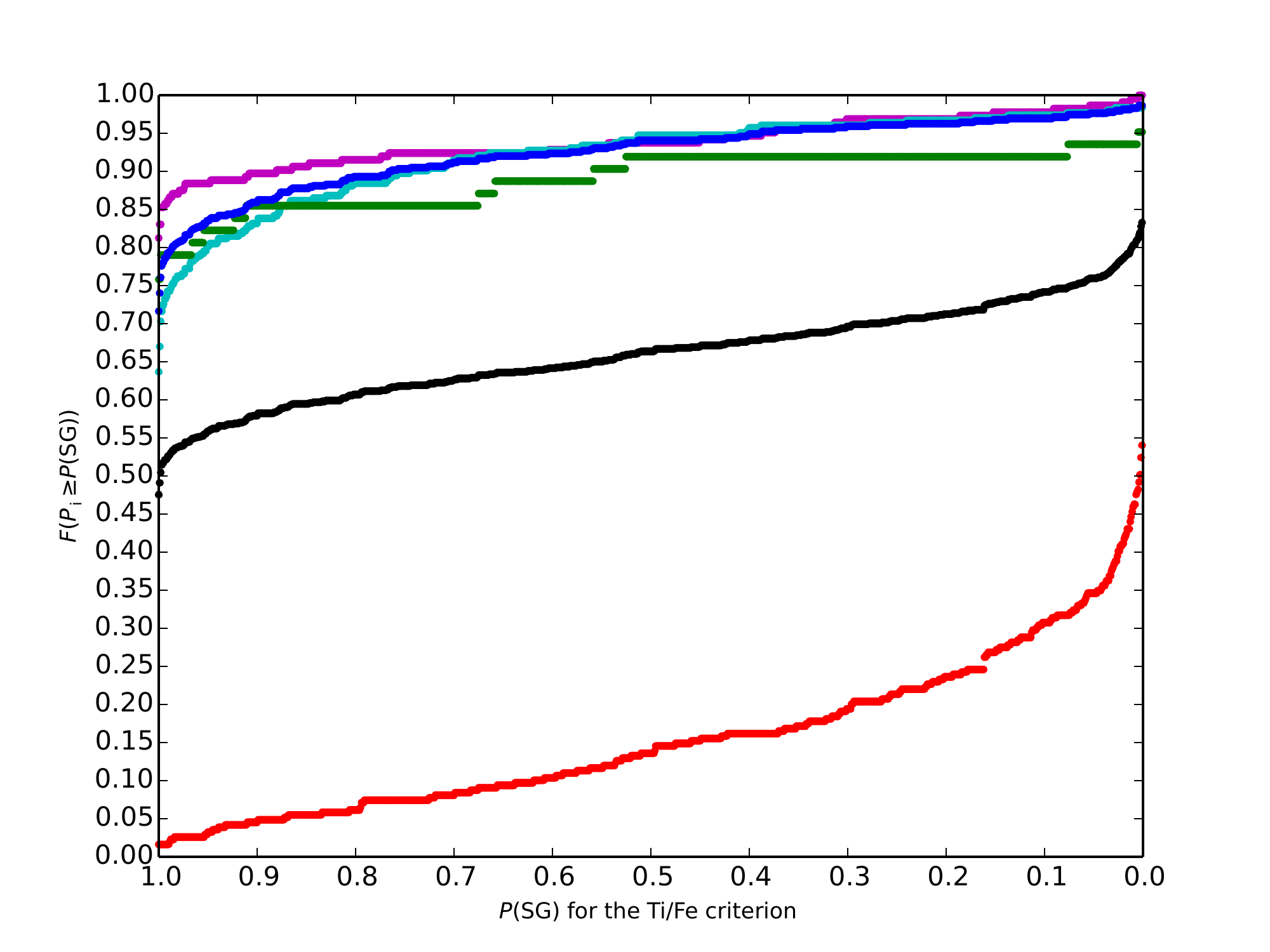}
   \caption{Fraction of the calibration sample that has a probability of being a SG (calculated through the ratio of the Fe\,{\sc{i}}~$8514\:$\AA{} to Ti\,{\sc{i}}~$8518\:$\AA{} lines) equal to or higher than the corresponding $x$-axis value. The colours indicate the subsample, as explained in Fig.~\ref{all_pca}. Each fraction is calculated with respect to the size of its own subsample.}
   \label{all_ti_fe}
\end{figure}

To illustrate this, and also to evaluate the application of this method to the identification of SGs, we calculated how many targets have their individual probability $P_{i}$ equal to or higher than a given $P(\mathrm{SG})$ value. As the SGs from each galaxy in the calibration sample have different typical SpTs (\citealt{lev2013}; \citetalias{dor2016a}), we performed this calculation for six different subsamples taken from the calibration sample: SGs from the SMC, from the LMC, from the MW, all SGs, all non-SGs, and the whole sample. We present the results for each of these subsamples as fractions ($F(P_{i}\geq P(\mathrm{SG}))$) with respect to their corresponding total size, in Figs.~\ref{all_pca}, \ref{all_cat}, and~\ref{all_ti_fe}. For all three classification criteria, the SGs from both MCs present very similar behaviours, but the SGs from the MW present slightly lower probabilities. This small difference is likely due to the lower efficiency of all criteria towards later subtypes, as it is well known that SGs in the MW tend to have later subtypes than those in the MCs \citep{lev2013}.

The CaT and the Ti/Fe criteria result in a large fraction of SGs with high values of $P(\mathrm{SG})$, but there are non-SGs with probabilities as high as $P(\mathrm{SG})=1$. Thus, these methods provide a quick way to identify most SGs in the sample, but at the price of having a a significant contamination. Of these two methods, the CaT one is less strict, finding more SGs, but also including more non-SGs with high $P(\mathrm{SG})$ values.

The PCA method finds a very small fraction of SGs with $P(\mathrm{SG})>0.9$ (and this fraction is significantly higher for SMC SGs than for MW ones, as can be seen in Fig.~\ref{all_pca}). However, non-SGs present significantly lower values of $P(\mathrm{SG})$, with none of them having $P(\mathrm{SG})>0.75$. For this value the fraction of SGs identified is about $0.90\pm0.04$ ($\sim0.80\pm0.13$ for the SGs from the MW). Therefore, using this value as a threshold, the vast majority of SGs can be identified without any contamination. In addition, it is also possible to identify a group of likely SGs with a relatively low contamination, by taking the targets whose $P(\mathrm{SG})$ lies within the interval between $P(\mathrm{SG})=0.75$ and a lower limit set at convenience (depending on the level of contamination that may be considered acceptable).

For a new sample, such as the Perseus arm sample in this paper, it is possible to estimate the value of this lower limit of $P(\mathrm{SG})$ that results in an optimal selection of potential SGs. In such a sample, the only information available will be the shape of the $P(\mathrm{SG})$ fraction curve (the black line in our figures). This curve, however, will always have an inflexion point at the $P(\mathrm{SG})$ value where most SGs have already been selected, while most non-SGs have lower values of $P(\mathrm{SG})$. Thus, from this point towards lower probabilities, the addition of extra targets to the selection becomes dominated by non-SGs. Therefore, this inflexion point can be used as a lower boundary for the group of potential SGs, and can be easily estimated for any sample under study, as we do for the Perseus sample in next Section.

In the calibration sample, the inflexion point is at $P(\mathrm{SG})\sim0.60$. Taking this value as a lower boundary, the efficiency of the resultant selection is higher than $0.95\pm0.04$ ($\sim0.90\pm0.13$ for SGs from the MW), while the contamination is only $0.03\pm0.04$ ($0.08\pm0.13$ in the case of the MW sample). Note that the contaminations were calculated for the total number of stars tagged as SGs, i.e. all those having $P(\mathrm{SG})\geq0.60$). For similar efficiencies in the CaT and Ti/Fe ratio criteria, the contaminations are slightly higher ($\sim0.08\pm0.04$ in both cases). These values become slightly worse in the case of MW SGs, with contaminations of $0.17\pm0.13$ for the Ti/Fe ratio criterion and $0.20\pm0.13$ for the CaT one. In \citetalias{dor2016b} we found that the PCA method provides a higher quality method to identify SGs than the other two, because it has a significantly lower contamination. In this work, we found another advantage: the possibility to identify a large fraction of SGs without any contamination.

\subsection{Probabilities for the Perseus sample}
\label{prob_per}

Before the analysis of our Perseus sample, we must stress that the SGs from the MW typically have M subtypes. We may thus expect our sample to be dominated by these subtypes. Moreover, most of the interlopers found in the manual classification are red giants with M types. In consequence, the diagrams obtained for the Perseus sample have their datapoints concentrated in the regions typical of M-type stars, and look quite different from the distributions seen in the calibration sample (see Figs.~\ref{PC1_PC3_prob}, \ref{CaT_prob}, and \ref{Ti_Fe_prob}), whose SpT range spans from G0 till late-M subtypes. For further details about the calibration sample and their SpT distribution, see \citetalias{dor2016b} and figs.~7a, 9, and 12a therein.

\begin{figure}
  \centering
  \includegraphics[trim=1cm 0.3cm 2.3cm 1.2cm,clip,width=\columnwidth]{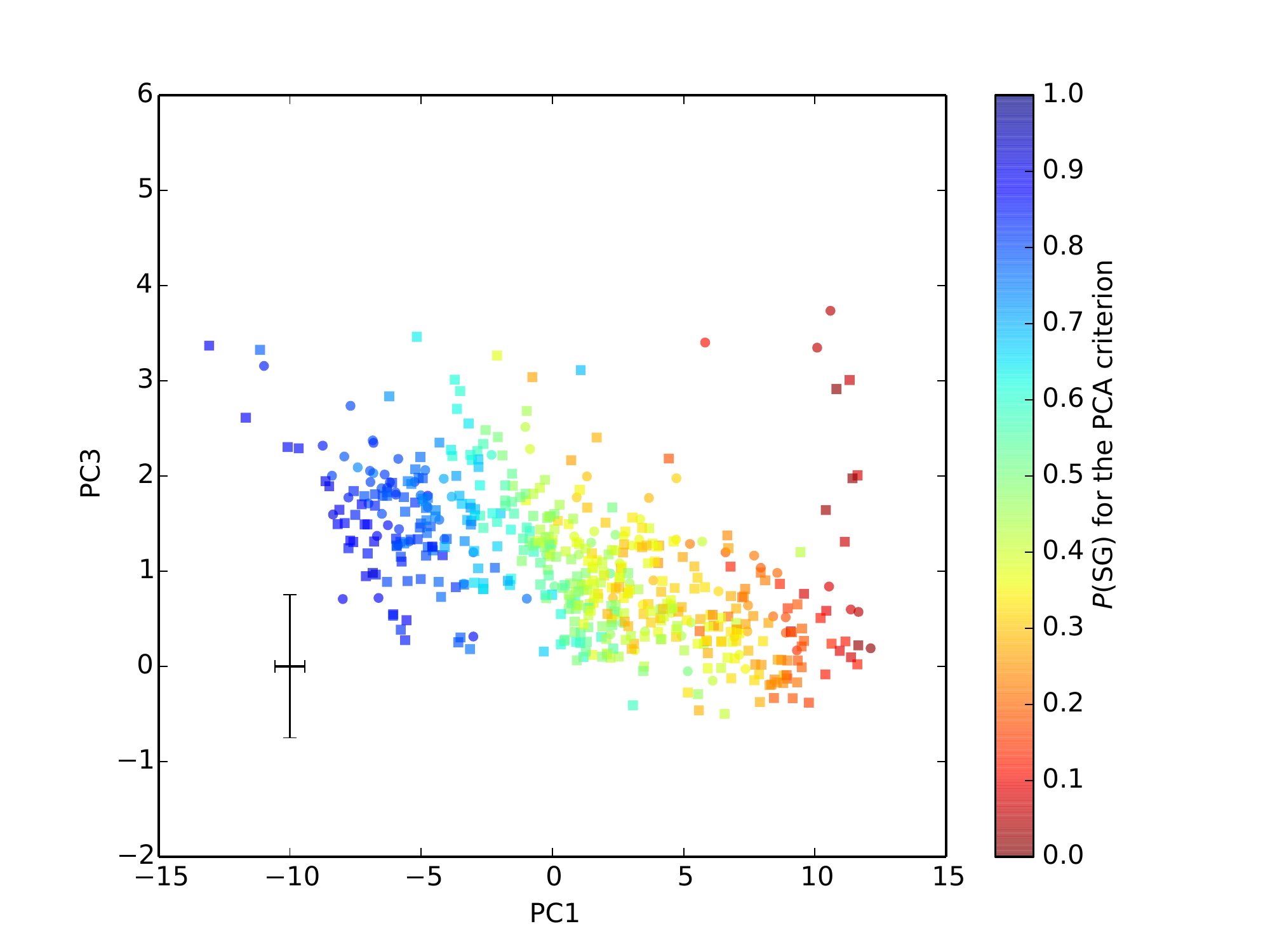}
 \caption{PC1 versus PC3 diagram for the Perseus sample. The shapes indicate epoch: 2011 circles, 2012 squares. The black cross indicates the median uncertainties, which have been calculated by propagating the uncertainties through the lineal combination of the input data (EWs and bandheads) with the coefficients calculated. The colour indicates $P(\mathrm{SG})_{\mathrm{PCA}}$. The plot is shown at the same scale as Fig.~\ref{PC1_PC3_prob}, to ease comparison. The differences in the target distribution with respect to the calibration sample are due to the different ranges of spectral types.}
  \label{PC1_PC3_prob2}
\end{figure}

We calculated the individual probabilities of being a SG for each target in the Perseus sample, following the same method described for the calibration sample (Section~\ref{individual_prob}). Using the PCs previously obtained for our targets, $P(\mathrm{SG})_{\mathrm{PCA}}$ was calculated through a Montecarlo process (generating 1\,000 new sets of PCs per target). The results are given in Table~\ref{cat_perseo} and represented in a PC1 to PC3 diagram in Fig.~\ref{PC1_PC3_prob2}.

\begin{figure}
  \centering
  \includegraphics[trim=0.8cm 0.2cm 1.2cm 1.2cm,clip,width=\columnwidth]{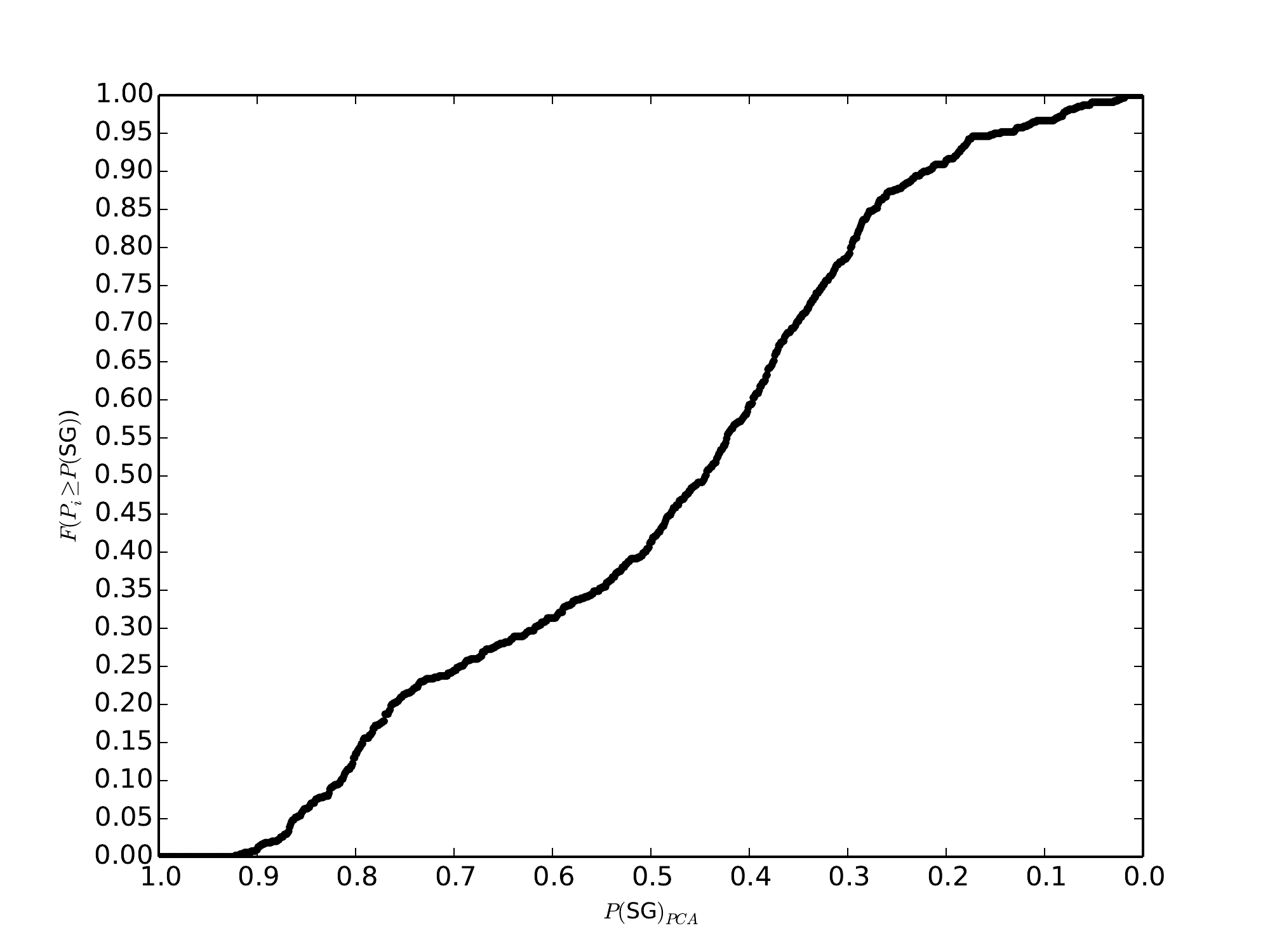}
  \caption{Fraction of the Perseus sample that has a probability of being a SG (calculated through the PCA method) equal to or higher than the corresponding $x$-axis value.}
  \label{prob_pca}
\end{figure}

\begin{figure}
  \centering
  \includegraphics[trim=1cm 0.4cm 2.3cm 1.2cm,clip,width=\columnwidth]{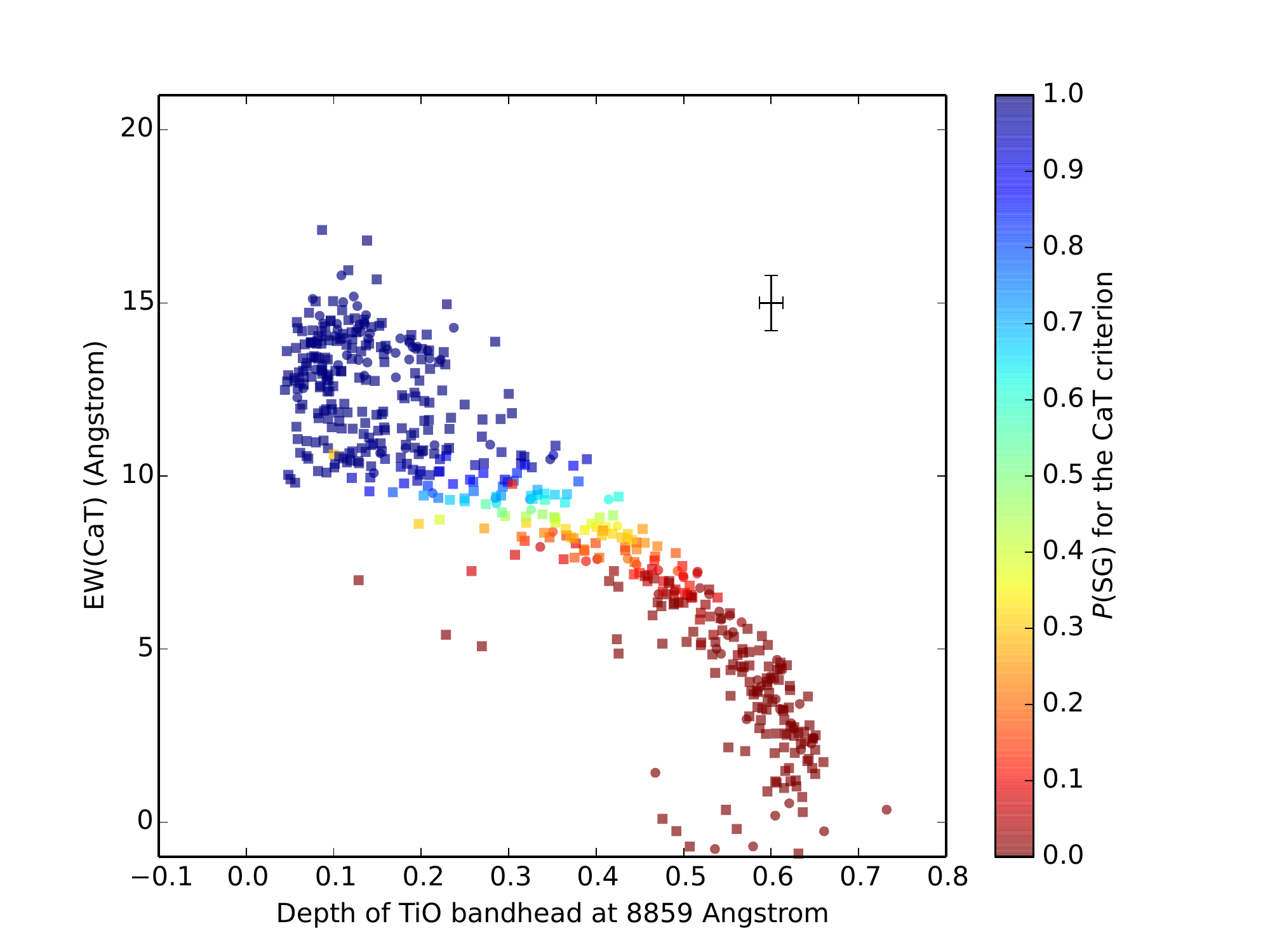}
  \caption{Depth of the TiO bandhead at $8859$\:\AA{} with respect to the sum of the EWs of the CaT lines. The shapes indicate epoch: 2011 circles, 2012 squares. The black cross indicates the median uncertainties. The colour indicates $P(\mathrm{SG})_{\mathrm{CaT}}$. Note again the difference in SpT distribution with respect to the calibration sample (Fig.~\ref{CaT_prob}).}
  \label{TiO_CaT_prob}
\end{figure}

\begin{figure}
  \centering
  \includegraphics[trim=0.8cm 0.4cm 2.3cm 1.2cm,clip,width=\columnwidth]{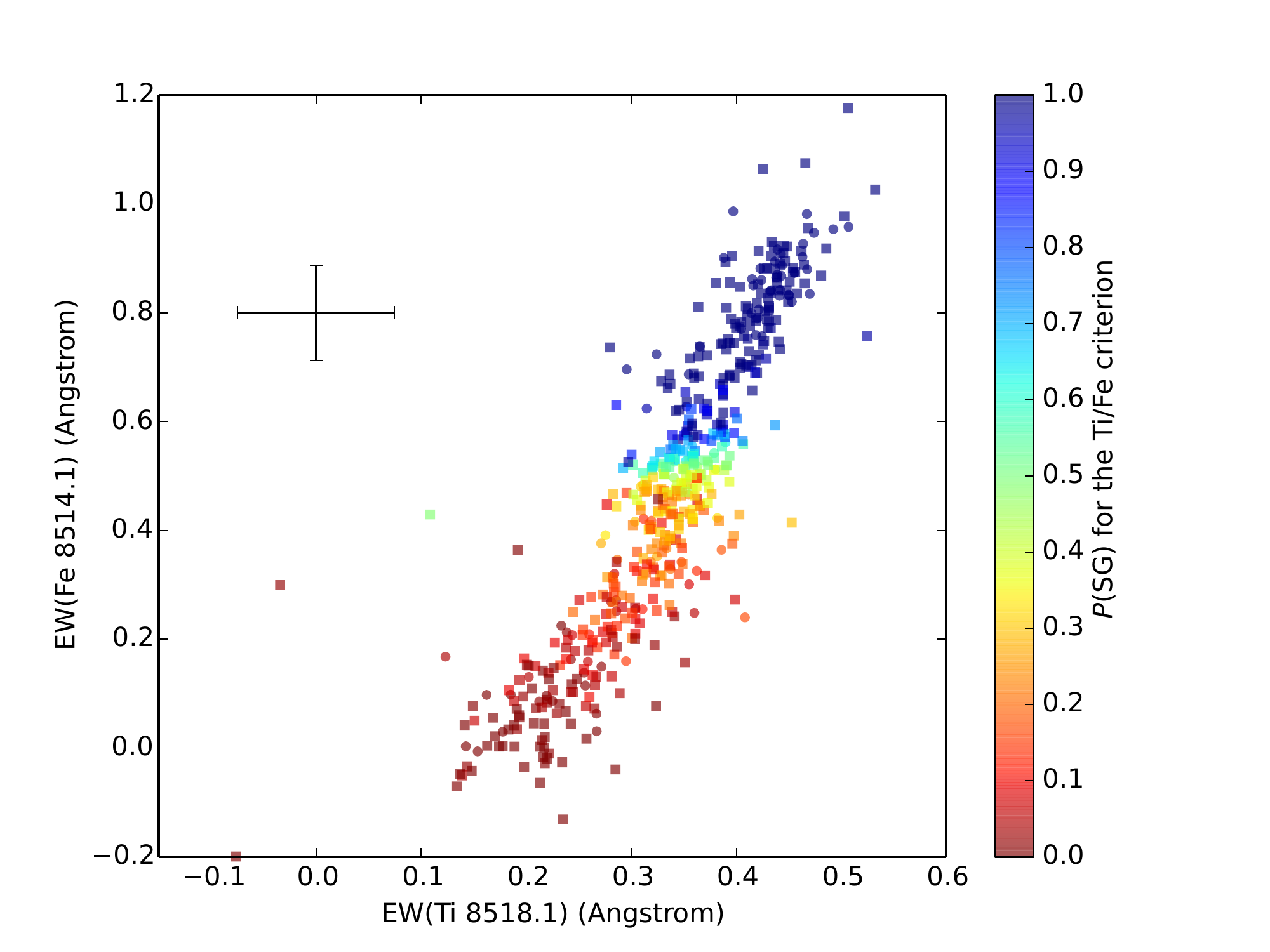}
 \caption{EWs of the Fe\,{\sc{i}}~$8514\:$\AA{} and Ti\,{\sc{i}}~$8518\:$\AA{} lines. The shapes indicate epoch: 2011 circles, 2012 squares. The black cross indicates the median uncertainties. The colour indicates $P(\mathrm{SG})_{\mathrm{Ti/Fe}}$. Comparison to Fig.~\ref{Ti_Fe_prob} highlights the lack of stars with G and K spectral types.}
  \label{Ti_Fe_prob2}
\end{figure}

Although the PCA method provides significantly better results than classical criteria, we also calculated the probabilities for them (CaT and Ti/Fe). We include these criteria because they are useful for a quick estimate despite their limitations. In addition, this is the first time that these criteria are systematically applied them to a very large sample at solar metallicity: more than 500 targets, instead of the $\sim100$ MW stars from the calibration sample. The results are given in Table~\ref{cat_perseo}, and presented in Figs.~\ref{TiO_CaT_prob} and \ref{Ti_Fe_prob2}.

\section{Results}

\subsection{Supergiants identified}
\label{SG_ident}

When we studied the distribution of $P(\mathrm{SG})_{\mathrm{PCA}}$ among the components of the calibration sample, we found that only true SGs present values higher than $P(\mathrm{SG})_{\mathrm{PCA}}=0.75$ (see Section~\ref{ident_indiv_prob}). Thus, we were able to obtain a group of SGs a priori free from any non-SG (the ``reliable SGs" set). In addition, it is possible to define an interval of probabilities between $P(\mathrm{SG})_{\mathrm{PCA}}=0.75$ and a lower limit, that increases the selection of SGs, while keeping the contamination very low (the ``probable SGs" set). The optimal lower limits for the Galactic samples were selected through the diagram shown in Fig.~\ref{prob_pca}, by the estimation of the inflexion point in the corresponding curve. For the Perseus sample we estimated it at $P(\mathrm{SG})_{\mathrm{PCA}}\sim0.55$. The number of SGs found by these cuts is indicated in Table~\ref{prob_pca_tabla}.

\begin{table}
\caption{Number of targets tagged as ``reliable SGs" or ``probable SGs" (see~\ref{SG_ident}) through the analysis of $P(\mathrm{SG})_{\mathrm{PCA}}$. The luminosity class was assigned through the manual classification. We also show the fraction that these groups represent with respect to the number of total targets in the sample (594). The 2-sigma uncertainties for the given fractions are equal to $1/\sqrt[]{n}$, where $n$ is the total number of targets. Thus, the uncertainty of both fractions is equal to $\pm0.04$.}
\label{prob_pca_tabla}
\centering
\begin{tabular}{c c c | c c c}
\hline\hline
\noalign{\smallskip}
\multicolumn{3}{c |}{Number}&\multicolumn{3}{c}{Fraction}\\
Reliable&Probable&&Reliable&Probable&\\
SGs&SGs&Total&SGs&SGs&Total\\
\noalign{\smallskip}
\hline
\noalign{\smallskip}
116&75&191&$0.20$&$0.13$&$0.33$\\
\noalign{\smallskip}
\hline
\end{tabular}
\end{table}

Classical methods are based on a linear boundary in a two-dimensional space. In consequence, when curves of $P(\mathrm{SG})$ are plotted for them (see Section~\ref{ident_indiv_prob}), there is no hint of a threshold value for ``realiable SGs" as in the case of $P(\mathrm{SG})_{\mathrm{PCA}}$. Thus, the only reasonable minimum value, given the two-dimensional nature of the boundary, is $P(\mathrm{SG})=0.5$. The number of targets identified as SGs are given in Table~\ref{SG_class}.

\begin{table}
\caption{Number of SGs found by different methods, and the fraction that they represent with respect to the total number of targets observed (594). For the classical criteria, we used a threshold of $P(\mathrm{SG})=0.5$; for the PCA method, we adopted a threshold of $P(\mathrm{SG})=0.55$ (see Sect.~\ref{prob_per}). The 2-sigma uncertainties of the fractions are equal to $1/\sqrt[]{n}$, where $n$ is the total number of targets.}
\label{SG_class}
\centering
\begin{tabular}{c c c }
\hline\hline
\noalign{\smallskip}
Criterion&Number of SGs&Fraction\\
\noalign{\smallskip}
\hline
\noalign{\smallskip}
CaT&304&$0.51\pm0.04$\\
Ti/Fe&238&$0.40\pm0.04$\\
PCA&193&$0.32\pm0.04$\\
\noalign{\smallskip}
\hline
\end{tabular}
\end{table}

The targets tagged as SGs through $P(\mathrm{SG})_{\mathrm{PCA}}$ represent a significant fraction (almost one third) of the Perseus sample. Moreover, most of them ($\sim66$\%) are tagged as ``reliable SGs"; we can thus consider this group in good confidence free of any interloper. The number of SGs found through the PCA method is, however, significantly lower than the numbers found through the CaT and Ti/Fe criteria. We must be cautious with the results obtained using these methods, as their contaminations were higher  ($0.17\pm0.13$ for Ti/Fe and $0.20\pm0.13$ for CaT) than for the PCA ($0.08\pm0.13$) among MW stars in the calibration sample (see \citetalias{dor2016b}). The difference in the expected contamination is not enough to explain the number of stars tagged as SG, but it seems clear that the higher the contamination is for a method, the larger number of stars it identifies as SGs. Moreover, we have to take into account that the Galactic set from the calibration sample is limited in two ways. Firstly, the subsample was relatively small, which causes high uncertainties in our fractions ($\pm0.13$). Secondly, this sample is not comparable to any observed sample, because it was intentionally created by assembling a similar number of well-known SGs and non-SGs. Thus, it will not be at all representative in terms of the number of non-SG stars that one may expect to find as interlopers when using photometric criteria to select SG candidates in the Galactic Plane. In view of these limitations, to study the efficiency and contamination of our methods in the Perseus sample, we resort to a direct calculation, in the next Section.

\subsection{Efficiency of the photometric selection}
\label{phot_eff}

The most important source of contaminants in the photometric selection comes from the magnitude/distance degeneracy. In this case, we are interested in structures relatively close to Earth, and in stars that are intrinsically bright, so we can enforce strict limits in apparent magnitude that will filter out most of the intrinsically dimmer populations along the line of sight. The overall efficiency of the selection criteria outlined in Sect.~\ref{targ_sel} is $47\%$. This includes the $43$~MK standards mentioned in Section~\ref{obs}, as these were not included a posteriori but picked up by the selection algorithm.

\begin{figure}
   \centering
   \includegraphics[trim=0.9cm 0cm 0cm 0cm,clip,width=8.8cm]{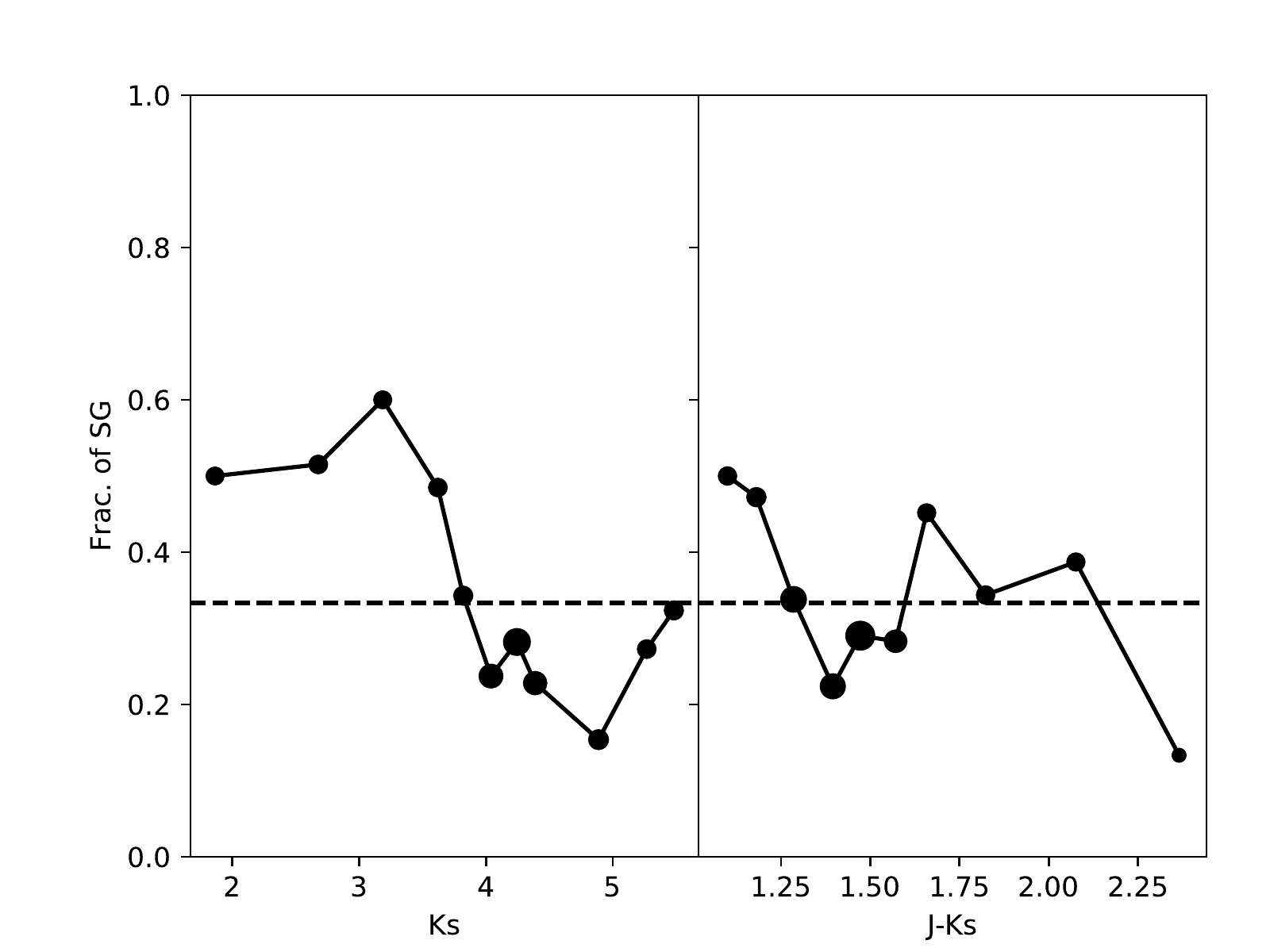}
   \caption{Fraction of SGs found in the target sample as a function of apparent $K\mathrm{\!s}$ magnitude and colour. The dashed line marks the total average fraction, $47\%$. Of these detected SGs, $\sim5\%$ where already known.}
   \label{selection_eff}
\end{figure}

As can be seen in Fig.~\ref{selection_eff}, the efficiency decays with magnitude: at $m_{K_{\mathrm{S}}}\sim4.5$ most of the observed stars turn out to be interlopers. This agrees roughly with Paper I, as at the low end of the brightness distribution of SGs the selected sample is dominated by bright giants. Similarly, while the fraction of SGs is more or less homogeneous with colour, the red end of the distribution (stars with $(J-K_{\mathrm{S}})\geq1.7$) is mostly composed of bright carbon stars. These results for a MW sample confirm those obtained in the MCs, in \citetalias{gon2015}, and will also be useful for future photometric selections. However, we must caution that such a red cut-off can only be used to discriminate carbon stars in fields of low (such as the MCs) or moderate (like the present sample) extinction. For the high extinctions ($A_{V}\ga5$~mag) found in many lines of sight towards the inner Milky Way, M-type stars would be shifted to very high values of $(J-K_{\mathrm{S}})$ and other discriminants must be found.

\subsection{Efficiency and contamination in the PCA method}
\subsubsection{Efficiency}
\label{efficiency}

To estimate directly the efficiency of our survey in the Perseus arm, we used the manual classification previously performed. We have to note that this classification is not \textit{a priori} more reliable than our automatised methods. Manual classification was done before we developed the automated process detailed in \citetalias{dor2016b}. For the manual classification we used classical criteria, such as the EW of the Calcium Triplet, the ratio between nearby Ti and Fe lines (Fe\,{\sc{i}}~$8514\:$\AA{} and Ti\,{\sc{i}}~$8518\:$\AA{} among others) and the EW of the blend at $8468\:$\AA{}. In \citetalias{dor2016b}, we demonstrated that the criteria based on these features have an efficiency slightly worse (at best) than our automated method. The manual classification can be somewhat better than these methods at identifying SGs, as it is a global process (like our PCA method), not based on any single spectral feature. Thus, the efficiency found in this work is useful to estimate the average quality of the classification methods under study with respect to a manual classification done following the classical criteria for the CaT range.

In the first place, we calculated the efficiency for each method (see Table~\ref{effi_r_per}). The efficiency in this case is the fraction of all SGs found through the manual classification, which were also tagged as such by a given automated criterion. The PCA method has the lowest global efficiency. It is similar to the value for the Ti/Fe criterion, but significantly lower than the efficiency of the CaT criterion. Nevertheless, when the LC of the targets is taken into account, the results can be seen in a very different light.

\begin{table*}
\caption{Number of targets from the Perseus sample tagged as SGs through the manual classification that were also identified as such by the different methods considered. Note that we found 241 SGs through the manual classification. Among them, 90 were classified as Ia or Iab, 85 as Ib, and 66 as Ib\,--\,II. Thus, the efficiencies, and their uncertainties (that are equal to $1/\sqrt[]{n}$) are calculated with respect to these values, and modified by the definition of efficiency (an efficiency $>1$ is not possible).}
\label{effi_r_per}
\centering
\begin{tabular}{c | c c c c | c c c c}
\hline\hline
\noalign{\smallskip}
&\multicolumn{4}{c |}{Number of SGs found}&\multicolumn{4}{c}{Efficiency}\\
Method&All&Ia to Iab&Ib&Ib\,--\,II&All&Ia to Iab&Ib&Ib\,--\,II\\
\noalign{\smallskip}
\hline
\noalign{\smallskip}
PCA&182&86&83&13&$0.76\pm0.06$&$0.96^{+0.04}_{-0.11}$&$0.98^{+0.02}_{-0.11}$&$0.20\pm0.12$\\
CaT&204&86&81&37&$0.85\pm0.06$&$0.96^{+0.04}_{-0.11}$&$0.95^{+0.02}_{-0.11}$&$0.56\pm0.12$\\
Ti/Fe&194&83&80&31&$0.80\pm0.06$&$0.92^{+0.08}_{-0.11}$&$0.94^{+0.06}_{-0.11}$&$0.47\pm0.12$\\
\noalign{\smallskip}
\hline
\end{tabular}
\end{table*}

The calibration sample (see \citetalias{dor2016b} for details) is dominated by high- and mid-luminosity SGs (Ia and Iab), with only a small fraction of Ib or less luminous SGs (LC~Ib\,--\,II). In consequence, our PCA method is optimized to find Ia and Iab SGs. In view of this, in the Perseus sample we considered the efficiency for different LCs separately. The efficiencies of the PCA and CaT criteria for high-luminosity SGs are the same, $0.96\pm0.11$, and comparable to those found for the calibration sample. The efficiencies for low luminosity SGs (Ib) are also similar in both methods, and compatible with the results obtained for Ia and Iab stars. However, for the Ib\,--\,II stars the efficiencies are significantly different depending on the criterion used. The higher efficiency of the CaT method in the Ib\,--\,II group stems from the fact that this criterion is much less strict than the PCA one, but at the price of being more susceptible to contamination of red giants (see the following subsection). As the Ib\,--\,II subclass is the boundary between SGs (LC~I) and bright giants (LC~II), the morphology of the objects with this tag is intermediate. Moreover, there are AGB stars, which are not high-mass stars, whose spectra are pretty similar to those of a low luminosity SGs (Ib). The perfect example of this is $\alpha$~Her. This star is the high-luminosity MK standard with the latest spectral type available \citep[M5\,Ib\,--\,II;][]{kee1989}. However, \citet{mor2013} show that this star is not a high-mass star ($M_{*}\ga10\:$M$_{\odot}$), but an AGB star with a mass around $3\:$M$_{\odot}$, even though its spectral morphology is very close to that of a SG. In view of this, through the manual classification we probably identified as SGs stars that are not really SGs, but pretty similar to them morphologically. The PCA criterion, instead, is more restrictive, and only selects as SGs those objects similar enough to the luminous (high-mass) SGs (those having LC Ia and Iab) used to calibrate it.

Our methods, and especially the PCA method, are very efficient for mid- to high-luminosity SGs (Iab to Ia), and also for lower luminosity supergiants (Ib). However, there is also a small number of stars ($6$) manually classified between Ia and Ib that were not identified as SGs by the PCA. All these 6 stars have mid- to late-M types. All but one of them are M5 or later, with most of them (four) having very late SpTs (M7 or M7.5). In fact, these stars are the majority of the RSGs with SpTs M5 or later in the whole Perseus sample, as there are only two other M5\,Ib stars (which were correctly identified by the PCA method). The only star earlier than M5 (it was classified as M3) which was not identified as a SG is S~Per, an extreme RSG (ERSG). The reason why this object was not correctly identified is clear: its lines are weakened by veiling, an effect that may appear in ERSG stars which has been reported before for S~Per \citep{hum1974a}. For more details about ERSGs and veiling, see Section~4.4 from \citetalias{dor2016b} and references therein. 

Just like the PCA method, the CaT and the Ti/Fe criteria fail for mid- to late-M SGs. They failed to identify the same true supergiants that were not found by the PCA. In addition, they also failed for a group of Ia to Ib stars with slightly earlier SpTs (M3 and M4). The obvious conclusion is that all methods fail almost completely in the identification of mid- to late-M RSGs. However, the PCA method provides significantly better results for mid-M SGs (up to M5) than the other criteria. This, in turn, cannot be considered a major drawback, as the number of mid- to late-M RSGs is very small, with only a handful of supergiants presenting spectral types later than M5 (and most of them presenting spectral variability).

\subsubsection{Contamination}

The three identification methods studied above have similar efficiencies for mid- to high-luminosity subsamples. The advantage of the PCA method over the other two is to provide significant lower contaminations, at least for the calibration sample. Therefore, we estimated the contamination obtained through each method for the Perseus sample. The contamination in this case is the fraction of the stars selected as SGs by a given automated criterion that were not identified as real SGs through the manual classification. The results are shown in Table~\ref{contamination_per}.

\begin{table*}
\caption{Contaminations obtained through different methods for the Perseus sample. As the contamination is the fraction of targets tagged as SGs that actually are not SGs, its 2-sigma uncertainty is equal to $1/\sqrt[]{n}$, where $n$ is the number of objects identified as SGs.}
\label{contamination_per}
\centering
\begin{tabular}{c c c c}
\hline\hline
\noalign{\smallskip}
&Number of targets&Number of non-SGs&\\
Method&tagged as SGs&wrongly identified&Contamination\\
\noalign{\smallskip}
\hline
\noalign{\smallskip}
PCA&193&11&$0.06\pm0.07$\\
CaT&304&100&$0.33\pm0.06$\\
Ti/Fe&238&43&$0.18\pm0.07$\\
\noalign{\smallskip}
\hline
\end{tabular}
\end{table*}

The method with the lowest contamination is by far PCA. All the non-SGs wrongly selected by the $P(\mathrm{SG})$ have LC~II in the manual classification, and therefore their spectra are very similar morphologically to those of low-luminosity RSGs. Indeed, we cannot dismiss a priori the possibility that they may be low-luminosity SGs wrongly identified in the manual classification. The Ti/Fe criterion has a significantly higher contamination, but the CaT criterion works significantly worse than the other two in this respect. This is not completely unexpected, as the strength of the CaT lines is not only a function of luminosity, but also effective temperature and metallicity \citep{dia1989}.

The contamination found in the Perseus sample through the PCA method ($0.06\pm0.07$) is compatible with those obtained for the calibration sample ($0.03\pm0.04$) and its MW subset ($0.08\pm0.13$) in \citetalias{dor2016b}. In the case of the CaT and Ti/Fe methods, their contaminations when applied to the MW subset of the calibration sample are $0.17\pm0.13$ for the Ti/Fe criterion and $0.20\pm0.13$ for the CaT criterion, which are again compatible with those obtained in this work for these methods (see Table~\ref{contamination_per}). Therefore the results for the Perseus sample corroborate the conclusions that we reached based on the subsample of MW stars in the calibration sample in \citetalias{dor2016b}, this time for a significantly larger sample.

\subsection{The population of cool supergiants in Perseus}
\label{gal_pol}

As explained in Section~\ref{prob_per}, with the values proposed for the $P(\mathrm{SG})_{\mathrm{PCA}}$ we identified 191 targets as SGs in Perseus (86 of them having LC~Ia or Iab according to the manual classification), while our manual identification found 258 (96 of them having LC~Ia or Iab), including all the 191 PCA SGs. The difference between both sets is mainly due to Ib\,--\,II stars, which, as discussed above, may in fact not be true SGs, but bright giants. The rest of the difference is due to the late-M stars, which are not correctly selected by any of the automated criteria studied, even though their SG nature is very likely. Thus, for the present analysis we decided to adopt the PCA selection, but also include the five SGs (Ia to Ib) with late subtypes (M5 to M7) that were identified through manual classification, as well as S~Per, which is a well-known ERSGs (see Sect.~\ref{efficiency}).

The supergiant content of the Perseus arm was studied by \cite{hum1970,hum1978}, who found more than 60 CSGs in this region. Later, \cite{lev2005} studied the RSG population of the Galaxy, adding a handful of new stars to the list of known RSGs in the Perseus arm. We also took into account a small number of CSG standards from \cite{kee1989} located in the Perseus arm. Using these works and crossing their lists, we obtained a list of 77 previously known CSGs in the Perseus arm. Among the 197 CSGs we found, there are only six that were included in this list. Thus, our work increases the number of CSGs known in Perseus in 191 stars, more than trebling the size of previous compilations (from 77 to 268 CSGs). 

This large number of CSGs allows us to study statistically the population of CSGs in the Perseus arm with unprecedented significance. Indeed, this sample permits a direct comparison of the CSG population in the Perseus arm and those in the MCs studied in \citetalias{dor2016a}. For this analysis, we used the SpT and LC given through the manual classification for the CSGs in our Perseus sample, and the classification given in the literature for the rest of the Perseus SGs that had gone to the calibration sample. Unfortunately, the distances to many of these stars still have significant uncertainties, which do not allow us to compare absolute magnitudes. However, in the near future \textit{Gaia} will provide reliable and homogeneous distances for almost all of them. We will then use these distances together with the radial velocities obtained from our spectra (which can be compared to the \textit{Gaia}/RVS radial velocities to detect binarity) to study in detail the spatial and luminosity distributions for the CSG population in the Perseus arm. In the present work we only analyse the SpT and LC distributions.

When previous works have analysed a given population of RSGs, they have typically found their SpTs to be distributed around a central subtype with maximum frequency. In all populations, the frequency of the subtypes is lower the farther away from the central value the subtype is. The central subtype is related to the typical metallicity of the population, with later types for higher metallicities \citep{hum1979a,eli1985}. This effect has been confirmed by recent works for different low-metallicity environments \citep{lev2012}. In \citetalias{dor2016a} we confirmed this effect for very large samples in both MCs.

\begin{figure}
   \centering
   \includegraphics[trim=1cm 0.4cm 1.8cm 1.3cm,clip,width=\columnwidth]{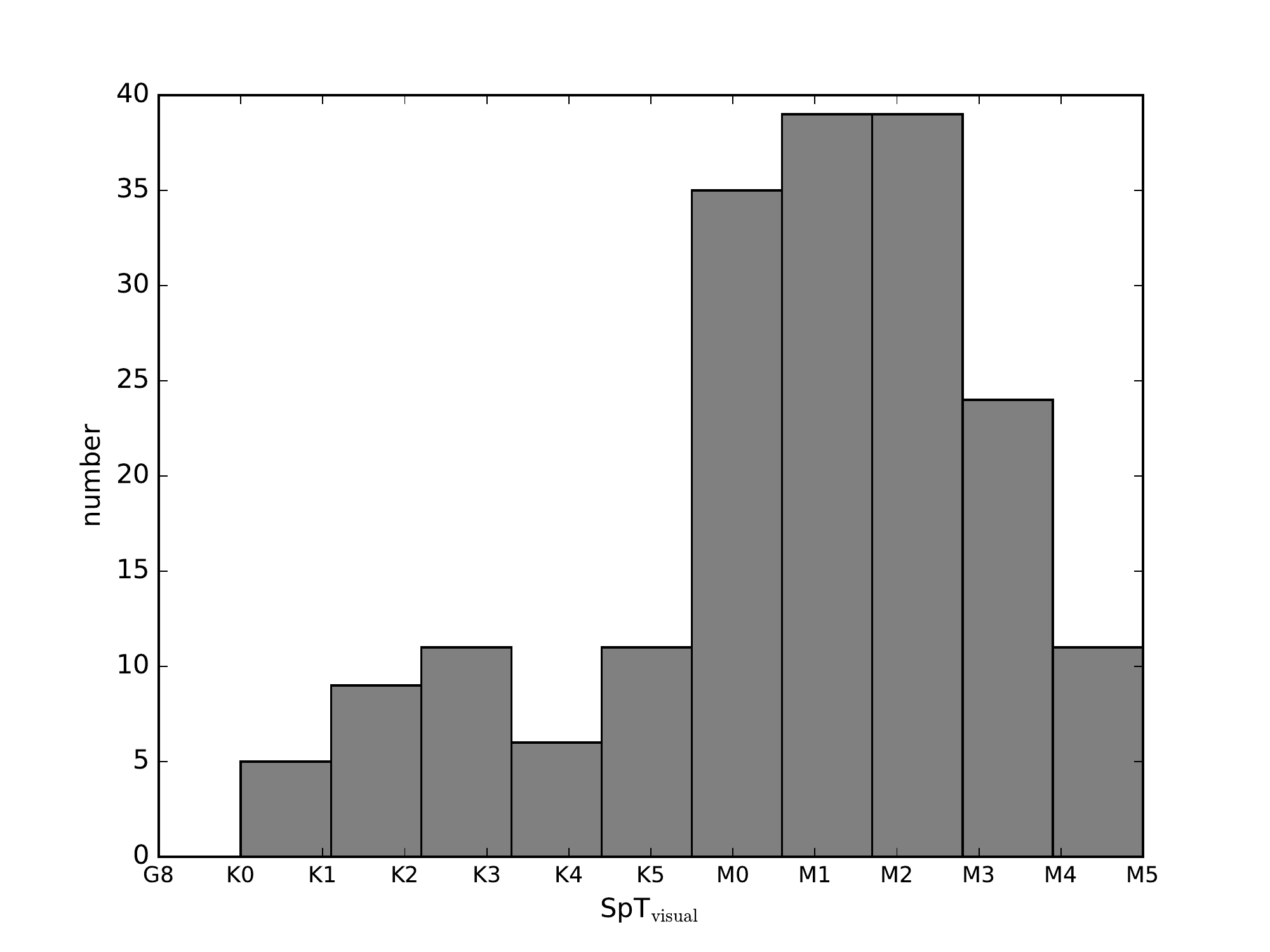}
   \caption{Distribution of SpTs for the targets identified as SGs using the PCA method.}
   \label{histo_spt}
\end{figure}

\begin{figure*}
   \centering
   \includegraphics[trim=1cm 0.35cm 1.8cm 1.35cm,clip,width=\columnwidth]{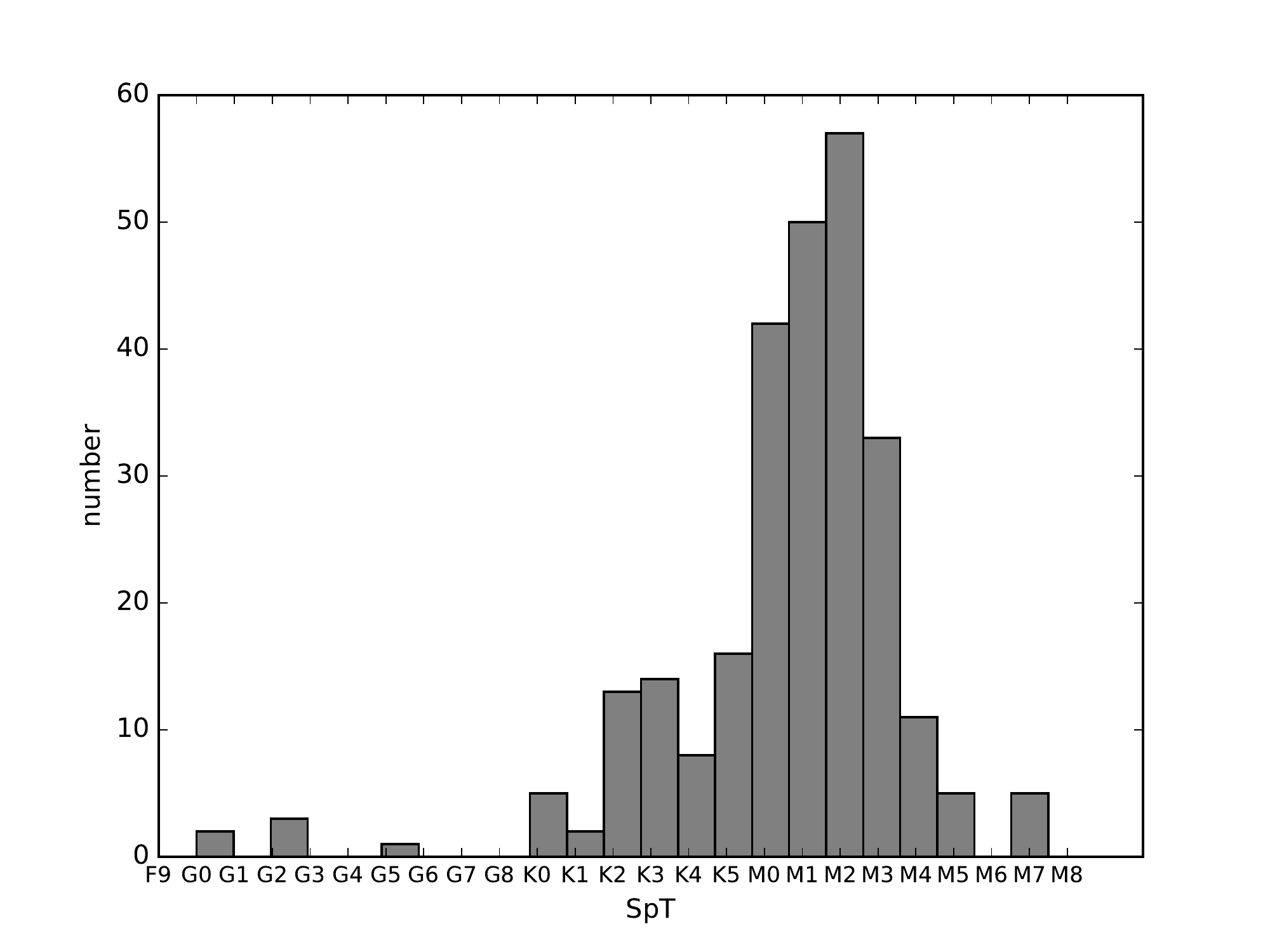}
   \includegraphics[trim=1cm 0.35cm 1.8cm 1.35cm,clip,width=\columnwidth]{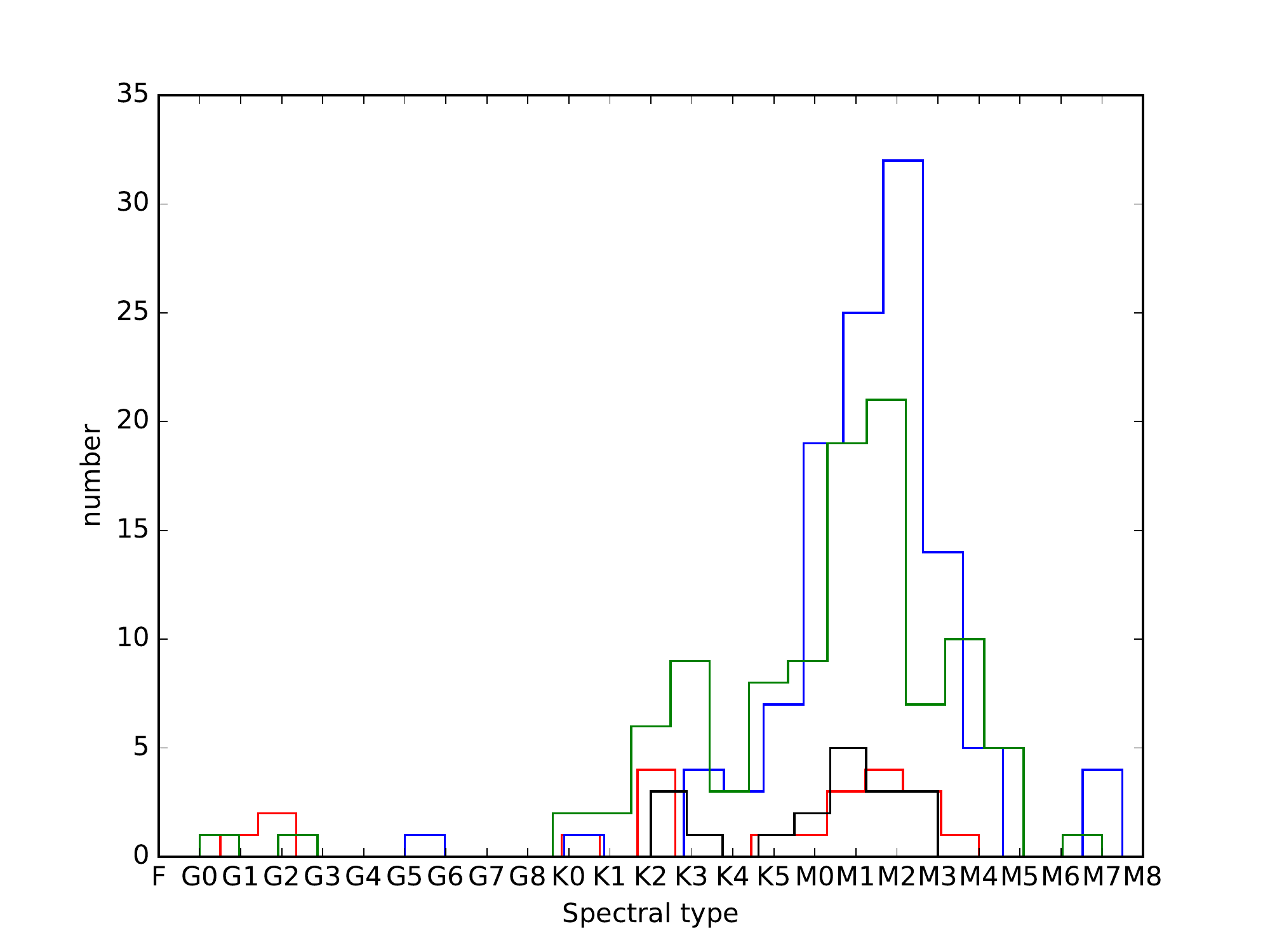}
   \caption{
   {\bf Left (\ref{histo_spt_all_per}a):} Distribution of SpTs for Perseus CSGs (our sample plus previous identifications).
   {\bf Right (\ref{histo_spt_all_per}b):} The same sample as in left panel, but split by luminosity class, with red for Ia, blue for Iab, green for Ib, and black for Ib\,--\,II.
   }
   \label{histo_spt_all_per}
\end{figure*}

The SpT distribution of the Perseus CSGs found in the present work (the PCA selection plus the six late RSGs visually identified) is shown in Fig.~\ref{histo_spt}. The median SpT of this sample is M1. We also studied the global population (268 CSGs), which includes all the previously known RSGs from the Perseus arm together with all our newly-found CSGs. Its histogram is shown in Fig.~\ref{histo_spt_all_per}a. Addition of the set of previously known RSGs not included in our own sample (see Fig.~\ref{histo_spt_all_per}a) shifts very slightly the median type to M1.5. Both median types are slightly earlier than values typically given for the MW according in the literature \citep[M2;][]{eli1985,lev2013}. However, the difference is not large enough to be truly significant, given the typical uncertainty of one subtype in our manual classifications. We can thus consider our results consistent with the value found in the literature. Despite this, we note that our sample is intrinsically different from any previous sample of Galactic RSGs. With the possible exception of a few background RSGs (which could be present given our magnitude cut, but should be very rare, because of the steeply falling density of young stars towards the outer Milky Way), our sample is volume-limited; it represents the total RSG population for a section of a Galactic arm. Previous works are mostly magnitude-limited and therefore tend to include an over-representation of later-type M supergiants, as these objects tend to be intrinsically brighter (see \citetalias{dor2016a} and references therein).

The SpT distribution shown presents a clear asymmetry due to the presence of a local maximum at early-K types. This local maximum was not detected by \cite{eli1985}, but is present in \cite{lev2013}, in their figure~1. The SGs considered in \cite{eli1985} were mainly of luminosity classes Ia and Iab, while most of the early-K SGs used in \cite{lev2013} are of Ib class. This is also the case in our sample; most early-K (K0\,--\,K3) supergiants present low luminosity classes (Ib or less, see Fig.~\ref{histo_spt_all_per}b). Studies of similar stars in open clusters \citep[e.g.][]{neg2012b,alo2017}, show that these low-luminosity supergiants with early-K types are in general intermediate-mass stars (of $6$\,--\,$8\:$M$_{\sun}$), with typical ages ($\sim50\:$Ma) much older than luminous RSGs (typically between $10$ and $25\:$Ma). Therefore, despite their morphological classification as SGs, these stars should not be considered as true supergiants, because they are not quite high-mass stars. These stars are not very numerous in our sample (we have 19~stars with early-K types and LC Ib or less luminous) nor in the total population (23~stars). Therefore, our median types do not change if we do not consider these stars as part of the CSG population. It is worthwhile stressing that there are very few K-type true supergiants in the Milky Way, to the point that the original list of MK standards contains only one such object (the K3\,Iab standard $o^{1}$~CMa, later moved to K2.5\,Iab; \citealt{mk73}), as opposed to five K\,Ib stars, representative of the lower-mass population discussed above \citep[see][]{johnson53}. This absence of K-type SGs represents the main difference between the present catalogue and those from the MCs, as illustrated by Fig.~\ref{histo_spt_all_per}b.

In \citetalias{dor2016a}, we found that RSGs in the MCs present a relation between SpTs and LCs, with later typical types for Ia than for Iab stars. As a consequence, we found an earlier typical SpT for each MC than in previous works by a few subtypes. This difference was caused by the inclusion in our survey of a large number of Iab CSGs, while previous studies studied were centred on the brightest RSGs, mostly Ia (see sect.~4.2 of \citetalias{dor2016a}). In contrast, when we analyse the different LC subsamples in Perseus, we do not find any significant difference between Ia and Iab stars, as both groups have the same median SpT: M1 (see Fig.~\ref{histo_spt_all_per}b). When we consider the global population, Iab supergiants have a median type of M1.5, but a difference of half a subtype cannot be considered significant. These results contrast strongly with the trends found in the MCs. It is unclear, though, if we can derive any reliable conclusions from this difference, because the number of Ia stars in the Perseus sample is too low compared to the number of Iab stars: seven~Ia against 83~Iab in our sample; 19~Ia against 116~Iab in the global population.

There are a number of factors to consider before attempting any interpretation. Firstly, there are four early K-type Ia SGs pushing the median type to early types. As mentioned, these spectral types are rare in the MW, and many of these objects present unusual characteristics, such as evidence for binary interaction or heavy mass loss. Due to the small size of the Ia sample, these rare objects may have a disproportionate impact on the average type. Moreover, we may be biasing our sample because of a classification issue: there are no MK SG standards for spectral types later than M4 (except for $\alpha$~Her, mentioned above, which is not a true SG). At these spectral types, luminosity indicators are strongly affected by the molecular bands, specially TiO bands. In fact, for types later than M3, many luminosity indicators (e.g.\ the Ca Triplet) do not separate RSGs from red giants (\citealt{dor2013}, \citetalias{dor2016a}, and \citetalias{dor2016b}). Our sample contains a number of RSGs with mid to late types, which were given a generic I classification, as it was not possible to give a more accurate luminosity subclass \citep[see discussion in][]{neg2012}. For calculation purposes, these objects have been assigned to the intermediate luminosity Iab. This could be incorrect, as the few late-M RSGs found in open clusters tend to have much higher luminosities than earlier RSGs in the same clusters \citep{neg2013,mar2013}.

Within our sample, we have an interesting example of the situation explained above in the cluster NGC~7419. This rich cluster contains five RSG members; four of them have M0 to M2\,Iab types, while the last one, MY~Cep, is M7.5\,I \citep{mar2013}. As can be seen in fig.~13 of \cite{mar2013}, MY~Cep is about one and a half magnitude more luminous than the other 4 RSGs. As MY~Cep was the only comparison star available for the manual classification of the late RSGs in our sample, it is reasonable to expect that the three stars classified as M7\,I could also be high luminosity RSGs, as MY~Cep is. Four other Ia stars present types M3 to M4. One of them is S~Per, a known spectral variable that can present types as late as M7, according to \cite{faw1977}. In view of this, it is highly likely that we are underestimating the number of late-M Ia~RSGs. Even though these are also rare objects, given the small size of the Ia sample, they could move the median to later types. In this context, it is important to note that the MC populations studied in \citetalias{dor2016a} include very few mid- or late-M supergiants. Most MC Ia RSGs were M3 or earlier, allowing their LC classification without the complications that affect luminosity indicators at later types. In addition, the distance to the RSGs in the MCs is well known, allowing a direct knowledge of the actual luminosity. In the Perseus sample, we have to resort only to morphological characteristics in most cases, at least until accurate distances are provided by \textit{Gaia}. 

The low number of Ia SGs may be meaningful in itself. On one side, magnitude-limited samples will always have a bias towards intrinsically bright stars that is not present in the Perseus sample. On the other side, the sample of CSGs in the SMC presented in \citetalias{gon2015}, which may not be complete, but is at least representative, has a much higher fraction of Ia supergiants with respect to the Iab cohort. As discussed in \citetalias{dor2016a}, there may be two different pathways leading to high-luminosity CSGs. Since stellar evolutionary models \citep{eks2012,geo2013,bro2011} indicate that evolution from the hot to the cool side of the HR diagram happens at approximately constant luminosity, the brightest CSGs should be descended from more massive stars (with masses $\sim25\:$M$_{\sun}$ and up to $\sim40\:$M$_{\sun}$). On the other hand, observations of open clusters \citep{neg2013,beasor16} suggest that less massive stars (with masses between $10$ and $\sim20\:$M$_{\odot}$) could evolve from typical Iab CSGs towards higher luminosities and cooler temperatures at some point in their lives. This idea is suggested by the presence in massive clusters of some RSGs with significantly later SpTs and much higher luminosities than most of the other RSGs in the same cluster (as in the example of NGC~7419 mentioned above).

The low fraction of Ia CSGs in the Perseus arms may shed some light on these issues. Although there are some very young star clusters and associations (mainly Cep~OB1 and Cas~OB6) in the area surveyed, most of the clusters and OB associations are not young enough to still have any RSGs with high masses ($\ga20\:$M$_{\sun}$). The most massive clusters included in the sample region have ages around $15\:$Ma, with main-sequence turn-offs at B1\,V. This is the case of NGC~7419 \citep{mar2013} or the double Perseus cluster, the core of the Perseus OB1 association \citep{sle2002}, while the clusters in Cas OB8 are even older. For an age $\sim15\:$Ma, according to Geneva evolutionary models \citep{eks2012}, RSGs should be descended from stars with an initial mass $\sim15\:$M$_{\odot}$ and not be much more luminous than $M_{\textrm{bol}}\sim-7$. As can be seen in fig.~16 of \citetalias{dor2016a}, most Ia RSGs are more luminous than this value. Therefore, the scarcity of Ia RSGs in Perseus can be interpreted as a straight consequence of the lack of high-mass RSGs, which supports the idea that Ia CSGs come mainly from stars with initial masses between $20$ and $40\:$M$_{\odot}$. However, there still is a significant fraction ($0.07\pm0.06$) of Ia stars, which are not directly related to any very young cluster. For example, following with the example of Per~OB1, this association contains the well known ERSG S~Per \citep{hum1978}, which has been observed to vary from M4 to M7\,Ia. This suggests that indeed some intermediate mass RSGs may increase their luminosity up to LC~Ia from lower luminosities. Their low number in the sample agrees with small fraction of very luminous RSGs found in massive open clusters. 

\subsection{Candidates to extreme red supergiants}

\begin{figure}
  \centering
  \includegraphics[trim=1cm 0.4cm 2.3cm 1.2cm,clip,width=\columnwidth]{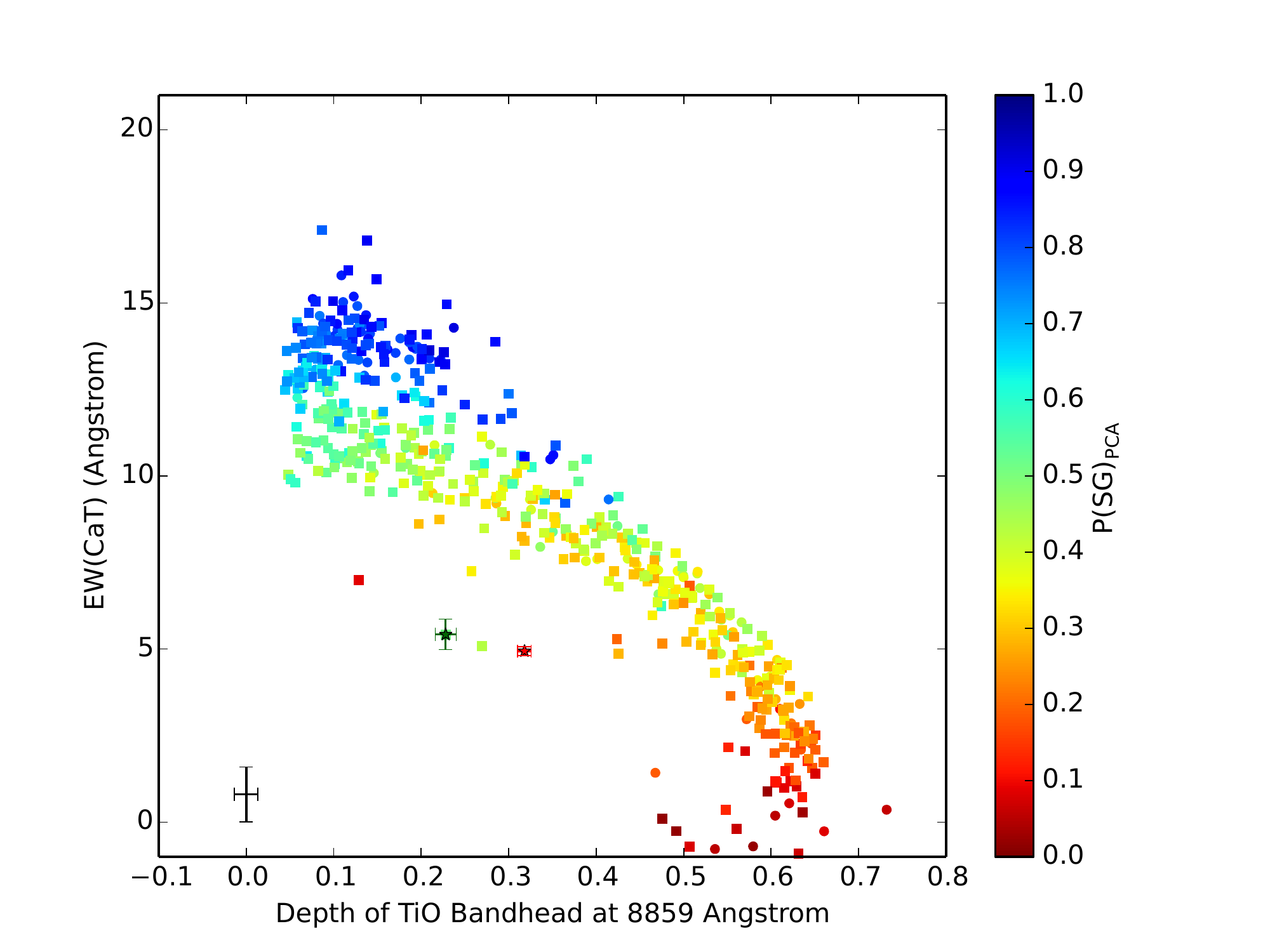}
  \caption{Depth of the TiO bandhead at $8859$\:\AA{} with respect to the sum of the EWs of the CaT lines, for the Perseus sample. The colour indicates $P(\mathrm{SG})_{\mathrm{PCA}}$, and the shapes indicate epoch (2011 circles, 2012 squares), except for the two stars, which are reference ERSGs. The green star is the S~Per and the red star is UY~Sct. Both ERSGs are represented with their own error bars. The black cross indicates the median uncertainties of the sample. The scale used in this Figure is the same as for Fig.~14a from \citetalias{dor2016b}, which show the same diagram for the calibration sample, to ease the comparison.}
  \label{TiO_CaT_ERSG}
\end{figure}

In \citetalias{dor2016b} we proposed the use of two diagrams to detect RSGs affected by veiling, a characteristic effect that ERSGs present at some points in their spectral variation \citep[for details about veiling see][and Section~4.4 in \citetalias{dor2016b}]{hum1974a}. In Fig.~\ref{TiO_CaT_ERSG} we include the location of the two veiled ERSGs, UY~Sct and S~Per (which indeed is one of the stars in the Perseus sample), that were available to us. They indicate the typical region where veiled ERSGs seem to lie. For the Perseus sample we found only one star close to them, outside the main band of giant and supergiant stars. This object, PER433, was rejected as a SG by the $P(\mathrm{SG})_{\mathrm{PCA}}$ (and also by the other methods), but given the effect of veiling on atomic lines, this rejection cannot be considered conclusive. In the bibliography this object, known as V627~Cas, has been identified as some kind of symbiotic star \citep{kol1996}. We checked its spectrum and found that it shows the O\,{\sc{i}} line at 8448\:\AA{} in emission, which is usual in Be stars, but not expected in ERSGs, since it requires higher temperatures. It also has its CaT lines in emission, partially filling them, which explains why this star shows EW(CaT) much smaller than expected for a giant star. Therefore, we can conclude that this star is not an ERSG.

\section{Conclusions and future work}

In \citetalias{dor2016b} we proposed a method for using PCA in the identification of CSGs. In the present work we have developed it further, obtaining a way to estimate the probability that a given spectrum is a CSG, instead of just giving a binary result (``SG" or ``non-SG"). We have then applied this method to a large sample of galactic stars selected to be part of the Perseus arm. We also compared the results obtained through the method using PCA with two other classical criteria studied in \citetalias{dor2016b} (those based on the CaT and Ti/Fe criteria). Summarising, from the analysis presented in this work we can conclude:

\begin{enumerate}

\item We find that the efficiencies of all three automated methods are similarly high ($>90\%$) for objects which were visually classified as certain CSGs (Ia to Ib), and compatible with those obtained for the calibration sample in \citetalias{dor2016b}. The results are much worse in the case of those targets visually classified as Ib\,--\,II for the three methods, and especially for the PCA one. However, this group of LC Ib\,--\,II objects is probably formed mostly by non-SGs and the automated methods could be simply pointing this out. Finally, we find that the efficiency is almost zero for stars visually identified as SGs having subtypes later than M5, independently of the method used.

\item Although the efficiencies are similarly good in the three cases, the contaminations are very different for each method, when manual classification is used as a reference. As in the case of the MCs, the PCA method provides the cleanest sample of SGs, with a contamination fraction as low as $0.06\pm0.07$, against $0.33\pm0.06$ and $0.18\pm0.07$ for the CaT and Ti/Fe criteria. The contamination found for the PCA method is compatible with that obtained for the calibration sample of \citetalias{dor2016b}. However, the other two methods result in values significantly higher, probably because the Perseus sample has a larger fraction of bright M giants than our MC samples, because in this case we are observing through the Galactic plane.

\item Using the PCA method, we identified 191 targets as CSGs, plus 6 RSGs with late SpTs which were identified through the manual classification. These 197 CSGs are a significant fraction of the total sample ($0.33\pm0.04$), demonstrating that the photometric selection criteria used have a very high efficiency at moderate reddenings. This sample represents the largest catalogue of CSGs in the MW observed homogeneously, increasing the census of catalogued CSGs in the Perseus arm dramatically: to the 77 CSGs contained in previous lists, this catalogue adds 191 more objects. The list of stars observed, with their corresponding probabilities of being a SG through different methods is given in Tables~\ref{cat_perseo}.

\end{enumerate}

The final catalogue, with almost 200~CSGs, is the largest coherent sample of CSGs observed to date in the Galaxy. In the future, we will use this sample to study both the CSG population and its relation to structure ofthe Perseus arm. We will use the radial velocities that we can obtain from our spectra, along with \textit{Gaia} distances (which will be available for these stars by mid 2018), to study the spatial distribution of the CSGs in the Perseus arm and their relation with nearby clusters and OB associations. In addition, we will also analyse the physical properties of these stars, deriving them from their spectra by using the method that we are developing (Tabernero et al. in prep.). Finally, it is our intention to extend the study of CSG populations toward the inner Galaxy, where we should find higher metallicities, but will also have to fight much higher extinction and stellar densities

\section*{Acknowledgements}

We thank the referee, Prof. Roberta Humphreys, for the swiftness of her response.
The INT is operated on the island of La Palma by the Isaac Newton Group in the Spanish Observatorio del Roque de Los Muchachos of the Instituto de Astrof\'{\i}sica de Canarias. This research is partially supported by the Spanish Government Ministerio de Econom\'{\i}a y Competitivad (MINECO/FEDER) under grant AYA2015-68012-C2-2-P. This research has made use of the Simbad, Vizier 
and Aladin services developed at the Centre de Donn\'ees Astronomiques de Strasbourg, France.  This research has made use of the WEBDA database, operated at the Department of Theoretical Physics and Astrophysics of the Masaryk University. It also
makes use of data products from the Two Micron All Sky Survey, which is a joint project of the University of
Massachusetts and the Infrared Processing and Analysis Center/California Institute of Technology, funded by the National
Aeronautics and Space Administration and the National Science
Foundation.
This research has made use of the SIMBAD database, operated at CDS, Strasbourg, France

\bibliographystyle{mnras}
\bibliography{general} 

\appendix
\onecolumn

\section{Sample observed}
\begin{longtable}[h]{c | c c | c c | c c | c c c }
\caption{Small sample of the stars we observed in the Perseus arm. We include for each target the observation epoch, the manual classification done, the calculated probabilities of being a SG (obtained through the PCA method as well as the CaT or Ti/Fe criteria). For more details see Section~\ref{analysis}. \label{cat_perseo}}\\
\hline\hline
\noalign{\smallskip}
&RA&DEC&l&b&&Visual&&&\\
ID&J2000&J2000&(deg)&(deg)&Epoch&Classification&P$_{\mathrm{PCA}}$&P$_{\mathrm{CaT}}$&P$_{\mathrm{Ti/Fe}}$\\
\noalign{\smallskip}
\hline
\noalign{\smallskip}
\endfirsthead
\caption{continued.}\\
\noalign{\smallskip}
&RA&DEC&l&b&&Visual&&&\\
ID&J2000&J2000&(deg)&(deg)&Epoch&Classification&P$_{\mathrm{PCA}}$&P$_{\mathrm{CaT}}$&P$_{\mathrm{Ti/Fe}}$\\
\noalign{\smallskip}
\hline
\noalign{\smallskip}
\endhead
PER001&0:00:10.00&+62:27:36.0&117.044&0.17579&2011&M6.0\:II&0.429&0.035&0.251\\
PER002&0:00:18.00&+60:21:02.0&116.644&-1.89536&2011&M4.5\:Ib\,--\,II&0.513&0.345&0.299\\
PER003&0:01:44.20&+62:11:23.8&117.171&-0.12456&2012&M3.0\:II&0.595&0.998&0.5\\
PER004&0:01:46.90&+64:16:36.8&117.575&1.92282&2012&M6.0\:II\,--\,III&0.364&0.0&0.073\\
PER005&0:02:20.00&+57:02:14.1&116.261&-5.19714&2012&M8.0\:III&0.117&0.0&0.0\\
PER006&0:02:59.00&+61:22:05.0&117.160&-0.95948&2011&M3.0\:Ib\,--\,II&0.39&0.998&0.636\\
PER007&0:04:10.80&+60:55:22.3&117.220&-1.42367&2012&C star&--&--&--\\
PER008&0:06:39.00&+58:02:18.0&117.015&-4.31797&2011&M5.0\:Ib\,--\,II&0.357&0.157&0.277\\
PER009&0:08:58.40&+62:42:57.0&118.087&0.24419&2012&C star&--&--&--\\
PER010&0:09:26.30&+63:57:14.0&118.341&1.45716&2012&M2.0\:Iab&0.879&1.0&1.0\\
PER011&0:10:49.60&+64:51:14.2&118.633&2.32174&2012&M6.5\:II&0.374&0.009&0.14\\
PER012&0:12:21.60&+62:53:33.6&118.497&0.35787&2012&K0.0\:Iab&0.64&1.0&1.0\\
PER013&0:13:23.20&+63:27:31.2&118.696&0.90028&2012&C star&--&--&--\\
PER014&0:14:57.50&+66:37:30.3&119.318&4.00967&2012&M1.0\:II&0.498&0.998&0.641\\
PER015&0:15:01.00&+66:06:50.2&119.251&3.50283&2012&K3.0\:Ib\,--\,II&0.501&1.0&0.939\\
PER016&0:16:42.30&+67:33:02.8&119.615&4.90300&2012&M4.0\:II&0.408&0.353&0.443\\
PER017&0:16:54.90&+57:31:51.1&118.293&-5.02962&2012&M7.5\:II&0.332&0.0&0.065\\
PER018&0:18:23.00&+61:52:28.0&119.047&-0.74769&2011&M6.0\:II&0.278&0.0&0.073\\
PER019&0:18:26.40&+60:54:08.9&118.929&-1.71255&2012&M1.0\:Iab&0.811&1.0&1.0\\
PER020&0:19:03.00&+57:43:42.2&118.603&-4.87090&2012&M6.0\:III&0.329&0.0&0.048\\
PER021&0:19:04.10&+66:22:13.1&119.691&3.70249&2012&C star&--&--&--\\
PER022&0:20:43.50&+61:52:46.5&119.321&-0.77665&2012&M1.0\:Iab&0.794&1.0&1.0\\
PER023&0:21:31.80&+61:31:13.5&119.374&-1.14453&2012&M3.0\:Ib\,--\,II&0.445&0.837&0.358\\
PER024&0:22:26.80&+59:11:33.4&119.220&-3.47006&2012&C star&--&--&--\\
PER025&0:23:17.00&+62:21:39.0&119.673&-0.33264&2011&M6.5\:III&0.215&0.0&0.012\\
PER026&0:26:18.00&+61:41:19.0&119.956&-1.03778&2011&M6.0\:II&0.311&0.007&0.077\\
PER027&0:26:18.70&+61:32:03.1&119.942&-1.19160&2012&M1.5\:II&0.543&1.0&0.755\\
PER028&0:27:11.00&+63:33:25.0&120.236&0.81225&2011&M6.0\:Ib\,--\,II&0.444&0.001&0.049\\
PER029&0:27:29.00&+59:19:47.7&119.876&-3.39962&2012&M5.0\:III&0.407&0.005&0.196\\
PER030&0:28:08.50&+60:29:29.5&120.065&-2.25047&2012&M6.0\:Ib\,--\,II&0.262&0.0&0.047\\
PER031&0:28:40.00&+63:27:40.0&120.392&0.70174&2011&M2.0\:Ib&0.786&1.0&1.0\\
PER032&0:29:48.50&+60:29:43.4&120.270&-2.26451&2012&K2.0\:Ib&0.626&1.0&1.0\\
PER033&0:30:26.20&+67:00:06.2&120.877&4.21378&2012&M1.5\:II&0.396&1.0&0.86\\
PER034&0:30:59.50&+61:26:19.0&120.490&-1.33615&2012&M0.0\:Ib&0.611&1.0&0.962\\
PER035&0:31:25.40&+60:15:19.5&120.449&-2.51982&2012&M0.0\:II&0.492&1.0&0.699\\
PER036&0:33:47.10&+58:15:37.3&120.605&-4.53160&2012&M5.0\:III&0.387&0.067&0.251\\
PER037&0:34:31.00&+61:56:42.0&120.944&-0.86140&2011&M7.0\:III&0.201&0.0&0.01\\
PER038&0:35:02.70&+61:19:02.1&120.966&-1.49189&2012&C star&--&--&--\\
PER039&0:35:26.00&+61:14:48.0&121.008&-1.56527&2011&M7.0\:II&0.249&0.0&0.024\\
PER040&0:35:37.10&+67:55:33.0&121.441&5.09990&2012&K3.0\:Iab&0.728&1.0&1.0\\
PER041&0:35:41.10&+64:09:07.9&121.216&1.33296&2012&C star&--&--&--\\
PER042&0:35:42.00&+63:07:47.0&121.155&0.31229&2011&M0.0\:Ib&0.756&1.0&1.0\\
PER043&0:37:16.50&+58:46:24.2&121.093&-4.04777&2012&M5.0\:III&0.313&0.066&0.158\\
PER044&0:38:28.00&+63:14:09.0&121.472&0.40086&2011&M2.0\:Iab&0.811&1.0&1.0\\
PER045&0:38:42.40&+61:43:57.4&121.425&-1.10188&2012&M2.0\:Ib\,--\,II&0.423&0.998&0.574\\
PER046&0:40:01.00&+62:19:41.0&121.606&-0.51429&2011&M0.5\:Ib&0.815&1.0&1.0\\
PER047&0:40:24.80&+59:30:49.7&121.532&-3.32797&2012&M2.0\:Ib&0.618&1.0&0.913\\
PER048&0:40:28.00&+64:17:33.0&121.742&1.44610&2011&M1.5\:Ib&0.797&1.0&1.0\\
PER049&0:41:24.10&+59:24:41.4&121.653&-3.43533&2012&M0.0\:Ib\,--\,II&0.53&1.0&0.803\\
PER050&0:43:51.00&+62:16:51.0&122.050&-0.57791&2011&M1.0\:Iab&0.868&1.0&1.0\\
PER051&0:44:00.90&+58:56:05.5&121.972&-3.92305&2012&M6.0\:II&0.443&0.0&0.081\\
PER052&0:44:32.00&+62:07:15.0&122.125&-0.74009&2011&M0.0\:Ib&0.706&1.0&1.0\\
PER053&0:45:51.50&+58:02:18.4&122.191&-4.82528&2012&M6.0\:II&0.358&0.025&0.133\\
PER054&0:48:34.00&+62:04:23.0&122.596&-0.79682&2011&M5.0\:Ib\,--\,II&0.368&0.086&0.213\\
PER055&0:49:11.00&+64:56:19.0&122.693&2.06791&2011&M3.0\:Ib&0.827&1.0&1.0\\
PER056&0:49:17.60&+63:10:05.3&122.690&0.29740&2012&M4.5\:II\,--\,III&0.498&0.407&0.246\\
PER057&0:50:25.00&+63:03:05.0&122.816&0.17987&2011&M4.5\:Ib\,--\,II&0.471&0.065&0.162\\
PER058&0:50:38.40&+60:13:07.1&122.833&-2.65296&2012&S star&0.509&0.864&0.966\\
PER059&0:52:49.70&+57:24:23.7&123.120&-5.46466&2012&M5.0\:III&0.297&0.113&0.196\\
PER060&0:53:38.00&+63:20:29.0&123.178&0.47069&2012&C star&--&--&--\\
PER061&0:54:53.80&+58:33:49.2&123.384&-4.30504&2012&C star&--&--&--\\
PER062&0:55:09.80&+57:16:34.1&123.438&-5.59195&2012&M1.0\:Ib&0.667&1.0&1.0\\
PER063&0:57:31.70&+60:20:12.2&123.686&-2.52611&2012&M6.0\:II\,--\,III&0.411&0.007&0.144\\
PER064&0:57:35.80&+61:28:08.1&123.667&-1.39403&2012&M7.0\:III&0.245&0.0&0.0\\
PER065&0:58:02.00&+62:49:32.0&123.685&-0.03655&2011&M2.0\:Iab&0.814&1.0&1.0\\
PER066&0:58:04.30&+67:41:51.8&123.563&4.83411&2012&M1.5\:II&0.457&1.0&0.519\\
PER067&0:58:12.30&+59:34:24.0&123.790&-3.28714&2012&M1.0\:Iab&0.8&1.0&1.0\\
PER068&1:00:26.00&+63:33:16.0&123.933&0.70019&2011&M3.0\:II&0.526&0.998&0.89\\
PER069&1:01:58.40&+57:59:48.2&124.332&-4.84637&2012&M7.5\:III&0.191&0.0&0.02\\
PER070&1:02:43.60&+61:51:43.0&124.263&-0.98063&2012&C star&--&--&--\\
PER071&1:02:44.00&+60:36:15.4&124.319&-2.23705&2012&M6.0\:III&0.28&0.019&0.198\\
PER072&1:02:55.70&+60:58:23.9&124.326&-1.86733&2012&M0.5\:Ib&0.7&1.0&1.0\\
PER073&1:03:15.00&+63:05:11.0&124.268&0.24529&2011&M5.0\:Ib\,--\,II&0.389&0.204&0.337\\
PER074&1:03:33.60&+61:12:31.5&124.392&-1.62864&2012&M2.0\:II&0.44&0.999&0.54\\
PER075&1:04:36.30&+61:22:44.9&124.509&-1.45229&2012&C star&--&--&--\\
PER076&1:05:16.00&+62:29:10.0&124.528&-0.34277&2011&M1.5\:Ib&0.77&1.0&1.0\\
PER077&1:05:23.90&+62:21:24.6&124.550&-0.47103&2012&M0.5\:Iab&0.845&1.0&1.0\\
PER078&1:06:30.00&+57:34:00.4&124.959&-5.24568&2012&M3.5\:Ib\,--\,II&0.477&0.947&0.406\\
PER079&1:06:59.70&+63:46:23.4&124.650&0.95332&2012&G2.0\:Ia&0.689&1.0&0.017\\
PER080&1:07:53.00&+63:25:11.6&124.770&0.60686&2012&K3.0\:Iab&0.815&1.0&1.0\\
PER081&1:08:55.70&+61:10:54.3&125.039&-1.61842&2012&M4.5\:II&0.295&0.109&0.193\\
PER082&1:09:13.00&+65:07:02.0&124.802&2.31005&2011&M2.0\:Ib&0.77&1.0&1.0\\
PER083&1:09:42.20&+62:25:09.8&125.044&-0.37742&2012&M8.0\:III&0.179&0.0&0.001\\
PER084&1:09:44.50&+57:03:52.6&125.430&-5.71832&2012&M2.0\:II&0.49&1.0&0.295\\
PER085&1:10:20.10&+62:30:39.8&125.111&-0.28073&2012&M2.0\:Iab&0.846&1.0&1.0\\
PER086&1:11:32.50&+56:56:20.7&125.685&-5.82514&2012&M7.5\:III&0.235&0.001&0.029\\
PER087&1:12:57.60&+59:52:39.6&125.633&-2.88118&2012&M5.0\:III&0.285&0.33&0.089\\
PER088&1:14:56.60&+59:42:13.3&125.897&-3.03268&2012&M9.0\:III&0.088&0.0&0.001\\
PER089&1:15:19.20&+57:21:40.4&126.162&-5.36067&2012&M4.0\:II&0.389&0.65&0.285\\
PER090&1:16:04.70&+58:33:48.4&126.149&-4.15432&2012&M4.0\:II\,--\,III&0.396&0.081&0.321\\
PER091&1:16:45.00&+63:28:27.0&125.754&0.74118&2011&M0.0\:Ib&0.765&1.0&1.0\\
PER092&1:17:48.10&+64:13:39.7&125.793&1.50252&2012&K5.0\:Ib\,--\,II&0.558&1.0&0.951\\
PER093&1:18:13.80&+57:48:11.3&126.508&-4.88246&2012&M4.0\:II&0.394&0.144&0.173\\
PER094&1:18:14.00&+57:48:11.0&126.508&-4.88250&2011&M5.0\:II&0.381&0.108&0.221\\
PER095&1:18:38.00&+58:02:12.8&126.537&-4.64438&2012&M1.0\:Ib\,--\,II&0.536&1.0&0.978\\
PER096&1:18:48.20&+59:46:41.4&126.374&-2.91057&2012&M4.0\:II&0.292&0.138&0.155\\
PER097&1:18:52.70&+58:09:30.9&126.556&-4.51994&2012&C star&--&--&--\\
PER098&1:20:01.40&+57:31:29.1&126.777&-5.13341&2012&M4.5\:II&0.297&0.02&0.091\\
PER099&1:21:09.10&+56:32:02.3&127.044&-6.10037&2012&M7.0\:II&0.211&0.0&0.001\\
PER100&1:21:55.00&+61:20:55.0&126.578&-1.30718&2011&M7.0\:III&0.085&0.0&0.04\\
PER101&1:22:04.10&+66:50:12.3&125.944&4.14433&2012&S star&0.347&0.072&0.451\\
PER102&1:22:56.30&+61:10:34.6&126.721&-1.46353&2012&K4.0\:Iab&0.826&1.0&1.0\\
PER103&1:23:01.80&+61:59:40.7&126.632&-0.64996&2012&M1.5\:Ib&0.876&1.0&1.0\\
PER104&1:24:25.20&+57:11:53.4&127.408&-5.38668&2012&M0.5\:Ib&0.643&1.0&1.0\\
PER105&1:25:02.10&+61:04:41.0&126.984&-1.52907&2012&M9.0\:III&0.066&0.0&0.0\\
PER106&1:25:09.40&+58:49:18.7&127.294&-3.76402&2012&M5.0\:Ib\,--\,II&0.384&0.078&0.217\\
PER107&1:25:10.40&+60:52:38.0&127.027&-1.72601&2012&M1.0\:Iab&0.867&1.0&1.0\\
PER108&1:25:22.80&+57:38:11.7&127.479&-4.93518&2012&M3.0\:II&0.383&0.825&0.324\\
PER109&1:25:58.30&+63:29:32.3&126.774&0.87846&2012&M3.0\:Iab&0.867&1.0&1.0\\
PER110&1:26:43.20&+62:52:31.7&126.941&0.27868&2012&M0.5\:Iab&0.796&1.0&1.0\\
PER111&1:29:20.60&+61:45:41.8&127.400&-0.78085&2012&M2.0\:Ib&0.764&1.0&0.947\\
PER112&1:29:47.90&+58:47:19.3&127.895&-3.71291&2012&C star&--&--&--\\
PER113&1:31:34.30&+59:57:48.1&127.940&-2.51758&2012&M0.0\:Ib&0.783&1.0&1.0\\
PER114&1:32:00.20&+62:19:44.5&127.622&-0.17273&2012&M2.0\:Iab&0.867&1.0&1.0\\
PER115&1:33:32.60&+57:45:05.5&128.545&-4.66064&2012&C star&--&--&--\\
PER116&1:33:33.10&+61:33:29.6&127.925&-0.90458&2012&M0.5\:Iab&0.869&1.0&1.0\\
PER117&1:34:07.20&+65:11:18.5&127.393&2.68709&2012&K5.0\:Iab&0.748&1.0&0.984\\
PER118&1:34:48.50&+65:47:51.8&127.362&3.29973&2012&M5.0\:III&0.381&0.087&0.237\\
PER119&1:34:52.20&+62:46:28.6&127.876&0.32052&2012&K5.0\:Iab\,--\,Ib&0.758&1.0&1.0\\
PER120&1:37:52.40&+62:47:48.8&128.210&0.40199&2012&M3.0\:Iab&0.922&1.0&1.0\\
PER121&1:38:03.50&+61:02:49.2&128.546&-1.31558&2012&K5.0\:Iab\,--\,Ib&0.752&1.0&1.0\\
PER122&1:38:35.90&+60:49:25.7&128.651&-1.52323&2012&K5.0\:Ib&0.607&1.0&0.998\\
PER123&1:39:19.00&+60:39:38.0&128.767&-1.66763&2011&M4.0\:II&0.381&0.745&0.459\\
PER124&1:39:46.80&+59:42:29.2&129.000&-2.59284&2012&S star&0.591&0.969&0.726\\
PER125&1:39:51.60&+60:54:08.1&128.787&-1.41783&2012&M2.0\:Iab&0.798&1.0&1.0\\
PER126&1:41:01.00&+61:31:01.0&128.808&-0.78790&2011&M4.0\:III&0.37&0.066&0.439\\
PER127&1:42:16.40&+61:25:16.4&128.974&-0.85278&2012&K5.0\:Iab\,--\,Ib&0.782&1.0&1.0\\
PER128&1:44:38.30&+61:37:43.0&129.208&-0.59314&2012&M0.0\:Iab\,--\,Ib&0.734&1.0&1.0\\
PER129&1:44:49.70&+57:42:01.4&130.039&-4.43273&2012&M4.5\:II&0.297&0.653&0.184\\
PER130&1:45:38.70&+61:02:22.7&129.448&-1.14445&2012&M0.0\:Iab&0.841&1.0&1.0\\
PER131&1:47:44.00&+62:06:36.0&129.463&-0.04573&2011&M6.0\:III&0.338&0.0&0.045\\
PER132&1:47:46.00&+63:50:22.0&129.093&1.64351&2012&S star&0.089&0.001&0.0\\
PER133&1:51:40.00&+61:20:59.0&130.088&-0.68246&2011&M2.0\:Ib&0.795&1.0&1.0\\
PER134&1:54:01.20&+64:39:50.9&129.566&2.60149&2012&M3.0\:II\,--\,III&0.382&0.911&0.272\\
PER135&1:55:53.20&+64:16:56.1&129.854&2.28013&2012&M4.0\:II&0.381&0.774&0.398\\
PER136&1:56:35.80&+62:04:13.0&130.481&0.15737&2012&M1.0\:Ib&0.794&1.0&1.0\\
PER137&1:56:41.00&+57:01:04.0&131.757&-4.73185&2011&M5.0\:II&0.452&0.106&0.224\\
PER138&1:56:45.40&+60:49:03.8&130.813&-1.05065&2012&M1.0\:Ib&0.653&1.0&1.0\\
PER139&1:57:40.10&+60:13:07.9&131.072&-1.60191&2012&K4.0\:Ib&0.629&1.0&1.0\\
PER140&1:58:14.50&+59:37:01.1&131.295&-2.16562&2012&M3.0\:II\,--\,III&0.426&0.765&0.122\\
PER141&1:58:18.50&+64:03:55.8&130.164&2.13674&2012&M5.0\:III&0.198&0.0&0.074\\
PER142&1:58:56.60&+61:00:04.0&131.023&-0.80569&2012&M0.5\:Iab&0.803&1.0&1.0\\
PER143&2:00:09.30&+55:45:14.1&132.550&-5.82933&2012&M5.5\:II&0.279&0.034&0.09\\
PER144&2:00:57.00&+58:36:58.0&131.893&-3.04103&2011&M7.5\:III&0.18&0.0&0.026\\
PER145&2:01:26.60&+64:08:37.8&130.474&2.30203&2012&M3.5\:Iab&0.805&1.0&0.997\\
PER146&2:02:42.00&+58:04:53.0&132.259&-3.49410&2011&M4.0\:III&0.185&0.0&0.021\\
PER147&2:03:08.20&+62:11:24.1&131.188&0.47303&2012&M0.0\:Ib&0.776&1.0&1.0\\
PER148&2:05:05.90&+58:16:23.1&132.510&-3.22228&2012&M9.0\:III&0.075&0.0&0.0\\
PER149&2:08:15.70&+59:15:56.0&132.617&-2.15386&2012&M7.0\:Iab&0.352&0.0&0.029\\
PER150&2:08:54.10&+58:42:28.6&132.860&-2.66254&2012&M7.5\:III&0.213&0.0&0.028\\
PER151&2:14:53.30&+66:29:56.6&131.127&4.96425&2012&M0.0\:Ib\,--\,II&0.508&1.0&0.965\\
PER152&2:16:19.50&+64:52:17.4&131.791&3.46990&2012&M3.0\:II&0.375&0.993&0.253\\
PER153&2:19:34.00&+58:23:57.0&134.279&-2.51757&2011&M7.5\:III&0.144&0.0&0.0\\
PER154&2:19:47.00&+58:38:48.0&134.223&-2.27487&2011&K0.0\:II&0.697&1.0&1.0\\
PER155&2:21:00.00&+57:09:30.0&134.878&-3.62148&2011&M1.5\:Iab&0.855&1.0&1.0\\
PER156&2:22:24.20&+57:06:34.0&135.074&-3.60227&2012&M4.0\:Iab&0.671&0.604&0.556\\
PER157&2:22:51.70&+58:35:11.2&134.621&-2.19506&2012&M3.0\:Ia&0.51&0.0&0.004\\
PER158&2:23:39.00&+61:24:58.0&133.731&0.49281&2011&M3.0\:Ib\,--\,II&0.798&1.0&1.0\\
PER159&2:24:29.80&+55:41:06.3&135.844&-4.83619&2012&M5.0\:III&0.319&0.053&0.204\\
PER160&2:24:41.10&+59:57:47.2&134.358&-0.82371&2012&M4.0\:II&0.501&0.987&0.669\\
PER161&2:26:02.70&+65:13:51.6&132.634&4.15678&2012&K0.0\:Ib&0.605&1.0&1.0\\
PER162&2:27:49.00&+57:05:52.0&135.767&-3.35271&2011&M5.0\:II&0.427&0.022&0.131\\
PER163&2:28:02.70&+59:46:10.0&134.821&-0.85346&2012&M2.0\:II&0.501&0.995&0.286\\
PER164&2:28:13.00&+58:37:09.0&135.261&-1.91656&2011&M7.0\:III&0.371&0.0&0.005\\
PER165&2:29:14.00&+61:25:53.7&134.348&0.74659&2012&M3.5\:II&0.323&0.444&0.351\\
PER166&2:29:51.00&+59:58:58.0&134.954&-0.57185&2011&M1.5\:Iab&0.802&1.0&1.0\\
PER167&2:30:27.50&+62:31:45.6&134.075&1.81948&2012&M7.0\:III&0.177&0.0&0.0\\
PER168&2:31:04.00&+56:50:26.0&136.274&-3.42795&2011&M0.0\:Ib&0.78&1.0&1.0\\
PER169&2:32:27.90&+54:18:12.9&137.421&-5.70136&2012&M7.5\:II&0.177&0.0&0.002\\
PER170&2:35:44.60&+65:08:58.7&133.610&4.46314&2012&M9.5\:III&0.017&0.0&0.0\\
PER171&2:36:11.30&+60:22:41.3&135.531&0.09430&2012&M7.0\:III&0.232&0.0&0.006\\
PER172&2:37:33.20&+54:27:48.0&138.044&-5.26506&2012&M4.0\:II&0.29&0.119&0.098\\
PER173&2:38:43.00&+55:45:59.4&137.672&-4.00450&2012&C star&--&--&--\\
PER174&2:39:22.90&+60:42:40.0&135.759&0.55608&2012&M7.0\:III&0.182&0.0&0.012\\
PER175&2:41:04.90&+62:17:31.9&135.300&2.08286&2012&M8.0\:II&0.114&0.0&0.001\\
PER176&2:41:07.80&+55:12:59.8&138.210&-4.36778&2012&M8.0\:III&0.081&0.0&0.0\\
PER177&2:42:39.20&+66:35:04.9&133.675&6.06114&2012&M4.5\:II&0.288&0.316&0.079\\
PER178&2:42:56.90&+60:12:16.2&136.367&0.27429&2012&K5.0\:Iab&0.735&1.0&1.0\\
PER179&2:43:45.10&+60:25:25.0&136.366&0.51492&2012&C star&--&--&--\\
PER180&2:44:19.00&+60:55:55.6&136.215&1.00546&2012&M7.0\:II\,--\,III&0.269&0.0&0.078\\
PER181&2:44:30.30&+65:42:52.7&134.213&5.35040&2012&M3.0\:Ib\,--\,II&0.611&0.97&0.552\\
PER182&2:45:12.20&+58:05:24.5&137.518&-1.52304&2012&M0.0\:Iab&0.751&1.0&1.0\\
PER183&2:45:39.10&+59:17:34.4&137.060&-0.40942&2012&M4.0\:II&0.406&0.416&0.231\\
PER184&2:46:00.70&+58:45:20.1&137.331&-0.87579&2012&M5.0\:Ib&0.351&0.169&0.364\\
PER185&2:46:21.10&+53:09:46.7&139.773&-5.90966&2012&K2.0\:II&0.504&1.0&0.841\\
PER186&2:46:23.00&+64:19:44.0&134.986&4.18130&2011&M8.0\:Ib\,--\,II&0.108&0.0&0.165\\
PER187&2:46:31.40&+59:35:23.9&137.033&-0.09347&2012&K2.0\:Ib&0.627&1.0&1.0\\
PER188&2:46:40.00&+63:00:19.8&135.583&2.99950&2012&M6.0\:III&0.338&0.001&0.093\\
PER189&2:47:52.50&+64:45:17.0&134.946&4.63461&2012&M3.0\:II&0.504&0.989&0.68\\
PER190&2:49:07.00&+60:13:11.0&137.054&0.61494&2011&M2.0\:Ib&0.697&1.0&0.999\\
PER191&2:49:54.00&+61:02:09.0&136.782&1.39051&2011&M3.0\:Ib&0.917&1.0&1.0\\
PER192&2:50:14.00&+62:25:14.0&136.207&2.65093&2011&M3.5\:Ib&0.892&1.0&1.0\\
PER193&2:50:39.50&+58:53:08.4&137.816&-0.49729&2012&C star&--&--&--\\
PER194&2:50:57.00&+60:44:27.0&137.027&1.18225&2011&M6.0\:Ib\,--\,II&0.401&0.084&0.177\\
PER195&2:52:42.00&+58:42:49.5&138.129&-0.53329&2012&M5.0\:Ib\,--\,II&0.237&0.0&0.091\\
PER196&2:52:52.90&+54:24:34.2&140.091&-4.36578&2012&C star&--&--&--\\
PER197&2:54:19.00&+59:29:14.1&137.965&0.25050&2012&C star&--&--&--\\
PER198&2:55:30.00&+60:13:59.0&137.756&0.98184&2011&M2.0\:Iab&0.802&1.0&1.0\\
PER199&2:56:19.60&+58:52:18.8&138.475&-0.17861&2012&M0.5\:II&0.524&1.0&0.981\\
PER200&2:56:53.20&+57:33:24.9&139.149&-1.31053&2012&M2.0\:Iab&0.841&1.0&1.0\\
PER201&2:59:17.80&+51:50:24.5&142.141&-6.19853&2012&M6.5\:III&0.213&0.006&0.175\\
PER202&21:26:36.8&+59:08:42.4&99.230&6.05462&2012&M8.0\:III&0.151&0.0&0.01\\
PER203&21:37:03.0&+54:55:40.8&97.384&2.00469&2012&M4.5\:II\,--\,III&0.479&0.439&0.453\\
PER204&21:40:39.0&+54:19:28.7&97.374&1.20759&2012&M7.0\:II\,--\,III&0.325&0.0&0.011\\
PER205&21:41:08.1&+58:15:56.8&100.023&4.12973&2012&M0.0\:II&0.648&0.995&0.71\\
PER206&21:42:08.8&+54:58:02.1&97.959&1.55022&2012&M4.5\:III&0.423&0.207&0.357\\
PER207&21:42:16.0&+54:38:43.7&97.762&1.29558&2012&C star&--&--&--\\
PER208&21:44:04.3&+53:42:11.6&97.348&0.40877&2012&K2.0\:Ia&0.78&1.0&1.0\\
PER209&21:47:17.2&+54:21:13.8&98.129&0.60149&2012&M0.0\:II&0.552&1.0&0.571\\
PER210&21:48:39.6&+52:54:07.0&97.358&-0.64564&2012&M4.0\:Iab&0.576&0.623&0.343\\
PER211&21:49:30.2&+53:22:11.5&97.753&-0.36488&2012&M3.0\:II&0.401&0.999&0.416\\
PER212&21:50:00.6&+53:23:41.5&97.827&-0.39344&2012&M5.0\:III&0.308&0.301&0.237\\
PER213&21:50:18.2&+52:38:15.6&97.382&-1.00762&2012&C star&--&--&--\\
PER214&21:51:30.3&+54:44:27.0&98.847&0.51209&2012&M6.0\:III&0.129&0.0&0.0\\
PER215&21:51:31.0&+54:49:10.8&98.898&0.57231&2012&M0.5\:II&0.541&1.0&0.659\\
PER216&21:51:38.0&+54:45:38.0&98.874&0.51578&2011&M6.0\:Ib\,--\,II&0.37&0.093&0.279\\
PER217&21:51:47.5&+55:08:10.5&99.127&0.79368&2012&M5.0\:III&0.393&0.03&0.181\\
PER218&21:51:48.3&+55:05:12.1&99.098&0.75393&2012&C star&--&--&--\\
PER219&21:52:12.1&+54:49:46.2&98.981&0.51799&2012&M2.0\:Ib\,--\,II&0.593&1.0&0.739\\
PER220&21:52:19.9&+52:53:14.6&97.778&-1.00687&2012&M2.0\:II&0.471&0.991&0.633\\
PER221&21:52:36.0&+55:58:38.0&99.744&1.37687&2011&M6.0\:Ib\,--\,II&0.5&0.001&0.057\\
PER222&21:52:44.9&+55:17:36.2&99.332&0.83052&2012&M4.0\:II&0.401&0.875&0.499\\
PER223&21:52:59.3&+52:54:56.0&97.873&-1.04690&2012&M2.0\:II\,--\,III&0.463&0.999&0.391\\
PER224&21:53:03.6&+62:02:14.6&103.599&6.06086&2012&M5.0\:III&0.341&0.029&0.193\\
PER225&21:53:41.3&+59:17:33.0&101.933&3.87295&2012&M5.0\:III&0.494&0.169&0.234\\
PER226&21:53:53.4&+53:18:45.6&98.226&-0.82097&2012&M0.5\:II&0.498&1.0&0.852\\
PER227&21:54:14.5&+52:40:50.6&97.874&-1.34862&2012&M4.5\:II&0.437&0.319&0.325\\
PER228&21:54:16.9&+58:33:20.8&101.533&3.24922&2012&M2.0\:II&0.523&1.0&0.207\\
PER229&21:54:24.6&+54:15:52.5&98.878&-0.12324&2012&M4.5\:II\,--\,III&0.325&0.395&0.317\\
PER230&21:54:28.0&+53:43:51.8&98.553&-0.54658&2012&M4.5\:II&0.3&0.178&0.258\\
PER231&21:54:30.8&+56:04:00.3&100.009&1.28014&2012&K3.0\:II&0.443&0.982&0.694\\
PER232&21:55:05.2&+54:29:01.5&99.091&-0.01242&2012&C star&--&--&--\\
PER233&21:55:07.9&+52:04:10.6&97.602&-1.91211&2012&M7.0\:II\,--\,III&0.315&0.0&0.051\\
PER234&21:55:23.2&+53:34:57.2&98.568&-0.74760&2012&M1.0\:Ib&0.858&1.0&1.0\\
PER235&21:55:40.0&+53:58:03.0&98.839&-0.47054&2011&K4.0\:Iab\,--\,Ib&0.873&1.0&1.0\\
PER236&21:55:43.0&+52:31:09.0&97.950&-1.61421&2011&M4.0\:II&0.299&0.289&0.33\\
PER237&21:55:49.0&+54:00:05.4&98.877&-0.45741&2012&M7.5\:III&0.185&0.0&0.003\\
PER238&21:55:54.7&+53:59:04.7&98.878&-0.47929&2012&K4.0\:Ib&0.683&1.0&0.997\\
PER239&21:56:09.8&+59:30:33.1&102.316&3.84717&2012&M3.5\:II&0.443&0.649&0.371\\
PER240&21:56:27.7&+53:46:19.4&98.811&-0.69667&2012&M4.0\:II&0.405&0.081&0.037\\
PER241&21:57:04.2&+59:06:17.7&102.157&3.45747&2012&M5.0\:Ib\,--\,II&0.426&0.01&0.145\\
PER242&21:57:50.2&+54:53:58.9&99.661&0.07092&2012&C star&--&--&--\\
PER243&21:57:52.9&+52:52:39.4&98.429&-1.53309&2012&M7.0\:Ib&0.431&0.007&0.093\\
PER244&21:58:15.9&+55:07:05.2&99.843&0.20633&2012&M1.0\:Ib\,--\,II&0.656&1.0&0.929\\
PER245&21:58:50.2&+52:00:58.8&98.019&-2.30413&2012&M5.5\:III&0.319&0.002&0.085\\
PER246&21:59:28.3&+56:14:17.8&100.659&0.99161&2012&M2.0\:II&0.605&0.996&0.723\\
PER247&21:59:40.7&+53:10:29.2&98.825&-1.46193&2012&M4.5\:II&0.485&0.378&0.345\\
PER248&22:00:03.1&+61:16:25.3&103.780&4.94768&2012&K2.0\:Ib&0.671&1.0&1.0\\
PER249&22:00:10.0&+52:54:16.0&98.719&-1.72147&2011&M6.0\:Ib\,--\,II&0.39&0.026&0.114\\
PER250&22:00:45.7&+54:54:47.0&100.003&-0.17381&2012&M2.0\:Ib\,--\,II&0.447&1.0&0.664\\
PER251&22:01:00.8&+51:17:20.6&97.846&-3.08719&2012&M0.5\:Ib&0.614&1.0&0.899\\
PER252&22:01:04.6&+54:29:16.1&99.783&-0.54055&2012&M6.0\:III&0.316&0.011&0.12\\
PER253&22:02:53.1&+56:50:09.9&101.395&1.18558&2012&M4.0\:II&0.465&0.32&0.313\\
PER254&22:02:55.7&+54:14:00.5&99.846&-0.90583&2012&M3.0\:II&0.521&0.971&0.572\\
PER255&22:03:04.9&+51:39:29.4&98.326&-2.98537&2012&M5.5\:III&0.268&0.001&0.083\\
PER256&22:03:12.1&+55:06:56.8&100.404&-0.22134&2012&M8.0\:III&0.13&0.0&0.016\\
PER257&22:05:09.0&+63:04:48.2&105.348&6.03446&2012&M5.5\:III&0.455&0.025&0.142\\
PER258&22:05:28.4&+62:30:10.3&105.034&5.54783&2012&M7.0\:III&0.189&0.0&0.04\\
PER259&22:05:32.6&+63:02:37.9&105.363&5.97889&2012&M5.0\:III&0.29&0.175&0.153\\
PER260&22:05:37.9&+53:55:59.9&99.987&-1.38265&2012&M7.5\:Ib\,--\,II&0.373&0.001&0.04\\
PER261&22:05:48.5&+58:45:40.0&102.851&2.50429&2012&M0.5\:II&0.585&0.998&0.143\\
PER262&22:05:49.5&+53:37:44.3&99.831&-1.64540&2012&M5.0\:III&0.459&0.156&0.189\\
PER263&22:06:10.7&+51:49:35.1&98.812&-3.13410&2012&M4.0\:II&0.342&0.675&0.323\\
PER264&22:06:17.0&+55:00:17.4&100.694&-0.57187&2012&M7.0\:II\,--\,III&0.404&0.0&0.024\\
PER265&22:06:17.1&+52:59:16.3&99.509&-2.20427&2012&M7.5\:Ib\,--\,II&0.258&0.0&0.015\\
PER266&22:06:20.9&+55:53:33.7&101.223&0.14141&2012&K3.0\:Ib&0.567&1.0&1.0\\
PER267&22:06:21.0&+59:39:38.1&103.438&3.19065&2012&M2.0\:Iab&0.852&1.0&1.0\\
PER268&22:06:36.5&+55:29:55.9&101.022&-0.19902&2012&M3.0\:Ib\,--\,II&0.589&1.0&0.724\\
PER269&22:06:37.7&+59:41:20.2&103.483&3.19296&2012&M4.0\:Ib&0.782&0.636&0.786\\
PER270&22:07:03.7&+57:22:58.0&102.176&1.29127&2012&M4.0\:III&0.318&0.703&0.249\\
PER271&22:07:30.0&+53:04:02.3&99.704&-2.24693&2012&M3.0\:II&0.445&0.998&0.358\\
PER272&22:08:12.0&+56:30:47.0&101.795&0.49434&2011&M6.0\:II&0.301&0.0&0.049\\
PER273&22:08:12.7&+55:55:03.7&101.450&0.00910&2012&M3.0\:II&0.529&0.956&0.393\\
PER274&22:08:38.4&+59:33:01.3&103.608&2.93190&2012&M1.0\:Ia&0.885&1.0&1.0\\
PER275&22:08:41.9&+51:45:14.1&99.086&-3.42081&2012&M6.0\:III&0.327&0.001&0.038\\
PER276&22:08:42.3&+53:26:04.9&100.065&-2.05323&2012&M7.0\:II\,--\,III&0.231&0.0&0.009\\
PER277&22:09:33.2&+60:53:54.2&104.484&3.96419&2012&M4.0\:II&0.4&0.25&0.195\\
PER278&22:09:37.0&+52:09:49.0&99.440&-3.16903&2012&M5.0\:III&0.37&0.031&0.08\\
PER279&22:09:43.7&+51:26:13.9&99.033&-3.77161&2012&M4.0\:II&0.332&0.558&0.127\\
PER280&22:09:49.0&+52:05:06.7&99.419&-3.25076&2012&M5.0\:III&0.274&0.022&0.047\\
PER281&22:10:20.0&+56:11:49.0&101.853&0.06556&2011&M1.5\:Ib&0.77&1.0&1.0\\
PER282&22:10:22.0&+56:28:54.0&102.020&0.29578&2011&M8.0\:III&0.078&0.0&0.0\\
PER283&22:11:00.5&+55:05:33.2&101.296&-0.89319&2012&M1.0\:Ib&0.673&1.0&1.0\\
PER284&22:11:11.0&+55:16:54.0&101.425&-0.75253&2011&M8.0\:III&0.022&0.0&0.0\\
PER285&22:11:25.0&+56:34:34.0&102.193&0.29008&2011&M2.0\:Ib&0.739&1.0&1.0\\
PER286&22:11:35.6&+55:16:04.4&101.465&-0.79725&2012&M2.0\:Ib&0.782&1.0&0.999\\
PER287&22:12:33.0&+57:17:03.0&102.724&0.78342&2011&M2.0\:Iab&0.848&1.0&1.0\\
PER288&22:12:53.8&+54:12:29.6&101.016&-1.77523&2012&M3.5\:III&0.263&0.672&0.263\\
PER289&22:13:00.0&+57:34:59.0&102.944&0.99498&2011&M3.0\:II\,--\,III&0.294&0.652&0.232\\
PER290&22:13:09.2&+54:37:15.7&101.281&-1.45655&2012&M6.5\:III&0.24&0.0&0.095\\
PER291&22:13:41.6&+54:19:35.4&101.179&-1.74372&2012&M2.0\:III&0.545&0.796&0.215\\
PER292&22:13:45.7&+52:19:21.4&100.052&-3.40175&2012&M3.0\:Ib\,--\,II&0.453&0.978&0.393\\
PER293&22:13:53.8&+55:22:01.2&101.791&-0.90188&2012&M1.0\:Iab&0.618&1.0&0.919\\
PER294&22:14:05.8&+55:04:28.2&101.650&-1.15942&2012&M4.5\:II&0.364&0.283&0.269\\
PER295&22:14:15.4&+59:22:00.8&104.089&2.37290&2012&M4.0\:II\,--\,III&0.284&0.19&0.283\\
PER296&22:14:29.4&+55:50:14.8&102.126&-0.56062&2012&M1.0\:Ib&0.579&1.0&1.0\\
PER297&22:14:33.9&+55:13:51.7&101.793&-1.06781&2012&M4.0\:II&0.348&0.346&0.235\\
PER298&22:14:43.2&+56:20:22.4&102.435&-0.16356&2012&M2.0\:Ib\,--\,II&0.479&0.997&0.589\\
PER299&22:15:17.6&+54:11:52.3&101.299&-1.98160&2012&M6.0\:III&0.314&0.001&0.089\\
PER300&22:15:42.0&+57:53:06.0&103.412&1.04139&2011&M0.0\:Iab&0.812&1.0&1.0\\
PER301&22:16:08.0&+52:29:47.4&100.450&-3.46146&2012&M4.5\:II\,--\,III&0.357&0.944&0.244\\
PER302&22:16:20.9&+53:34:32.9&101.080&-2.58418&2012&M8.0\:III&0.188&0.0&0.001\\
PER303&22:16:24.5&+57:23:59.5&103.219&0.58598&2012&K4.0\:Iab\,--\,Ib&0.792&1.0&1.0\\
PER304&22:16:25.4&+54:01:18.4&101.338&-2.22018&2012&C star&--&--&--\\
PER305&22:16:55.4&+50:05:23.0&99.204&-5.52681&2012&M5.0\:Ib\,--\,II&0.433&0.026&0.15\\
PER306&22:17:16.5&+56:47:15.4&102.977&0.01160&2012&M4.0\:II\,--\,III&0.396&0.671&0.268\\
PER307&22:18:31.1&+58:39:40.9&104.151&1.48141&2012&M1.5\:II&0.49&0.998&0.623\\
PER308&22:19:27.6&+51:09:43.9&100.137&-4.85686&2012&M5.0\:III&0.397&0.006&0.103\\
PER309&22:19:40.6&+53:09:45.9&101.266&-3.20287&2012&M2.0\:II&0.378&0.971&0.212\\
PER310&22:19:41.3&+50:33:20.0&99.832&-5.38342&2012&M5.0\:Ib\,--\,II&0.383&0.062&0.198\\
PER311&22:19:45.7&+54:42:16.3&102.121&-1.91968&2012&M7.0\:II&0.289&0.002&0.062\\
PER312&22:19:55.5&+53:39:45.2&101.571&-2.80491&2012&M5.5\:III&0.362&0.067&0.143\\
PER313&22:19:56.5&+50:39:18.2&99.921&-5.32234&2012&M3.0\:II&0.269&1.0&0.563\\
PER314&22:20:10.8&+56:02:24.2&102.901&-0.83367&2012&M3.5\:Ib&0.786&1.0&0.987\\
PER315&22:20:22.0&+54:15:09.6&101.948&-2.34614&2012&M7.5\:II\,--\,III&0.225&0.0&0.003\\
PER316&22:20:37.7&+55:42:59.0&102.777&-1.13931&2012&M4.0\:III&0.432&0.463&0.173\\
PER317&22:20:49.0&+52:50:49.6&101.237&-3.56093&2012&M8.0\:II\,--\,III&0.218&0.0&0.013\\
PER318&22:21:08.8&+54:27:06.4&102.152&-2.24097&2012&M4.0\:II\,--\,III&0.353&0.778&0.365\\
PER319&22:21:11.3&+51:19:54.2&100.457&-4.86296&2012&M8.0\:II&0.155&0.0&0.001\\
PER320&22:21:29.1&+60:43:11.6&105.591&3.00055&2012&C star&--&--&--\\
PER321&22:21:59.0&+55:18:03.0&102.713&-1.59247&2011&M6.0\:II\,--\,III&0.352&0.0&0.018\\
PER322&22:22:26.9&+57:11:32.2&103.788&-0.03470&2012&M10.0\:III&0.081&0.0&0.0\\
PER323&22:22:30.1&+55:10:55.4&102.711&-1.73234&2012&M5.0\:III&0.305&0.188&0.255\\
PER324&22:22:34.0&+57:15:04.0&103.833&0.00625&2011&M2.0\:Iab&0.808&1.0&1.0\\
PER325&22:23:13.6&+56:08:53.3&103.317&-0.97265&2012&M4.0\:Iab&0.581&0.944&0.411\\
PER326&22:23:34.1&+57:39:03.0&104.161&0.27119&2012&M4.0\:II&0.363&0.709&0.19\\
PER327&22:23:54.4&+58:14:28.6&104.515&0.74614&2012&K3.0\:Ib&0.72&1.0&0.999\\
PER328&22:24:06.0&+57:33:37.8&104.173&0.15679&2012&M6.0\:II&0.366&0.004&0.073\\
PER329&22:24:56.5&+58:39:03.8&104.848&1.02084&2012&M2.0\:Iab&0.774&1.0&1.0\\
PER330&22:25:04.8&+58:36:13.2&104.838&0.97112&2012&M3.0\:Ib&0.642&1.0&0.966\\
PER331&22:25:15.8&+57:10:15.3&104.099&-0.25655&2012&K5.0\:Ib&0.736&1.0&1.0\\
PER332&22:25:34.9&+54:38:50.3&102.800&-2.42081&2012&M3.5\:II&0.423&0.873&0.254\\
PER333&22:25:55.6&+56:38:52.5&103.899&-0.74830&2012&M0.0\:Iab&0.764&1.0&1.0\\
PER334&22:26:28.5&+58:15:41.0&104.812&0.58406&2012&M3.0\:II&0.4&0.963&0.079\\
PER335&22:26:28.8&+58:42:27.8&105.048&0.96327&2012&M5.0\:III&0.431&0.07&0.174\\
PER336&22:26:40.0&+58:31:34.0&104.973&0.79602&2011&M7.0\:III&0.187&0.002&0.147\\
PER337&22:26:52.7&+60:09:54.4&105.857&2.17639&2012&K3.0\:Ib&0.687&1.0&0.999\\
PER338&22:27:05.3&+53:29:22.0&102.377&-3.52117&2012&M2.0\:Ib&0.661&1.0&0.965\\
PER339&22:27:22.6&+57:15:59.0&104.392&-0.32604&2012&M9.0\:III&0.069&0.0&0.0\\
PER340&22:27:29.4&+59:26:01.7&105.539&1.51329&2012&M1.0\:Iab&0.899&1.0&1.0\\
PER341&22:27:55.9&+52:49:15.6&102.133&-4.15672&2012&M2.0\:III&0.474&0.948&0.229\\
PER342&22:28:13.4&+57:42:44.2&104.722&-0.00510&2012&M3.0\:II&0.486&0.997&0.56\\
PER343&22:28:17.4&+59:14:04.1&105.522&1.29002&2012&M3.0\:Iab&0.913&1.0&1.0\\
PER344&22:28:53.6&+58:01:07.3&104.958&0.21041&2012&M4.5\:II&0.387&0.184&0.193\\
PER345&22:29:11.2&+57:12:46.7&104.574&-0.49910&2012&M1.0\:Ib\,--\,II&0.56&1.0&0.905\\
PER346&22:29:26.7&+59:06:24.4&105.582&1.10409&2012&M5.0\:III&0.334&0.051&0.285\\
PER347&22:29:37.8&+59:30:15.6&105.808&1.43228&2012&M2.0\:Iab&0.899&1.0&1.0\\
PER348&22:29:54.1&+56:27:25.5&104.267&-1.19684&2012&M4.0\:II&0.334&0.172&0.209\\
PER349&22:30:02.4&+57:03:12.8&104.591&-0.69539&2012&M4.5\:II&0.424&0.311&0.375\\
PER350&22:30:27.3&+57:21:51.0&104.799&-0.45790&2012&M2.0\:Ib&0.671&1.0&0.941\\
PER351&22:30:41.7&+54:53:51.5&103.560&-2.59168&2012&M2.0\:Ib\,--\,II&0.563&1.0&0.864\\
PER352&22:31:28.9&+64:41:50.6&108.672&5.77415&2012&M3.0\:II&0.478&0.999&0.531\\
PER353&22:31:41.0&+59:00:43.6&105.781&0.87516&2012&M4.0\:II&0.413&0.261&0.221\\
PER354&22:31:50.1&+56:59:48.9&104.771&-0.86912&2012&M6.0\:II&0.286&0.002&0.026\\
PER355&22:32:25.5&+59:32:58.4&106.136&1.28973&2012&C star&--&--&--\\
PER356&22:32:26.8&+58:37:05.8&105.666&0.48580&2012&C star&--&--&--\\
PER357&22:32:31.0&+59:36:23.4&106.175&1.33291&2012&M3.0\:Iab&0.906&1.0&1.0\\
PER358&22:33:00.0&+57:38:04.3&105.230&-0.39948&2012&M2.0\:Ib\,--\,II&0.376&1.0&0.745\\
PER359&22:33:01.2&+55:16:14.9&104.037&-2.44016&2012&M3.0\:III&0.354&0.674&0.194\\
PER360&22:33:05.0&+58:28:42.0&105.666&0.32307&2011&K1.0\:Ib&0.596&1.0&0.959\\
PER361&22:33:23.8&+55:27:41.0&104.180&-2.30273&2012&M4.0\:II&0.489&0.891&0.423\\
PER362&22:33:34.6&+58:53:47.0&105.933&0.65170&2012&M3.5\:Ib&0.825&1.0&1.0\\
PER363&22:33:46.0&+58:16:59.3&105.645&0.10935&2012&M7.5\:II\,--\,III&0.234&0.0&0.0\\
PER364&22:33:49.3&+57:29:48.8&105.256&-0.57389&2012&M3.0\:III&0.29&0.301&0.204\\
PER365&22:34:09.3&+58:59:26.1&106.045&0.69563&2012&K5.0\:Iab&0.853&1.0&1.0\\
PER366&22:34:10.4&+56:59:27.4&105.043&-1.03526&2012&M5.0\:II&0.286&0.003&0.056\\
PER367&22:34:58.0&+55:55:53.0&104.607&-2.00732&2011&M6.0\:Ib\,--\,II&0.416&0.003&0.113\\
PER368&22:35:15.3&+58:17:13.3&105.817&0.01478&2012&M1.0\:II&0.588&0.99&0.272\\
PER369&22:35:29.7&+56:19:56.5&104.871&-1.69645&2012&S star&0.62&1.0&0.287\\
PER370&22:35:40.5&+55:29:25.1&104.474&-2.43944&2012&M5.0\:II\,--\,III&0.529&0.255&0.224\\
PER371&22:35:54.4&+58:39:28.6&106.075&0.29424&2012&M5.0\:II\,--\,III&0.346&0.202&0.083\\
PER372&22:36:13.0&+52:59:04.1&103.295&-4.65318&2012&M9.0\:III&0.029&0.0&0.008\\
PER373&22:36:51.8&+52:37:05.0&103.198&-5.01984&2012&M9.0\:III&0.025&0.0&0.0\\
PER374&22:37:35.9&+61:16:09.1&107.548&2.46072&2012&M5.0\:Ib&0.579&0.002&0.104\\
PER375&22:37:44.2&+60:22:22.5&107.123&1.67154&2012&M2.0\:Iab&0.877&1.0&1.0\\
PER376&22:38:04.4&+55:36:27.2&104.827&-2.50485&2012&M0.5\:Ib\,--\,II&0.558&1.0&0.969\\
PER377&22:38:24.7&+58:28:54.8&106.273&-0.01973&2012&C star&--&--&--\\
PER378&22:38:51.3&+56:26:58.2&105.334&-1.82339&2012&M0.5\:Ib&0.688&1.0&1.0\\
PER379&22:38:51.8&+54:21:07.7&104.312&-3.65559&2012&M6.0\:III&0.26&0.0&0.004\\
PER380&22:39:20.6&+60:10:28.8&107.200&1.40132&2012&M5.0\:III&0.309&0.092&0.125\\
PER381&22:39:42.9&+55:30:38.6&104.983&-2.70246&2012&M2.0\:II&0.617&1.0&0.763\\
PER382&22:39:45.3&+57:36:52.8&106.007&-0.86404&2012&M5.0\:III&0.278&0.111&0.18\\
PER383&22:39:58.5&+57:20:19.1&105.899&-1.11997&2012&M1.0\:II&0.2&0.0&0.0\\
PER384&22:40:10.8&+59:57:18.1&107.185&1.15860&2012&M1.0\:II&0.546&0.29&0.562\\
PER385&22:40:12.1&+59:24:55.6&106.927&0.68461&2012&M7.5\:III&0.276&0.0&0.02\\
PER386&22:41:00.0&+58:44:20.1&106.692&0.04214&2012&M6.0\:III&0.314&0.0&0.02\\
PER387&22:41:11.7&+59:27:59.0&107.063&0.66860&2012&K2.0\:Ib&0.73&1.0&0.994\\
PER388&22:41:27.0&+56:39:09.0&105.745&-1.81818&2011&M8.0\:III&0.052&0.0&0.02\\
PER389&22:42:02.8&+58:04:05.4&106.492&-0.61257&2012&M3.5\:II&0.471&0.933&0.442\\
PER390&22:42:20.0&+56:47:52.0&105.921&-1.74828&2011&M2.0\:Iab\,--\,Ib&0.862&1.0&1.0\\
PER391&22:43:03.7&+59:44:03.4&107.398&0.79180&2012&M0.0\:II&0.469&1.0&0.32\\
PER392&22:43:14.0&+59:45:09.0&107.425&0.79764&2011&M5.0\:Ib&0.864&0.937&0.904\\
PER393&22:44:09.7&+54:58:18.9&105.283&-3.47840&2012&M7.0\:II&0.312&0.0&0.006\\
PER394&22:45:04.2&+56:37:18.6&106.170&-2.08036&2012&C star&--&--&--\\
PER395&22:45:23.4&+55:27:27.5&105.666&-3.13114&2012&M3.0\:II&0.484&1.0&0.681\\
PER396&22:45:32.3&+55:12:47.9&105.571&-3.35718&2012&M9.0\:III&0.09&0.0&0.0\\
PER397&22:45:44.3&+58:57:02.6&107.332&-0.06029&2012&M5.0\:III&0.389&0.012&0.059\\
PER398&22:45:46.9&+60:35:20.4&108.097&1.38842&2012&K5.0\:Ib&0.766&1.0&1.0\\
PER399&22:47:26.6&+58:18:31.6&107.233&-0.73222&2012&M0.0\:II\,--\,III&0.589&1.0&0.09\\
PER400&22:47:46.1&+55:18:13.0&105.895&-3.42380&2012&C star&--&--&--\\
PER401&22:48:58.1&+55:38:57.0&106.204&-3.19380&2012&M1.0\:Ib\,--\,II&0.539&1.0&0.9\\
PER402&22:49:03.6&+58:52:04.0&107.675&-0.33062&2012&M3.0\:Ib\,--\,II&0.533&0.807&0.386\\
PER403&22:49:10.4&+59:18:12.9&107.886&0.05136&2012&M2.0\:Iab&0.865&1.0&0.997\\
PER404&22:49:33.1&+58:58:09.6&107.778&-0.26873&2012&M2.0\:II&0.545&0.987&0.695\\
PER405&22:49:38.9&+57:33:31.7&107.153&-1.53324&2012&M4.0\:II&0.415&0.762&0.485\\
PER406&22:49:59.0&+60:17:57.0&108.426&0.89385&2011&K2.0\:Ia&0.846&1.0&1.0\\
PER407&22:50:06.9&+57:31:37.6&107.194&-1.58973&2012&C star&--&--&--\\
PER408&22:50:08.3&+55:36:48.6&106.335&-3.30026&2012&M7.5\:Ib\,--\,II&0.336&0.002&0.103\\
PER409&22:50:22.9&+57:29:08.6&107.208&-1.64280&2012&K3.0\:Ib&0.675&1.0&1.0\\
PER410&22:50:50.9&+53:21:14.8&105.409&-5.36490&2012&M4.5\:II\,--\,III&0.297&0.248&0.239\\
PER411&22:50:53.1&+61:45:57.8&109.182&2.15594&2012&M4.0\:Ib&0.792&0.98&0.938\\
PER412&22:51:02.1&+55:45:19.9&106.512&-3.23001&2012&M0.0\:Ib\,--\,II&0.594&1.0&0.987\\
PER413&22:51:04.1&+57:57:37.0&107.502&-1.25945&2012&M3.5\:II&0.418&0.72&0.37\\
PER414&22:51:17.9&+57:25:58.4&107.294&-1.74518&2012&M1.5\:II&0.465&0.999&0.766\\
PER415&22:51:29.7&+52:23:52.1&105.066&-6.26351&2012&M0.5\:II&0.466&1.0&0.633\\
PER416&22:51:34.5&+58:00:10.3&107.581&-1.25124&2012&M3.5\:II&0.289&0.436&0.234\\
PER417&22:51:59.1&+56:55:44.2&107.154&-2.23795&2012&M2.0\:II&0.419&1.0&0.724\\
PER418&22:51:59.3&+63:22:56.1&110.016&3.54623&2012&M2.0\:II&0.43&0.975&0.639\\
PER419&22:53:12.3&+61:17:00.3&109.216&1.60074&2012&M3.0\:Ib&0.837&1.0&1.0\\
PER420&22:53:18.1&+58:58:33.7&108.214&-0.47779&2012&K2.0\:Ib&0.694&1.0&1.0\\
PER421&22:54:01.2&+60:47:41.7&109.090&1.11835&2012&M1.5\:Iab&0.792&1.0&0.999\\
PER422&22:54:16.0&+60:49:28.9&109.130&1.13204&2012&M3.0\:Iab\,--\,Ib&0.77&1.0&0.987\\
PER423&22:54:30.4&+60:47:50.5&109.145&1.09469&2012&M0.5\:Ib&0.81&1.0&1.0\\
PER424&22:54:31.2&+57:25:58.2&107.684&-1.93610&2012&M7.0\:Ib\,--\,II&0.267&0.0&0.066\\
PER425&22:54:45.1&+60:46:42.2&109.164&1.06462&2012&C star&--&--&--\\
PER426&22:55:38.6&+55:50:43.9&107.135&-3.43365&2012&M3.5\:II\,--\,III&0.296&0.391&0.377\\
PER427&22:56:07.3&+54:13:45.5&106.498&-4.92150&2012&C star&--&--&--\\
PER428&22:56:36.8&+61:31:08.1&109.685&1.63618&2012&M4.0\:Iab&0.69&0.97&0.82\\
PER429&22:57:00.3&+57:39:59.6&108.085&-1.86870&2012&M7.5\:III&0.183&0.0&0.094\\
PER430&22:57:04.9&+57:40:43.6&108.100&-1.86202&2012&K3.0\:Ib&0.736&1.0&1.0\\
PER431&22:57:16.3&+58:17:16.8&108.382&-1.32148&2012&M0.0\:II&0.526&0.992&0.972\\
PER432&22:57:20.2&+56:22:04.2&107.573&-3.06293&2012&C star&--&--&--\\
PER433&22:57:40.9&+58:49:12.7&108.657&-0.86235&2012&M2.0\:II\,--\,III&0.433&0.0&0.489\\
PER434&22:58:16.0&+56:58:33.0&107.946&-2.56642&2011&M5.0\:Ib\,--\,II&0.341&0.147&0.27\\
PER435&22:59:50.9&+66:21:19.2&112.058&5.86840&2012&M4.0\:II&0.339&0.162&0.262\\
PER436&23:00:17.4&+56:08:49.3&107.852&-3.43576&2012&M9.0\:III&0.121&0.0&0.004\\
PER437&23:01:03.2&+56:53:33.2&108.257&-2.80079&2012&M0.5\:Ib\,--\,II&0.493&1.0&0.976\\
PER438&23:01:04.2&+56:58:30.1&108.294&-2.72661&2012&M3.0\:Ib\,--\,II&0.433&0.853&0.534\\
PER439&23:01:07.0&+61:02:52.0&109.979&0.98146&2011&M2.0\:Iab&0.8&1.0&1.0\\
PER440&23:02:24.0&+58:14:12.4&108.974&-1.64900&2012&K2.0\:Ib&0.589&1.0&0.998\\
PER441&23:02:59.5&+64:12:44.3&111.472&3.78070&2012&M1.5\:Ib\,--\,II&0.483&0.985&0.401\\
PER442&23:04:31.7&+64:08:44.3&111.598&3.65215&2012&M2.0\:Ib&0.84&1.0&1.0\\
PER443&23:04:38.3&+58:34:41.4&109.380&-1.45511&2012&M7.5\:Iab&0.294&0.0&0.004\\
PER444&23:04:49.6&+56:32:57.6&108.591&-3.32473&2012&M7.5\:II&0.269&0.0&0.04\\
PER445&23:06:07.2&+59:25:05.1&109.889&-0.76018&2012&M2.0\:Ib\,--\,II&0.439&1.0&0.547\\
PER446&23:06:27.2&+60:53:05.6&110.506&0.57152&2012&M1.0\:Iab&0.802&1.0&1.0\\
PER447&23:06:54.8&+60:33:26.0&110.429&0.24810&2012&M0.5\:Iab&0.901&1.0&1.0\\
PER448&23:06:56.4&+60:54:16.4&110.568&0.56637&2012&M3.0\:II&0.429&0.967&0.404\\
PER449&23:08:04.7&+55:42:56.6&108.679&-4.26942&2012&M5.5\:III&0.402&0.006&0.05\\
PER450&23:08:39.8&+58:18:09.9&109.757&-1.91535&2012&M7.0\:Ib\,--\,II&0.327&0.001&0.053\\
PER451&23:10:43.5&+64:28:52.3&112.349&3.69901&2012&M5.0\:III&0.349&0.076&0.169\\
PER452&23:10:47.7&+64:32:46.5&112.380&3.75635&2012&M3.0\:II\,--\,III&0.421&0.52&0.177\\
PER453&23:11:06.0&+57:16:17.8&109.663&-2.99233&2012&M5.0\:III&0.38&0.017&0.061\\
PER454&23:12:22.5&+57:04:44.2&109.751&-3.23579&2012&M4.0\:II&0.361&0.684&0.281\\
PER455&23:12:30.3&+59:58:22.5&110.846&-0.55680&2012&M6.0\:II\,--\,III&0.474&0.079&0.153\\
PER456&23:12:42.7&+63:56:10.0&112.344&3.11267&2012&M6.0\:III&0.264&0.006&0.099\\
PER457&23:12:56.1&+59:08:13.5&110.586&-1.35301&2012&M7.5\:III&0.268&0.001&0.091\\
PER458&23:13:13.3&+56:36:12.4&109.681&-3.72023&2012&M8.0\:II&0.178&0.0&0.0\\
PER459&23:13:25.8&+59:36:36.0&110.820&-0.93699&2012&M0.0\:Ib&0.772&1.0&1.0\\
PER460&23:14:37.7&+60:55:16.8&111.439&0.22831&2012&M1.5\:Iab&0.866&1.0&1.0\\
PER461&23:14:42.6&+64:40:05.2&112.816&3.71320&2012&M8.0\:III&0.174&0.0&0.043\\
PER462&23:15:03.6&+59:31:13.7&110.979&-1.09591&2012&M1.5\:Iab&0.802&1.0&1.0\\
PER463&23:15:26.1&+57:27:05.0&110.273&-3.04170&2012&M7.5\:II&0.128&0.0&0.211\\
PER464&23:16:02.0&+62:21:19.0&112.114&1.50429&2011&M5.0\:Ib&0.865&0.998&1.0\\
PER465&23:16:04.5&+57:01:56.9&110.202&-3.46359&2012&M4.0\:II&0.346&0.038&0.212\\
PER466&23:16:29.2&+60:57:45.2&111.664&0.18515&2012&M3.0\:Iab&0.854&1.0&1.0\\
PER467&23:16:47.3&+59:12:31.4&111.072&-1.46605&2012&M3.0\:II&0.364&0.989&0.514\\
PER468&23:17:29.1&+58:40:57.9&110.968&-1.98954&2012&M1.0\:Iab\,--\,Ib&0.754&1.0&0.998\\
PER469&23:17:58.4&+62:24:20.4&112.342&1.47120&2012&M4.0\:II&0.322&0.84&0.498\\
PER470&23:18:19.2&+60:16:22.0&111.630&-0.53952&2012&M1.0\:Iab\,--\,Ib&0.77&1.0&1.0\\
PER471&23:18:30.4&+58:33:10.8&111.047&-2.15790&2012&M1.0\:Ia&0.866&1.0&1.0\\
PER472&23:18:39.5&+61:53:13.9&112.235&0.95773&2012&K3.0\:Ib&0.658&1.0&1.0\\
PER473&23:18:47.9&+58:07:41.3&110.934&-2.56918&2012&M8.0\:III&0.231&0.0&0.029\\
PER474&23:19:26.7&+58:02:24.2&110.983&-2.68156&2012&M3.5\:Ib&0.568&0.106&0.494\\
PER475&23:19:52.4&+60:47:40.5&111.991&-0.11719&2012&M5.0\:III&0.481&0.105&0.223\\
PER476&23:22:30.7&+59:18:26.0&111.792&-1.62756&2012&M1.0\:Ia&0.895&1.0&1.0\\
PER477&23:23:28.1&+56:10:01.0&110.857&-4.62860&2012&C star&--&--&--\\
PER478&23:23:39.8&+60:20:00.6&112.272&-0.70893&2012&M1.5\:Iab&0.8&1.0&1.0\\
PER479&23:23:58.4&+55:39:28.6&110.754&-5.13205&2012&S star&0.528&0.997&0.294\\
PER480&23:24:30.0&+62:14:48.0&113.001&1.06306&2011&M6.0\:Ib\,--\,II&0.295&0.012&0.081\\
PER481&23:24:44.8&+61:20:38.4&112.731&0.20110&2012&K3.0\:Ib&0.639&1.0&0.999\\
PER482&23:24:57.2&+62:18:50.8&113.073&1.10942&2012&M4.0\:II&0.376&0.239&0.326\\
PER483&23:25:09.0&+61:22:01.0&112.784&0.20692&2011&M0.5\:Iab&0.791&1.0&1.0\\
PER484&23:25:33.0&+57:49:43.5&111.676&-3.15371&2012&M3.0\:Ib\,--\,II&0.435&0.929&0.404\\
PER485&23:26:43.5&+60:23:08.9&112.647&-0.78332&2012&M0.0\:Iab&0.766&1.0&1.0\\
PER486&23:27:38.9&+61:17:27.7&113.043&0.03837&2012&M4.0\:II\,--\,III&0.373&0.904&0.339\\
PER487&23:27:51.4&+62:45:37.3&113.533&1.42374&2012&M7.5\:III&0.192&0.0&0.002\\
PER488&23:28:17.8&+57:28:56.9&111.913&-3.59967&2012&C star&--&--&--\\
PER489&23:29:13.1&+56:39:33.1&111.772&-4.42050&2012&M7.5\:Iab&0.349&0.0&0.009\\
PER490&23:29:29.9&+58:57:11.5&112.526&-2.25306&2012&M5.0\:II&0.374&0.02&0.101\\
PER491&23:30:11.0&+60:16:45.0&113.020&-1.01944&2011&M5.0\:Ib&0.764&0.626&0.458\\
PER492&23:30:44.1&+60:15:20.5&113.078&-1.06284&2012&M4.0\:Ib&0.872&0.974&0.878\\
PER493&23:30:53.0&+62:07:22.0&113.667&0.70878&2011&M0.0\:Iab&0.812&1.0&1.0\\
PER494&23:32:03.1&+59:23:15.1&112.971&-1.94015&2012&M1.0\:Ib\,--\,II&0.552&1.0&0.956\\
PER495&23:32:16.4&+61:58:08.3&113.776&0.51274&2012&M0.0\:Iab&0.755&1.0&0.998\\
PER496&23:32:20.8&+62:06:32.2&113.826&0.64363&2012&C star&--&--&--\\
PER497&23:33:46.5&+61:32:22.6&113.817&0.04992&2012&M1.0\:Ib\,--\,II&0.539&1.0&1.0\\
PER498&23:34:21.0&+58:53:05.4&113.102&-2.50786&2012&M9.0\:III&0.08&0.0&0.0\\
PER499&23:35:02.3&+58:34:16.0&113.096&-2.83399&2012&M3.0\:Iab&0.897&1.0&1.0\\
PER500&23:35:27.4&+59:16:18.5&113.351&-2.17925&2012&M1.0\:Iab\,--\,Ib&0.803&1.0&1.0\\
PER501&23:35:46.5&+61:07:47.3&113.927&-0.41208&2012&M2.0\:II&0.485&0.91&0.265\\
PER502&23:35:50.4&+58:44:19.0&113.244&-2.70393&2012&M7.0\:II&0.226&0.0&0.009\\
PER503&23:37:20.4&+61:50:14.4&114.308&0.21262&2012&M6.0\:II&0.461&0.006&0.048\\
PER504&23:37:31.2&+59:42:13.5&113.726&-1.84029&2012&K0.0\:Ib&0.715&1.0&1.0\\
PER505&23:38:09.8&+56:01:46.9&112.772&-5.38983&2012&C star&--&--&--\\
PER506&23:38:23.7&+61:54:20.6&114.446&0.24333&2012&M7.5\:II\,--\,III&0.242&0.0&0.009\\
PER507&23:39:19.0&+60:13:02.0&114.085&-1.40990&2011&M2.0\:Iab&0.861&1.0&1.0\\
PER508&23:39:35.1&+59:35:09.8&113.944&-2.02597&2012&C star&--&--&--\\
PER509&23:40:40.6&+65:35:00.5&115.701&3.71005&2012&M3.0\:II\,--\,III&0.383&0.871&0.251\\
PER510&23:42:12.3&+65:39:09.1&115.871&3.73428&2012&M2.0\:Ib\,--\,II&0.706&1.0&0.899\\
PER511&23:43:06.0&+60:02:47.0&114.493&-1.70096&2011&M8.0\:III&0.053&0.0&0.0\\
PER512&23:43:06.8&+57:52:49.9&113.927&-3.79176&2012&M5.0\:II&0.42&0.029&0.143\\
PER513&23:44:25.0&+59:40:08.0&114.555&-2.10840&2011&M7.0\:III&0.2&0.0&0.002\\
PER514&23:45:05.0&+60:26:51.0&114.835&-1.37687&2011&M4.0\:II&0.422&0.987&0.56\\
PER515&23:45:36.9&+62:20:56.3&115.378&0.44658&2012&M3.0\:Ib&0.817&1.0&1.0\\
PER516&23:45:52.0&+56:55:58.7&114.043&-4.80277&2012&M5.0\:III&0.374&0.024&0.168\\
PER517&23:46:07.0&+60:27:54.0&114.962&-1.39206&2011&M6.5\:II&0.339&0.064&0.159\\
PER518&23:46:11.0&+62:40:05.0&115.521&0.73897&2011&M1.0\:Iab\,--\,Ib&0.826&1.0&1.0\\
PER519&23:46:48.1&+60:05:11.6&114.950&-1.77973&2012&M1.0\:Iab&0.784&1.0&1.0\\
PER520&23:47:12.4&+58:54:13.3&114.708&-2.93860&2012&M7.5\:II\,--\,III&0.112&0.0&0.003\\
PER521&23:47:21.0&+58:13:16.0&114.557&-3.60476&2011&M6.0\:II&0.407&0.003&0.066\\
PER522&23:47:25.0&+62:24:38.0&115.595&0.45428&2011&M2.0\:II&0.446&0.976&0.296\\
PER523&23:47:41.0&+60:42:37.0&115.209&-1.20199&2011&M5.0\:II&0.477&0.009&0.066\\
PER524&23:47:46.0&+57:42:07.0&114.483&-4.12161&2011&M2.0\:II&0.508&1.0&0.644\\
PER525&23:48:24.0&+58:50:27.0&114.842&-3.03723&2011&M6.0\:III&0.323&0.0&0.002\\
PER526&23:49:19.0&+58:12:00.0&114.804&-3.68804&2011&M5.0\:Ib\,--\,II&0.287&0.073&0.205\\
PER527&23:49:38.2&+56:39:25.5&114.477&-5.19673&2012&M1.5\:II&0.422&1.0&0.008\\
PER528&23:50:12.0&+61:06:16.0&115.601&-0.89267&2011&M3.0\:Ib&0.826&1.0&1.0\\
PER529&23:50:43.0&+61:52:33.0&115.841&-0.15713&2011&M2.0\:Iab&0.832&1.0&1.0\\
PER530&23:50:54.3&+65:38:36.9&116.740&3.50222&2012&M2.0\:II&0.496&1.0&0.424\\
PER531&23:51:15.1&+56:50:40.9&114.737&-5.06640&2012&M3.0\:II&0.484&0.957&0.149\\
PER532&23:51:29.0&+62:16:34.0&116.022&0.21148&2011&K3.0\:Iab&0.823&1.0&1.0\\
PER533&23:51:32.0&+57:37:40.0&114.956&-4.31336&2011&M7.0\:III&0.173&0.0&0.008\\
PER534&23:52:04.9&+61:48:12.4&115.982&-0.26469&2012&M9.5\:III&0.018&0.0&0.0\\
PER535&23:56:40.0&+62:11:23.0&116.591&-0.00586&2011&M2.0\:Ib&0.805&1.0&1.0\\
PER536&23:56:44.4&+58:49:01.2&115.889&-3.30480&2012&M7.5\:II&0.281&0.0&0.026\\
PER537&23:56:49.7&+66:05:07.9&117.429&3.79892&2012&M1.0\:Ib&0.742&1.0&1.0\\
PER538&23:57:21.2&+58:25:04.0&115.884&-3.71189&2012&S star&0.187&0.112&0.294\\
PER539&23:57:44.1&+56:56:46.1&115.627&-5.16196&2012&M4.5\:II\,--\,III&0.386&0.012&0.075\\
PER540&23:58:38.0&+60:53:42.0&116.553&-1.32103&2011&M8.0\:III&0.061&0.0&0.0\\
PER541&23:59:05.6&+56:58:15.0&115.814&-5.17573&2012&C star&--&--&--\\
PER542&3:00:50.60&+58:09:13.7&139.333&-0.53638&2012&M1.0\:II&0.536&0.998&0.508\\
PER543&3:00:50.70&+58:56:29.7&138.958&0.15643&2012&M5.0\:Ib\,--\,II&0.295&0.019&0.204\\
PER544&3:03:35.20&+57:52:01.3&139.789&-0.61316&2012&M5.0\:III&0.502&0.453&0.29\\
PER545&3:05:10.40&+59:54:00.3&138.977&1.26157&2012&M6.0\:II\,--\,III&0.368&0.09&0.143\\
PER546&3:06:45.00&+55:10:14.3&141.493&-2.74873&2012&M6.5\:III&0.28&0.003&0.118\\
PER547&3:08:44.40&+58:04:37.7&140.280&-0.09321&2012&K3.0\:Ib&0.763&1.0&1.0\\
PER548&3:10:40.80&+54:13:19.9&142.462&-3.28343&2012&M4.5\:II&0.545&0.307&0.34\\
PER549&3:11:12.20&+64:06:57.8&137.474&5.26437&2012&M1.0\:Iab&0.809&1.0&1.0\\
PER550&3:15:03.00&+56:30:30.1&141.815&-1.00436&2012&M1.0\:Ib\,--\,II&0.549&1.0&0.801\\
PER551&3:16:30.90&+59:56:00.3&140.184&2.01333&2012&M3.0\:II&0.459&0.969&0.454\\
PER552&3:16:40.80&+58:23:53.2&141.011&0.71959&2012&C star&--&--&--\\
PER553&3:18:55.10&+54:57:32.8&143.096&-2.03228&2012&M7.0\:II&0.26&0.001&0.063\\
PER554&3:19:07.30&+50:20:12.7&145.605&-5.91882&2012&M3.5\:II\,--\,III&0.424&0.671&0.223\\
PER555&3:21:16.70&+54:08:28.5&143.825&-2.53752&2012&K2.0\:Ib\,--\,II&0.668&1.0&0.999\\
PER556&3:21:40.60&+52:43:33.6&144.645&-3.69401&2012&M6.0\:III&0.244&0.019&0.072\\
PER557&3:21:59.70&+51:20:29.6&145.442&-4.82854&2012&M4.5\:II&0.402&0.23&0.173\\
PER558&3:22:49.30&+56:01:10.6&142.989&-0.84121&2012&C star&--&--&--\\
PER559&3:24:13.20&+56:13:33.0&143.039&-0.56191&2012&M4.0\:II&0.377&0.725&0.382\\
PER560&3:24:38.70&+58:22:25.5&141.903&1.26216&2012&M2.0\:Ib&0.821&1.0&1.0\\
PER561&3:28:01.00&+63:49:14.9&139.195&6.00936&2012&M5.0\:II&0.298&0.001&0.022\\
PER562&3:28:08.50&+57:19:26.2&142.874&0.64942&2012&M1.0\:Ib&0.702&1.0&1.0\\
PER563&3:31:22.50&+49:00:58.6&147.999&-5.91299&2012&M5.0\:III&0.488&0.225&0.423\\
PER564&3:31:35.10&+59:33:27.3&141.977&2.74399&2012&M6.5\:II&0.329&0.001&0.132\\
PER565&3:32:55.80&+52:44:13.7&146.051&-2.72989&2012&C star&--&--&--\\
PER566&3:34:29.50&+51:40:36.1&146.862&-3.45532&2012&M6.0\:III&0.287&0.0&0.044\\
PER567&3:39:42.50&+52:08:15.0&147.241&-2.60878&2012&C star&--&--&--\\
PER568&3:39:50.40&+50:16:43.7&148.370&-4.08685&2012&K5.0\:Ib&0.744&1.0&1.0\\
PER569&3:39:50.70&+51:06:30.3&147.873&-3.42143&2012&M3.0\:Ib&0.571&1.0&0.999\\
PER570&3:40:25.10&+60:46:51.7&142.149&4.37905&2012&M6.0\:III&0.336&0.001&0.14\\
PER571&3:41:48.10&+62:38:54.2&141.150&5.96777&2012&C star&--&--&--\\
PER572&3:41:56.00&+53:57:10.9&146.419&-0.95553&2012&M4.0\:Ib\,--\,II&0.437&0.23&0.367\\
PER573&3:42:49.50&+60:54:08.5&142.310&4.65275&2012&M7.0\:II&0.304&0.0&0.04\\
PER574&3:45:14.30&+55:55:42.9&145.596&0.90206&2012&M0.0\:Iab\,--\,Ib&0.827&1.0&1.0\\
PER575&3:45:27.40&+52:10:59.2&147.916&-2.03909&2012&M3.0\:II&0.416&0.959&0.708\\
PER576&3:45:55.70&+60:11:46.6&143.046&4.32734&2012&M6.0\:III&0.374&0.007&0.179\\
PER577&3:47:06.70&+53:03:53.0&147.571&-1.18956&2012&M2.0\:Ib&0.624&0.999&0.878\\
PER578&3:47:07.60&+52:40:41.5&147.812&-1.49219&2012&M3.5\:Ib&0.761&1.0&0.96\\
PER579&3:47:23.50&+54:40:50.0&146.605&0.10657&2012&M1.0\:II&0.486&1.0&0.542\\
PER580&3:47:43.20&+60:06:32.7&143.275&4.39664&2012&M4.5\:II&0.285&0.0&0.151\\
PER581&3:48:40.10&+58:17:11.9&144.506&3.04463&2012&M7.0\:II&0.253&0.0&0.017\\
PER582&3:51:34.00&+56:15:19.0&146.085&1.70624&2011&M4.5\:Ib\,--\,II&0.394&0.523&0.541\\
PER583&3:54:06.40&+60:21:00.1&143.741&5.08321&2012&M1.5\:Ib\,--\,II&0.506&1.0&0.985\\
PER584&3:58:36.00&+55:41:57.0&147.197&1.90796&2011&M6.5\:III&0.223&0.0&0.054\\
PER585&3:58:38.00&+55:14:27.0&147.499&1.56234&2011&M4.5\:Ib\,--\,II&0.52&0.155&0.13\\
PER586&4:03:44.00&+56:34:47.0&147.159&3.04159&2011&M4.0\:III&0.461&0.743&0.309\\
PER587&4:04:21.00&+55:04:20.7&148.225&1.97146&2012&M2.0\:Iab&0.855&1.0&1.0\\
PER588&4:04:27.80&+55:55:26.9&147.671&2.61865&2012&M5.0\:II&0.26&0.0&0.021\\
PER589&4:06:41.10&+58:40:53.2&146.048&4.87007&2012&M4.5\:II\,--\,III&0.344&0.009&0.203\\
PER590&4:07:11.60&+55:12:33.3&148.437&2.34465&2012&K4.0\:Ib&0.741&1.0&1.0\\
PER591&4:07:45.40&+60:12:58.5&145.109&6.09537&2012&M5.0\:III&0.442&0.297&0.307\\
PER592&4:10:28.60&+57:51:04.8&146.980&4.59653&2012&M3.0\:II&0.396&0.989&0.502\\
PER593&4:11:26.00&+57:22:29.0&147.400&4.33569&2011&M3.5\:II&0.312&0.807&0.687\\
PER594&4:16:54.30&+57:14:28.0&148.028&4.74939&2012&M0.5\:Ib\,--\,II&0.535&1.0&0.991\\
\noalign{\smallskip}
\hline
\end{longtable}

\bsp	
\label{lastpage}
\end{document}